\documentclass[a4paper,twoside,12pt]{article}

\usepackage{amsfonts}

\newcommand{\ZZ}{\mathbb{Z}}

\newcommand{\RR}{\mathbb{R}}
\newcommand{\CC}{\mathbb{C}}

\newcommand{\cgc}[6]{\left[
\begin{array}{ccc}#1&#2&#3 \\ #4&#5&#6\end{array}\right]}

\newcommand{\sixj}[6]{\left\{
\begin{array}{ccc}#1&#2&#5 \\ #3&#4&#6\end{array}\right\}}

\newcommand{\ttp}{\hat{\otimes}}

\newcommand{\shalf}{{\textstyle\frac{1}{2}}}

\newcommand{\ket}[1]{|\,#1\,\rangle}

\newcommand{\qnr}[1]{\lfloor #1 \rfloor_{q}}
\newcommand{\qbin}[2]{\left\lfloor  \begin{array}{c}  #1\\#2
                      \end{array}\right\rfloor_{q}} 

\newcommand{\yodi}[3]{\begin{picture}(#2,2) 
		      \setlength{\unitlength}{#1}
		      \multiput(0,1)(1,0){#2}{\framebox(1,1){}}
		      \multiput(0,0)(1,0){#3}{\framebox(1,1){}}
		      \end{picture}}
\newcommand{\thrryodi}[4]{\begin{picture}(#2,2) 
		      \setlength{\unitlength}{#1}
		      \multiput(0,2)(1,0){#2}{\framebox(1,1){}}
		      \multiput(0,1)(1,0){#3}{\framebox(1,1){}}
		      \multiput(0,0)(1,0){#4}{\framebox(1,1){}}
		      \end{picture}}
\newcommand{\yodit}[3]{\begin{picture}(2,#2) 
		      \setlength{\unitlength}{#1}
		      \multiput(0,#2)(0,-1){#2}{\framebox(1,1){}}
		      \multiput(1,#2)(0,-1){#3}{\framebox(1,1){}}
		      \end{picture}}

\begin{document}

\begin{flushright} {\bf ITFA-01-12} \vspace*{7mm}\end{flushright}

\begin{center}
{\LARGE \bf Quantum groups and nonabelian braiding in
quantum Hall systems} \\[2ex]
{\Large J.K. Slingerland \footnotemark[1], F.A. Bais \footnotemark[2]} \\[1ex]
{\em Institute for Theoretical Physics, University of Amsterdam,\\  
     Valckenierstraat 65, 1068XE Amsterdam,}\\
April 2, 2001 
\end{center}
\footnotetext[1]{e-mail: {\tt slinger@science.uva.nl}}
\footnotetext[2]{e-mail: {\tt bais@science.uva.nl}} 
\begin{abstract}
Wave functions describing quasiholes and electrons
in nonabelian quantum Hall states are well known to correspond to
conformal blocks of certain coset conformal field theories. In this
paper we explicitly analyse the algebraic structure underlying the
braiding properties of these conformal blocks. We treat the electrons
and the quasihole excitations as localised particles carrying charges
related to a quantum group that is determined explicitly for the cases
of interest. The quantum group description naturally allows one to
analyse the braid group representations carried by the multi-particle
wave functions.  As an application, we construct the nonabelian braid
group representations which govern the exchange of quasiholes in the
fractional quantum Hall effect states that have been proposed by
N. Read and E. Rezayi \cite{readrez}, recovering the results of
C. Nayak and F. Wilczek \cite{naywil} for the Pfaffian state as a
special case.
\end{abstract}

\section{Introduction}
 In a $2+1$ dimensional setting, quantum mechanics leaves room for
particles with exchange properties other than those of bosons and
fermions and the exchanges of such $n$ such particles are governed by a
representation of the braid group $B_{n}$. These representations may
be abelian, or, more excitingly, nonabelian. Quasihole excitations of
fractional quantum Hall plateaus have already provided us with
examples of the former possibility and may possibly reveal the latter
as well. Several (series of) candidate nonabelian states have been
proposed in the literature. Examples are the HR-state \cite{haldrez},
the Pfaffian state \cite{moorread}, the spin singlet states of Ardonne
and Schoutens \cite{eddyenkjs} and the parafermionic generalisations
of the Pfaffian proposed by Read and Rezayi \cite{readrez}.  It is the
last series of states that we will focus on in this paper, although
the methods we use will also be applicable to the other cases. It has been
suggested that the Read-Rezayi states should give a good description
of quantum Hall plateaus which occur at several filling fractions
\cite{gewewi91,gewewi92,readrez}. In particular, the Pfaffian is
thought to describe the plateau observed \cite{westge} at filling
fraction $\nu=\frac{5}{2}$. Numerical support for these claims has
been provided in \cite{morf,readrez,haldrez99}, where it was shown
that some of the RR-states (among which the Pfaffian state) have large
overlaps with the exact ground states for electrons with Coulomb
interactions at the same filling fractions. Many aspects of the
Read-Rezayi states have already been well-studied.  For example, one
may show (see \cite{moorread,readrez,gurrez}) that they are exact
ground states of certain ultra-local Hamiltonians with $k+1$-body
interactions, which gives hope that they will indeed represent new
universality classes of two dimensional physical systems. Also, their
zero modes have been counted and a suitable basis for the description
of the quasiholes has been obtained \cite{gurrez,kjs}. Finally, there
is recent work which explains how the RR-states may be obtained as
projections of abelian theories \cite{cageto}. Still, the braiding of
the quasiholes has been described explicitly only for the Pfaffian
state \cite{naywil}.  

One of the general aims of this paper is to analyse some of the
physical properties of Hall systems, not by studying the explicit form of
the wave functions but rather by exploiting the underlying algebraic
structure, which in turn derives from the associated conformal field
theories. This allows us for example to give an explicit description
of the braid group representations that govern the exchange properties
of the quasiholes for all of the RR-states. In order to do this, we
first describe the electrons and quasiholes of the RR-states as
particles that carry a representation of a certain quantum group. That
such a description is possible is a logical consequence of the well
known relation between quantum groups and conformal field theories and
in fact, we expect that a similar description is possible for all the
nonabelian quantum Hall states that have been proposed.  We believe
that the quantum group description of quantum Hall states will prove a
useful complement to the existing conformal field theory and wave
function methods, both technically, because it makes braiding
calculations much easier, and conceptually. The reason that braiding
calculations are so much simplified, is that the quantum group picture
allows one to deal with quasiholes and electrons without dealing with
their exact spatial coordinates. Exchanging two particles becomes a
purely algebraic operation, simple enough to be carried out explicitly
for large numbers of particles.

One of the useful features of the present paper is that it does bring
together a number of sophisticated physical and mathematical
ingredients, everybody knows to exist in principle, and integrates
them into a rather self-contained toolbox to analyse the problems at
hand.  The outline of the paper is as follows.  In section
\ref{cftsec}, we review the usual description of the Read-Rezayi
states in terms of conformal blocks of parafermionic conformal field
theories. We also count the number of independent states with a fixed
number of quasiholes in fixed positions. In section \ref{qgsec}, we
give the motivation for the use of quantum groups in the description
of particles with nonabelian braiding and provide the necessary
background. In particular we describe the braid group representations
that describe the exchanges in a system of localised particles with a
hidden quantum group symmetry. In section \ref{qgcftsec}, we recall
the connection between quantum groups and conformal field theories and
in particular, we obtain the quantum groups which can be used to
describe the braiding of the parafermion CFTs which are important for
the Read-Rezayi states. In section \ref{rrbraidsec} we describe the
RR-states as systems of point particles with a hidden quantum group
symmetry and give the explicit form of the associated braid group
representations. We also check that the results of Nayak and Wilczek
for the case of the Pfaffian are recovered. A discussion of the
results, including questions for future research can be found in
section \nolinebreak \ref{discosec}.

\section{The CFT description of the Read-Rezayi states}
\label{cftsec}

In this section, we will review the conformal field theoretic
description of the Read-Rezayi states. The definition of these states
can be found in subsection \ref{rrdefsec}, after the required
preliminaries on parafermionic conformal field theories in subsection
\ref{parafsec}. In subsection \ref{brattelisec}, we describe the
fusion of the Read-Rezayi quasiholes in terms of paths on a Bratteli
diagram and in the following subsection, we use this description to
calculate the dimensions of the braid group representations which
govern the exchanges of these quasiholes. The final subsection is
devoted to a description of the results of Nayak and Wilczek on the
braiding of the quasiholes of Pfaffian or Moore-Read state, which is
the simplest state in the Read-Rezayi series.

\subsection{The Parafermion CFT}
\label{parafsec}

Like many other (candidate) quantum Hall states, the RR-states can be
described using conformal blocks of a rational conformal field
theory. Here, the theory in question is the $\ZZ_k$ parafermionic
theory of Zamolodchikov and Fateev. \cite{fatzam,gepqiu} Before we 
write down any explicit expression for the RR-states, we 
recall some well known facts about this CFT. The $\ZZ_k$ parafermionic
CFT may be described completely in terms of a chiral algebra generated
by the modes of $k$ parafermionic currents (see \cite{fatzam,gepqiu}
and also \cite{jacmat} for some more recent work in this vein), but it
also has two different coset descriptions. The cosets involved are
$\widehat{sl(2)}_{k}/\widehat{U(1)}_{k}$ and
$\widehat{sl(k)}_{1}\times\widehat{sl(k)}_{1}/\widehat{sl(k)}_{2}.$
The first of these descriptions was used extensively already in
\cite{fatzam,gepqiu}, to determine fusion rules, characters and
partition functions for the parafermions. The treatment of the
parafermions in most of the literature on the RR-states has been
influenced by this description. The second coset was introduced by
Bais et al. in \cite{bais1, bais2} and used in \cite{grifnem} to construct a
Coulomb gas representation of the theory which led to alternative
character formulae \cite{nemes}. This coset description has recently
also been used in the work of Cappelli, Georgiev and Todorov on the
RR-states \cite{cageto}. In the rest of this section, we will give a
quick description of both pictures and indicate how they are
connected.

\subsubsection{The coset $\widehat{sl(2)}_{k}/\widehat{U(1)}_{k}$.}
\label{sl2cossec}

We start with the coset $\widehat{sl(2)}_{k}/\widehat{U(1)}_{k}$. For this coset, we
have more information about the coset primaries than usual. In
particular, one can decompose certain fields of the parent
$\widehat{sl(2)}$ WZW-theory as a product of a coset primary field and
a $U(1)$ primary field (see formula (\ref{parafwzw}) below). In order
to describe this decomposition, it is convenient to start with a
short description of the fields and fusion rules of the $\widehat{sl(2)}$
theory before moving on to the parafermions (for much more detail on
WZW-theories, see for example \cite{dsm}). Note that when we speak of
primary fields in the sequel, we will always mean chiral primary
fields.

Recall that the spectrum generating algebra of the
$\widehat{sl(2)}_{k}$ model is the affine Lie algebra
$\widehat{sl(2)}$ at level $k.$ The Virasoro algebra is embedded in
the enveloping algebra of the affine algebra through the Sugawara
construction.When discussing primary fields of the
$\widehat{sl(2)}_{k}$ model, we need to distinguish between primary
fields of the affine algebra (affine primaries) and primary fields of
the Virasoro algebra (Virasoro primaries). Each affine primary field
is necessarily also a Virasoro primary, but not vice versa. In fact,
one can always find infinitely many Virasoro primaries among the affine
descendants of an affine primary. 

Let us be more explicit. If $\theta$ is the highest root of a simple
Lie algebra $g$, then the affine primaries of the $\hat{g}_{k}$ model
are labelled by the dominant integral weights $\Lambda$ of $g$ for
which $(\Lambda,\theta)\le k$ For $g=sl(2)$ this just means
$0\le\Lambda\le k.$ Let us call these fields $G^{\Lambda}.$ the
conformal dimension $h_{\Lambda}$ of $G^{\Lambda}$ is given by

\begin{equation}
\label{wzwdim}
h_{\Lambda}=\frac{\Lambda(\Lambda+2)}{4(k+2)}
\end{equation}
The fusion rules of the $G^{\Lambda}$ are 

\begin{equation}
\label{wzwfus}
G^{\Lambda}{\times}G^{\Lambda'}=
\bigoplus_{\Lambda''=|\Lambda-\Lambda'|
}^{{\rm min}\{\Lambda+\Lambda',2k-\Lambda-\Lambda'\}}G^{\Lambda''},
\end{equation}
There is an affine descendant field of $G^{\Lambda}$ for each of the
states in the $\widehat{sl(2)}$ module with highest weight $\Lambda.$
Among these descendants, there are infinitely many Virasoro primaries,
which we may name $G^{\Lambda}_{\lambda}.$ The field
$G^{\Lambda}_{\lambda}$ is by definition the field of lowest conformal
dimension among the affine descendants of $G^{\Lambda}$ which carry
$sl(2)$-weight $\lambda.$ Naturally, we have
$G^{\Lambda}=G^{\Lambda}_{\Lambda}.$ Also, we have to demand that
$(\Lambda-\lambda)=0 ~({\rm mod ~} 2),$ otherwise the weight $\lambda$
will not appear in the representation with highest weight $\Lambda.$
One may check (see for instance \cite{dsm}) that all the
$G^{\Lambda}_{\lambda}.$ defined this way are indeed Virasoro
primary. Their conformal weights are given by

\begin{equation}
h^{\Lambda}_{\lambda}=\frac{\Lambda(\Lambda+2)}{4(k+2)} +n^{\Lambda}_{\lambda}
\end{equation}
where $n_{\Lambda,\lambda}$ is the lowest grade at which the weight
$\lambda$ appears in the affine Lie algebra representation of highest
weight $\Lambda.$ If $\lambda$ is a weight in the (ordinary) Lie
algebra representation of highest weight $\Lambda,$ then
$n_{\Lambda,\lambda}$ will be zero and we will have
$h_{\Lambda}=h^{\Lambda}_{\lambda}$.  The fusion rules of the
$G^{\Lambda}_{\lambda}$ are easily obtained from (\ref{wzwfus}) and the
sum rule for weights in operator products. They are

\begin{equation}
\label{wzwvirfus}
G^{\Lambda}_{\lambda}{\times}G^{\Lambda'}_{\lambda'}=
\bigoplus_{\Lambda''=|\Lambda-\Lambda'|
}^{{\rm min}\{\Lambda+\Lambda',2k-\Lambda-\Lambda'\}}
G^{\Lambda''}_{\lambda+\lambda'},
\end{equation}
Now let us turn to the $\ZZ_k$ parafermion theory, as described by the
coset $\widehat{sl(2)}_k/\widehat{U(1)}_{k}.$ As usual for cosets, the
Virasoro primary fields of the parafermion CFT may be labelled by a
highest weight $\Lambda$ of the horizontal algebra of the parent
theory ($sl(2)$) and by a similar weight $\mathcal{P}\lambda$ of the
embedded theory ($U(1)$), which is obtained by a projection matrix
$\mathcal{P}$ from a weight $\lambda$ of the parent theory. These
weights moreover have to satisfy a branching condition, which ensures
that the representation $\mathcal{P}\lambda$ of the embedded algebra
can occur as a summand in the decomposition of the representation
$\Lambda$ of the parent algebra into representations of the embedded
algebra. If we denote by $M$ the root lattice of the horizontal
algebra of the parent algebra, then this branching condition is

\begin{equation}
\label{branchcond}
\mathcal{P}\Lambda-\mathcal{P}\lambda\in\mathcal{P}M.
\end{equation}
In the case of $\widehat{sl(2)}_k/\widehat{U(1)}_{k}$, the projection
matrix is trivial and the branching rule just says that the difference
of the weights $\Lambda$ and $\lambda$ has to be an element of the
root lattice of $sl(2)$ i.e. the difference of $\Lambda$ and $\lambda$
has to be an even number. Thus, the parafermion theory has Virasoro
primaries $\Phi^{\Lambda}_{\lambda}$ labelled by a highest weight
$0\le\Lambda\le k$ of $sl(2)$ and a weight $\lambda$ of $sl(2)$ for
which we have $\Lambda-\lambda=0 ({\rm mod~} 2).$

Since the parafermion fields $\Phi^{\Lambda}_{\lambda}$ are now
labelled in the same way as the Virasoro
primary fields $G^{\Lambda}_{\lambda}$ of the $\widehat{sl(2)}_k$
theory, one might hope that there is a simple relation between these fields.
In fact, it was pointed out already in \cite{fatzam} that each of the
fields $G^{\Lambda}_{\lambda}$ may be written as the product of a
field $\Phi^{\Lambda}_{\lambda}$ from the parafermion theory and a
vertex operator of the $\widehat{U(1)}_{k}$ theory, which is just the
theory of a free boson on a circle of radius $\sqrt{2k}.$ This was
further clarified in \cite{gepqiu}, using the results of
\cite{gepfuchs}. One has

\begin{equation}
\label{parafwzw}
G^{\Lambda}_{\lambda}=\Phi^{\Lambda}_{\lambda}e^{i\lambda\phi/\sqrt{2k}}
\end{equation}
From this relation, one immediately reads off that the
field $\Phi^{\Lambda}_{\lambda}$ must have conformal weight
$(h')^{\Lambda}_{\lambda}$ given by

\begin{equation}
\label{parafwts}
(h')^{\Lambda}_{\lambda}=h^{\Lambda}_{\lambda}-\frac{\lambda^{2}}{4k}
=\frac{\Lambda(\Lambda+2)}{4(k+2)}
-\frac{\lambda^{2}}{4k}+n^{\Lambda}_{\lambda}
\end{equation}
As in other coset theories, the labelling of the fields
$\Phi^{\Lambda}_{\lambda}$ as we introduced it above is
redundant. First of all, the $U(1)$ label $\lambda$ is usually taken
to be defined modulo $2k,$ since the (extended) $\widehat{U(1)}_{k}$
characters $\chi_{\lambda}$ and $\chi_{\lambda+2k}$, that correspond
to the vertex operators $e^{i\lambda\phi/\sqrt{2k}}$ and
$e^{i(\lambda+2k)\phi/\sqrt{2k}}$, are equal (see for example
\cite{dsm}). Because of this and because of the fusion rules
(\ref{paraffus}) below, the label $\lambda$ is called the $\ZZ_{2k}$
charge of the field $\Phi^{\Lambda}_{\lambda}$.
\footnote{Note that in the original parafermion theory of
\cite{fatzam}, there was a $\ZZ_{k}\times\tilde{\ZZ}_{k}$
symmetry. The $\ZZ_k\times\tilde{\ZZ}_{k}$ charge $(l,\tilde{l})$ of
the field $\Phi^{\Lambda}_{\lambda}(z)
\Phi^{\bar{\Lambda}}_{\bar{\lambda}}(\bar{z})$ was given by
$l=\frac{1}{2}(\Lambda+\bar{\Lambda}),
\tilde{l}=\frac{1}{2}(\Lambda-\bar{\Lambda}),$ so that clearly in this
theory, one needed $\Lambda+\bar{\Lambda}$ to be even. Here, we will
not require this and thus allow chiral fields like $\Phi^{1}_{1}$.}
Also, in order to get proper behaviour of the fields' characters under
modular transformations, one has to identify fields whose labels are
sent onto each other by an external automorphism of the parent algebra
\cite{gepner89}.In the case at hand, this means that we have to
identify $\Phi^{\Lambda}_{\lambda}$ with
$\Phi^{k-\Lambda}_{\lambda-k}.$ Collecting, we get the field identifications 

\begin{eqnarray}
\label{fieldident}
\Phi^{\Lambda}_{\lambda}&\equiv&\Phi^{\Lambda}_{\lambda+2k} 
\nonumber\\
\Phi^{\Lambda}_{\lambda}&\equiv&\Phi^{k-\Lambda}_{\lambda-k}.
\end{eqnarray}
Using these identifications, we can choose a labelling of the
primaries such that $\lambda$ is a weight in the representation
$\Lambda$ of $sl(2),$ i.e. $-\Lambda\le\lambda\le\Lambda.$ In fact, we
may require $-\Lambda<\lambda\le\Lambda$ and if we do this then every
set of labels corresponds uniquely to a Virasoro primary Thus, the
number of Virasoro primaries is $\frac{1}{2}k(k+1)$ (note: there are
only $k$ primaries of the full parafermion algebra: the fields
$\Phi^{\Lambda}_{\Lambda}$).

One may check that the conformal weights given in (\ref{parafwts}) are
equal for identified fields. Also, note that the grade
$n^{\Lambda}_{\lambda}$ in (\ref{parafwts}) is zero if the labels
$(\Lambda,\lambda)$ are in the range chosen above. Using the
factorisation (\ref{parafwzw}) and the field identifications, we may
now also write down the fusion rules for the parafermion fields. They
are

\begin{equation}
\label{paraffus}
\Phi^{\Lambda}_{\lambda}{\times}\Phi^{\Lambda'}_{\lambda'}=
\bigoplus_{\Lambda''=|\Lambda-\Lambda'|
}^{{\rm min}\{\Lambda+\Lambda',2k-\Lambda-\Lambda'\}}
\Phi^{\Lambda''}_{\lambda+\lambda'},
\end{equation}
In other words, they are the same as the fusion rules for the
$G^{\Lambda}_{\lambda},$ except that
the labels on the right hand side have to be brought back into the set
chosen above, using the field identifications (\ref{fieldident}).

\subsubsection{The coset $\widehat{sl(k)}_{1}\times\widehat{sl(k)}_{1}/\widehat{sl(k)}_{2}$}
\label{slkcossec}

We now turn to the coset
$\widehat{sl(k)}_{1}\times\widehat{sl(k)}_{1}/\widehat{sl(k)}_{2}.$
This coset is a special case of the general class considered in
\cite{bais1,bais2}. Its current algebra is denoted as a $W$-algebra
and much is known about them. In the quantum Hall application however,
the parafermion analyses is more directly relevant and applicable as
is described in some detail in \cite{cageto}. Nevertheless one should
keep in mind that the discussion of the braid group representations
that feature in these models and which will be extensively analysed
later on, readily extend to the $W$ representation theory. The
Virasoro primaries of the coset may be labelled by an
$\widehat{sl(k)}_1\times\widehat{sl(k)}_1$ weight (or, equivalently,
two $\widehat{sl(k)}_1$ weights) and an $\widehat{sl(k)}_2$
weight. Let us call the $\widehat{sl(k)}_1$ weights $\mu_1$ and
$\mu_2$ and the $\widehat{sl(k)}_2$ weight $\mu,$ then we can write
$\Phi^{\mu_1,\mu_2}_{\mu}.$ The weights $\mu_1,\mu_2$ and $\mu$ once
again have to satisfy the branching condition (\ref{branchcond}). In
this case, the projection $\mathcal{P}$ maps $(\mu_1,\mu_2)$ onto
$\mu_1+\mu_2$ and it maps the root lattice of $sl(k)\times sl(k)$ onto
the root lattice of $sl(k).$ Hence we have the following requirement

\begin{equation}
\label{slkbranch}
\mu_1+\mu_2-\mu \in M_{sl(k)} 
\end{equation}
where $M_{sl(k)}$ is the root lattice of $sl(k)$. In other words, the
weights $\mu_1+\mu_2$ and $\mu$ should be in the same conjugacy class
(for details on this concept see for example \cite{fuchschweig,dsm}). In
terms of the Dynkin labels of the weights, this means that one has

\begin{equation}
\sum_{j=1}^{k-1}j(\mu_{1}^{(j)}+\mu_{2}^{(j)}-\mu^{(j)}) =
0 ~{\rm mod ~k} 
\end{equation}
Now denote by $e_i$ the $sl(k)$ weight whose Dynkin labels
$e_{i}^{(j)}$ are given by $e_{i}^{(j)}=\delta_{ij}$ (These correspond
to the fundamental representations of $sl(k)$). Then $\mu_1$ is either
zero or equal to one of the $e_i,$ since it is a level one weight. The
same goes for $\mu_2.$ For the level two weight $\mu,$ there are three
possibilities: it can be zero, equal to one of the $e_i$ or equal to
the sum of two of the $e_i$ (which may be the same). If we define
$e_0=0$, then we may simplify this description and say that $\mu_1$
and $\mu_2$ will equal one of the $e_i$ and $\mu$ will equal the sum
of two of the $e_i$ (where $i\in\{0,\ldots,k-1\}$) The branching rule
above then states that only triples $(\mu_1,\mu_2,\mu_3)$ of the form
$(e_l,e_{m+n-l~{\rm mod ~k}},e_m+e_n)$ are admissible. This leaves
$\frac{1}{2}k^2(k+1)$ admissible triples.  However, there are also
field identifications, induced by the external automorphisms of
$\widehat{sl(k)}_{1}\times\widehat{sl(k)}_{1}$. These identifications
take the form

\begin{equation}
\label{slkidents}
\Phi^{e_i,e_j}_{e_l+e_m}\equiv \Phi^{e_{i+s},e_{j+s}}_{e_{l+s}+e_{m+s}}
\end{equation}
for $s\in\{1,\ldots,k-1\}$. The sums in the indices on the right hand
side have to be taken modulo $k.$ Using these identifications, we can
choose to set either $\mu_1$ or $\mu_2$ to zero. Say we set $\mu_1$ to
zero. Then we are left with the triples $(0,e_{m+n~{\rm mod
k}},e_m+e_n)$. Clearly, $\mu_2$ is now uniquely determined by $\mu$
and we may choose to label the fields by the $\widehat{sl(k)}_{2}$ weight
only: $\Phi_{\mu}$. Every $\widehat{sl(k)}_{2}$ weight is
admissible and we are left with as many Virasoro primary fields as
there are $\widehat{sl(k)}_{2}$ weights: $\frac{1}{2}k(k+1)$. This is
just a reduction of the number of fields before identification by a
factor of $k,$ as should be expected. Also, we get the same number of
fields that we got in the other coset description of the
parafermions. 

The fractional part of the conformal weight of the field
$\Phi^{\mu_1,\mu_2}_{\mu}$ can be calculated directly from the coset
description; it is the same as the fractional part of the difference
between the conformal weight of field with labels $(\mu_1,\mu_2)$ in
the parent theory and the conformal weight of the field with label
$\mu$ in the embedded theory. One may show that this recipe always
yields the same fractional part, independently of the labels
$\mu_1,\mu_2,\mu$ that are chosen to represent a certain field
(i.e. labels that are identified through (\ref{slkidents}) yield the
same fractional part). Let us look at the field $\Phi_{e_m+e_n},$ with
$m \le n$. A particularly convenient choice of labels for this field,
made in \cite{cageto}, is $(e_{k-n},e_{m},e_{k+m-n}).$ The conformal
dimension of the WZW-field labelled by the weight $e_{m}$ is given by

\begin{equation}
\Delta_{p}(e_m)=\frac{(e_m,e_m+2\rho)}{2(p+k)}=\frac{m(k-m)(k+1)}{2k(k+p)},
\end{equation}
where $\rho$ is the Weyl-vector of $sl(k)$ and $p$ is the level (here,
we have p=1 or p=2). {F}rom this, we find 

\begin{equation}
\begin{array}{r}
{\displaystyle \Delta_{1}(e_{k-n})+\Delta_{1}(e_{m})-\Delta_2(e_{k+m-n})=
\frac{m(k-n)}{k}+\frac{(n-m)(k+m-n)}{2k(k+2)}} \\
{\displaystyle = \frac{(k+m-n)(k+m-n+2)}{4(k+2)}-\frac{(m+n-k)^2}{4k}} 
\end{array}
\end{equation}
The middle expression is the one given in \cite{cageto} and from the
last expression, we see that it is equal to the weight of the field
$\Phi^{\Lambda}_{\lambda}$ with $\Lambda=k+m-n$ and $\lambda=m+n-k$
(cf. formula (\ref{parafwts})). Thus, we have the correspondence

\begin{equation}
\label{sl2sl3corr}
\Phi^{k+m-n}_{m+n-k}\equiv\Phi_{e_{m}+e_{n}}
\Longleftrightarrow
\Phi^{\Lambda}_{\lambda}\equiv
\Phi_{e_{\frac{\Lambda+\lambda}{2}}+e_{\frac{2k-\Lambda+\lambda}{2}}}
\end{equation}
which is further supported by the fact that these fields have the same
fusion rules.
\footnote{Note that we could also identify the field
$\Phi_{e_{m}+e_{n}}$ with the field $\Phi^{k+m-n}_{k-m-n}$, which is
the conjugate of the field $\Phi^{k+m-n}_{m+n-k}$. It is impossible to
decide between these identifications on the level of conformal weights
and fusion rules}
In fact, the $\Phi_{\mu}$ fusion rules are the same as
the fusion rules for the corresponding $\widehat{sl(k)}_2$
representations and these are the same as the fusion rules of the
$\widehat{sl(2)}_{k}/\widehat{U(1)}_{k}$ coset as a consequence of level-rank
duality (see \cite{dsm} and references therein). One may also find the
equality of the fusion rules directly by looking at the fusion rules
of the field $\Phi^{1}_{1}\equiv\Phi_{e_{1}}$ with an arbitrary
field. These fusion rules are easily seen to be the same and since the
field $\Phi^{1}_{1}$ generates all the fields in the theory by
repeated fusion, it follows that the fusion rules of all the fields
that are identified through (\ref{sl2sl3corr}) are the same in both
cosets.

\subsection{Definition of the RR-states}
\label{rrdefsec}

As we mentioned before, the wave functions for the RR-states can be
described (or even defined) as conformal blocks in a certain Conformal
field theory. The CFT in question is the tensor product of the
parafermion theory with the theory of a chiral boson.

From the parafermion theory, one needs the chiral vertex operators
$\sigma:=\Phi^{1}_{1}=\Phi_{e_1}$ and $\psi:=\Phi^{0}_{2}=\Phi_{2e_1}$
defined above and from the bosonic theory one uses the usual vertex
operators $e^{i\alpha\zeta},$ of weight $\frac{1}{2}\alpha^2.$ Here,
$\zeta$ is the bosonic field (which should not be confused with the
bosonic field $\phi$ in the factorisation formula
(\ref{parafwzw})). Electrons and quasiholes are represented by
products of these bosonic and parafermionic vertex
operators. Specifically, the electron is represented by the product
$\psi e^{i\sqrt{\frac{kM+2}{k}}\zeta}$, where $M\in \ZZ$ and the
quasihole is represented by the product $\sigma e^{\frac{i
\zeta}{\sqrt{k(kM+2)}}}$. These combinations of bosonic and
parafermionic fields are not arbitrary; if the parafermion factor is
given, then the bosonic factor for the electron can be fixed by
requiring that electrons are mutually local (that is, the OPE of two
electron operators does not have a branch cut). This requirement is
needed to make sure that the wave functions defined below are single
valued in the electrons' coordinates. If the electron is to have
half-integer spin, then one should also require that $M$ be odd. The
exponent of the bosonic factor for the quasihole is fixed up to
integer times $\sqrt{\frac{k}{kM+2}}\zeta$ by the requirement that the
quasihole and the electron are mutually local.

The linear space of RR-states $\Psi^{k}_{N,n}$ which have $N$
electrons with coordinates $z_1,\ldots z_{N}$ and $n$ quasiholes
located at positions $w_1,\ldots w_n$ is now generated by the
conformal blocks of a correlator of $N$ electron fields and $n$
quasihole fields inserted at these positions and supplemented by a
homogeneously spread positive background charge, which ensures overall
charge neutrality \cite{moorread,readrez}. This correlator may be
factorized into parafermion and boson correlators, the latter of which
may be evaluated explicitly (paying due attention to the background
charge \cite{moorread}), after which one obtains

\begin{eqnarray}
\label{rrwf}
\Psi^{k}_{N,n}(z_1,\ldots,z_N,w_1,\ldots w_n)&=&
\left<\sigma(w_1)\ldots\sigma(w_n)\psi(z_1)\ldots\psi(z_N)\right>
\nonumber \\
~& \times & 
\prod_{i<j}(z_i-z_j)^{M+2/k}\prod_{i=1}^{N}\prod_{j=1}^{n}(z_i-w_j)^{1/k}
\nonumber \\
~& \times &
\prod_{i<j}(w_i-w_j)^{\frac{1}{k(kM+2)}} F_g(z_1,\ldots,z_N,w_1,\ldots w_n)
\end{eqnarray}
Here, the $z_i$ and $w_i$ are complex coordinates which parametrise
the sample. $F_g$ is a factor which depends on the geometry of the
sample. If the sample is a disc then this factor just implements the
usual Gaussian factors which confine the electrons and quasiholes to
the disc. If the sample is a sphere of radius $R$, then the
$z_i$ and $w_i$ are related to the usual coordinates on the sphere
through stereographic projection (see \cite{readrez} for
details) and one has

\begin{equation}
F_g(z_1,\ldots,z_N,w_1,\ldots w_n)=
\prod_{i}(1+|z_i|^{2}/4R^{2})^{-N_{\phi}/2-1}
\prod_{j}(1+|w_j|^{2}/4(kM+2)R^{2})^{-N_{\phi}/2-1}
\end{equation}
where $N_{\phi}=N/\nu-M-2+n$ is the number of flux quanta that go
through the surface of the sphere and $\nu=\frac{k}{kM+2}$ is the
filling fraction.

The construction of the Pfaffian and $RR$-states through conformal
field theory is analogous to the construction of the Laughlin state
and the Pfaffian state through conformal field theory as given in
\cite{moorread}. Indeed,
the $k=2$ case of the above wave function just describes the Pfaffian
state of \cite{moorread} with $N$ electrons and $n$ quasiholes. 

\subsection{Fusion of quasiholes and the Bratteli diagram}
\label{brattelisec}

It is interesting to know the number of independent states which the
formula (\ref{rrwf}) encodes, i.e. the number of independent states
with $N$ electrons that have $n$ quasiholes at fixed positions
$w_1,\ldots,w_n$. This interest is twofold. First of all, we want to
know which combinations $(N,n)$ are allowed. Second, the number of
independent states is also the dimension of the braid group
representation that governs the exchanges of electrons and
quasiholes. Hence a necessary condition for nonabelian braiding is
that it be larger than one.  A basis for the space of states that we
are looking for is given by the states we obtain if we replace the
parafermion correlator in (\ref{rrwf}) by its respective conformal
blocks. The number of such blocks is equal to to the number of fusion
channels that make the correlator in (\ref{rrwf})
non-vanishing. Hence, the number we are looking for is just the number
of ways in which $N$ electron fields $\psi$ and $n$ quasihole $\sigma$
fields may fuse into the vacuum.

Now the fusion of the $\psi$ fields is very simple;
it just corresponds to addition of the $\ZZ_{2k}$ charges. Hence the
$N$ electron fields fuse into the $\Phi^{0}_{2N}=\Phi_{2e_N}$ sector.
The fusion rules of the sigma fields, as given in equation
(\ref{paraffus}), are a bit more complicated, but they have a nice
graphical description in terms of a Bratteli diagram (see figures
\ref{sl2fusdiag},\ref{slkfusdiag}):

\begin{figure}[h,t,b]

\setlength{\unitlength}{1.2pt}
\begin{picture}(300,140)(-40,-20)
\put(-10,-10){\line(0,1){110}}
\multiput(-10,0)(0,20){5}{\line(1,0){5}}
\put(-20,0){\footnotesize 0}
\put(10,20){\yodi{5pt}{0}{1}}
\put(-20,20){\footnotesize 1}
\put(25,40){\yodi{5pt}{0}{2}}
\put(-20,40){\footnotesize 2}
\put(40,60){\yodi{5pt}{0}{3}}
\put(-20,60){\footnotesize 3}
\put(-35,90){\mbox{$\Lambda \uparrow$}}

\put(-10,-10){\line(1,0){260}}
\multiput(0,-10)(20,0){12}{\line(0,1){5}}
\put(0,-20){\footnotesize 0}
\put(20,-20){\footnotesize 1}
\put(40,-20){\footnotesize 2}
\put(60,-20){\footnotesize 3}
\put(80,-20){\footnotesize 4}
\put(235,-20){\mbox{$\lambda \rightarrow$}}

\multiput(0,0)(40,0){5}{\vector(1,1){20}}
\multiput(20,20)(40,0){5}{\vector(1,1){20}}
\multiput(40,40)(40,0){4}{\vector(1,1){20}}
\multiput(20,20)(40,0){5}{\vector(1,-1){20}}
\multiput(40,40)(40,0){4}{\vector(1,-1){20}}
\multiput(60,60)(40,0){4}{\vector(1,-1){20}}
\end{picture}
\caption{\footnotesize fusion diagram for the field
\mbox{$\sigma$}. The diagram must be thought extended indefinitely in
the $\lambda$-direction and up to $\Lambda=k$ in the
$\Lambda$-direction (the case $k=3$ is as drawn here). On each line,
we have drawn the Young diagram of the $sl(2)$ representation that
resides on that line.}
\label{sl2fusdiag}
\end{figure}
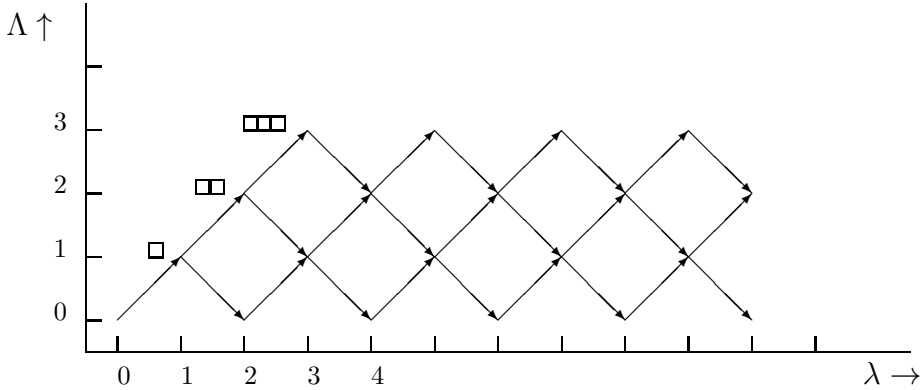
\begin{figure}[h,t,b]

\begin{picture}(400,160)(-40,-20)

\put(-10,-10){\line(0,1){150}}
\multiput(-10,0)(0,30){4}{\line(1,0){5}}
\put(-20,0){\footnotesize 0}
\put(-20,30){\footnotesize 1}
\put(-20,60){\footnotesize 2}
\put(-20,90){\footnotesize 3}
\put(-35,135){\mbox{$\Lambda \uparrow$}}

\put(-10,-10){\line(1,0){320}}
\multiput(0,-10)(30,0){10}{\line(0,1){5}}
\put(0,-20){\footnotesize 0}
\put(30,-20){\footnotesize 1}
\put(60,-20){\footnotesize 2}
\put(90,-20){\footnotesize 3}
\put(120,-20){\footnotesize 4}
\put(280,-20){\mbox{$\lambda \rightarrow$}}

\multiput(8,8)(60,0){4}{\vector(1,1){14}}
\multiput(38,38)(60,0){4}{\vector(1,1){14}}
\multiput(68,68)(60,0){3}{\vector(1,1){14}}
\multiput(38,22)(60,0){4}{\vector(1,-1){14}}
\multiput(68,52)(60,0){3}{\vector(1,-1){14}}
\multiput(98,82)(60,0){3}{\vector(1,-1){14}}

\put(0,2){\circle*{5}}
\put(27,22){\yodit{5pt}{1}{0}}
\put(57,50){\yodit{5pt}{2}{0}}
\put(90,90){\circle*{5}}

\put(55,-6){\yodit{5pt}{1}{1}}
\put(85,20){\yodit{5pt}{2}{1}}
\put(117,52){\yodit{5pt}{1}{0}}
\put(147,85){\yodit{5pt}{1}{1}}

\put(115,-6){\yodit{5pt}{2}{2}}
\put(147,20){\yodit{5pt}{2}{0}}
\put(175,47){\yodit{5pt}{2}{1}}
\put(207,83){\yodit{5pt}{2}{2}}

\put(180,2){\circle*{5}}
\put(207,20){\yodit{5pt}{1}{0}}
\put(237,47){\yodit{5pt}{2}{0}}

\put(235,-6){\yodit{5pt}{1}{1}}
\end{picture}

\caption{\footnotesize The same diagram as in figure \ref{sl2fusdiag},
but this time each site in the diagram is labelled by the Young
diagram for the $\widehat{sl(k=3)}_{2}$ weight of the field that
resides there. The dot represents the empty diagram. Again,
generalisation to arbitrary $k$ is straightforward. Note that in this
picture, the weights label the fields unambiguously, whereas in figure
\ref{sl2fusdiag}, one still has to take the field identifications
(\ref{fieldident}) into account}
\label{slkfusdiag}
\end{figure}
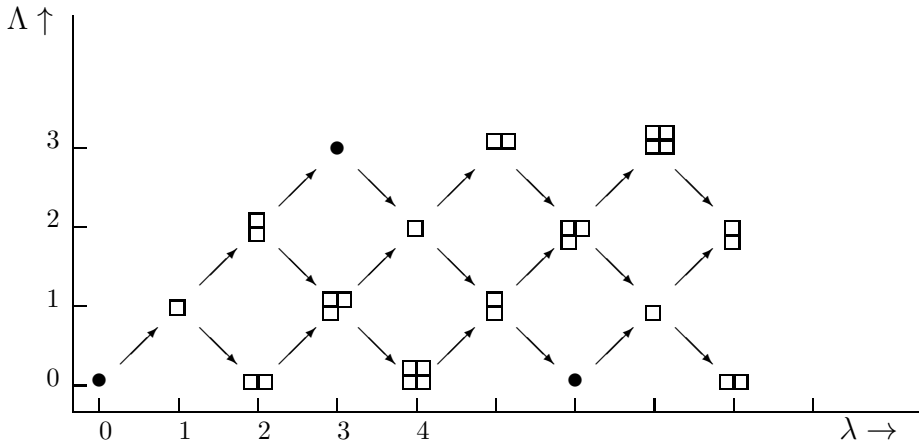
These diagrams must be read as follows: Each starting point or end
point of an arrow has coordinates $(\Lambda,\lambda)$ and represents
the $\Phi^{\Lambda}_{\lambda}$ sector of the parafermion CFT. Note
that this means that coordinates related by the identifications
(\ref{fieldident}) represent the same sector. In figure
\ref{slkfusdiag}, one may see this explicitly for $k=3.$ Here, we have
at each node of the diagram inserted the Young diagram for the
$\widehat{sl(3)}_2$ weight of the field which resides there. The
correspondence between the fields of the parafermionic theory and such
weights or diagrams is one to one and we see that the same diagram
appears in different places. The fusion rules of the sigma field are
encoded in the arrows; we start in the lower left corner, i.e. the
$\Phi^{0}_{0}$ sector which is the vacuum sector of the theory. Then
we take the operator product expansion with the field
$\sigma=\Phi^{1}_{1},$ which naturally, following the arrow, lands us
in the $\Phi^{1}_{1}$ sector. Once more taking the OPE with $\sigma,$
we end up, following the arrows, in the $\Phi^{2}_{2}$ or in the
$\Phi^{0}_{2}$ sector. In this way, each path of length $n$ through
the diagram represents a fusion channel for $n$ $\sigma$-fields.

\sloppypar To make the correlator in the wave function non vanishing, all the
quasihole and electron fields need to fuse into the vacuum sector.
Now since the electron fields $\psi(z_1),\dots,\psi(z_N)$ in the
correlator above fuse to $\Phi^{0}_{2N},$ the quasihole fields
$\sigma(w_1),\ldots,\sigma(z_n)$ have to fuse to the field
$\Phi^{0}_{-2N}=\Phi^{k}_{k-2N}.$ The number of ways to do this is
just the number of paths of length $n$ through the diagram of figure
\ref{sl2fusdiag} which end up at a point whose coordinates
$(\Lambda,\lambda)$ satisfy either
$(\Lambda,\lambda)=(0,-2N{\rm~mod~2k})$ or
$(\Lambda,\lambda)=(k,k-2N{\rm~mod~2k}).$ Clearly, for fixed $N,$ such
paths occur only for values of $n$ which are a multiple of $k$ apart,
so quasiholes can only be created in multiples of $k$ at a time (maybe
with the exception of the first few quasiholes if $N$ is not a
multiple of $k$). Note that, although the same fields (or sectors)
occur in different heights in the diagram, the same field never occurs
more than once at given $\lambda$ and hence different paths are never
identified by the field identifications. Thus, the number of fusion
channels for the parafermion CFTs is the same as that for the
corresponding WZW-theories.

\subsection{Counting the independent $n$-quasihole states}
\label{telsec}

Let us denote the number of paths through the Bratteli diagram which
end up at the point $(\Lambda,n)$ by $D(\Lambda,n).$ Also, let us
define $D(\Lambda,n)=0$ if there is no point with coordinates
$(\Lambda,n).$ The number of independent $n$-quasihole states encoded
by (\ref{rrwf}) is then $D(0,n)$ in case $2N+n=0 ~({\rm ~mod~} 2k),$
$D(k,n)$ in case $2N+n=k ~({\rm ~mod~} 2k),$ and zero otherwise.  It
should be obvious from looking at the Bratteli diagram that the
$D(\Lambda,n)$ satisfy the following recursion relation:

\begin{equation}
\label{dimrec}
D(\Lambda,n)=D(\Lambda-1,n-1)+D(\Lambda+1,n-1)
\end{equation}
Using this relation and the fact that $D(1,1)$ equals one,
$D(\Lambda,n)$ can be easily calculated in each particular case. At
least for low $k,$ the recursion relation can also be used to prove
simple closed expressions for the $D(\Lambda,n).$ In particular, we
find for $k=2,k=3$ and $k=4$

\begin{eqnarray}
\label{234dims}
D_2(0,2n)&=&D_2(1,2n-1)=2^{n-1} \nonumber\\
D_3(0,2n)&=&D_3(1,2n-1)={\rm Fib}(2n-2)\nonumber\\
D_3(2,2n)&=&D_3(3,2n+1)={\rm Fib}(2n-1)\nonumber\\
D_4(0,2n)&=&D_4(1,2n-1)=\frac{3^{n-1}+1}{2}\nonumber\\
D_4(2,2n)&=&3^{n-1} \nonumber \\
D_4(3,2n+1)&=&D_4(4,2n+2)=\frac{3^{n-1}-1}{2}
\end{eqnarray}
In these equations, we have written $D_{k}$ in stead of $D$ for clarity and we
have used the notation ${\rm Fib}(n)$ to denote the $n^{\rm th}$ Fibonacci
number, defined by

\begin{eqnarray}
{\rm Fib}(0)&=&{\rm Fib}(1)=1 \nonumber \\
{\rm Fib}(n+1)&=&{\rm Fib}(n)+{\rm Fib}(n-1)
\end{eqnarray}
It is also not that difficult to find and prove a closed formula for
infinite $k.$ We have

\begin{equation}
D_{\infty}(\Lambda,n)=\frac{\Lambda+1}{n+1}\left(
\begin{array}{c}
n+1\\
\frac{n-\Lambda}{2}
\end{array}\right) ~~~~(n+\Lambda=0 ~({\rm mod~}2))
\end{equation}
Of course this formula is valid for all $k$ as long as $n+\Lambda \le
2k.$ 

To get formulae for other values of $k$ it is more convenient to
rewrite the recursion relation (\ref{dimrec}) in matrix form. We
consider the $D(\Lambda,n)$ at a fixed $n$ together as a $k$-vector
and write the step from $n$ to $n+1$ as multiplication with a
$(k+1)\times (k+1)$ matrix $M_k.$ that is, we have

\begin{equation}
\left(\begin{array}{c}
D(0,n+1)\\
\vdots \\
D(k,n+1)
\end{array}\right)=
M_k
\left(\begin{array}{c}
D(0,n)\\
\vdots \\
D(k,n)
\end{array}\right)
\end{equation} 
where $M_k$ is given by

\begin{equation}
(M_k)_{ij}=\delta_{i,j+1}+\delta_{i+1,j}
\end{equation}
The asymptotic behaviour of the $D(\Lambda,n)$ for large $n$ will be
related to the largest eigenvalue of the matrix $M_k.$ The eigenvalues
of the $M_i$ are just the zeros of their characteristic
polynomials $P_k$. For these, we can easily deduce a recursion relation and
``initial conditions:''

\begin{eqnarray}
P_2(\lambda)&=&\lambda^{2}-1 \nonumber \\
P_3(\lambda)&=&\lambda^{3}-2\lambda\nonumber \\
P_{i+1}(\lambda)&=&\lambda P_i(\lambda)-P_{i-1}(\lambda)
\end{eqnarray}
but these are just the defining relations for the Chebyshev
polynomials, whose zeros are given by (see for example \cite{absteg})

\begin{equation}
\lambda_{k,m}=2\cos\left(\frac{(m+1)\pi}{k+2}\right)
\end{equation}
Since we know all the eigenvalues of $M_k$, we can now in principle
solve for the eigenvectors and using the solution, give explicit
formulae for the $D_{k}(\Lambda,n)$ for any $k.$ We will however
content ourselves with giving the asymptotic behaviour of the
$D_{k}(\Lambda,n)$ at large $n.$ The largest eigenvalues (in absolute
value) of the matrix $M_{k}$ are clearly $\lambda_0$ and
$\lambda_{k}=-\lambda_0.$ Hence, the asymptotic behaviour of the
$D_k(\Lambda,n)$ is given by

\begin{eqnarray}
D_{k}(\Lambda,n) &\sim& \left(2\cos\left(\frac{\pi}{k+2}\right)\right)^{n}
~~~(\Lambda+n {\rm~even}). \nonumber \\ 
D_{k}(\Lambda,n) &=& 0
~~~(\Lambda+n {\rm~odd}). 
\end{eqnarray}
This conforms with the closed formulae we gave for $k=2,3,4.$ 

\subsection{Braiding for $k=2$}

In the previous section, we have calculated the dimensions of the
braid group representations that govern the exchanges of the electrons
and the quasiholes of the RR-states. We have seen that these
dimensions increase with the number of quasiholes, which is an
indication for nonabelian braiding. However, this indication is not
conclusive evidence. To be sure, one needs to calculate the actual
matrices that describe the braiding of the $\sigma$-fields in the
conformal block in formula (\ref{rrwf}) above. Nayak and Wilczek
\cite{naywil} have done this calculation for the case $k=2$ (the
Pfaffian). The method they used was basically to compute the conformal
block for four quasihole fields explicitly and then to extend the
resulting braid group representation to a braid group representation
for any even number of quasiholes. \footnote{Note that the four point
blocks in the case $k=2$ are just the four point blocks for the chiral
Ising model, which have, within a different context, been known for a
long time (see for instance \cite{ags90} for explicit
expressions). The same is true for the corresponding braid group
representations. However, the embedding of the resulting braid group
representation into a rotation group, as given by Nayak and Wilczek
(see below) seems to be new} For general $k$, it is quite difficult to
calculate conformal blocks for four, let alone for arbitrary numbers
of quasiholes in the case. Fortunately it turns out that we can
circumvent this problem by using the known duality between conformal
field theory and quantum groups and using this, we will give a nice
description of the braiding for arbitrary $k$. However, we will first
briefly recall the results of Nayak and Wilczek for $k=2,$ for later
reference.

The braid group representation for $n=2m$ quasiholes has dimension
$2^{m-1}$ (cf. (\ref{234dims})). Nayak and Wilczek describe this space
as a subspace of a tensor product of $m$ two dimensional spaces. Each
of the two dimensional spaces has basis vectors $\{ \ket{+},\ket{-}
\}$ and the physical subspace of the tensor product is the space
generated by the vectors whose overall sign is positive (so for
$m=2,~\ket{--}$ is physical, but $\ket{+-}$ is not). On the tensor
product space, there is a spinor representation of $SO(2m)\times
U(1).$ The $U(1)$ acts as multiplicative factor, while the generators
$\sigma_{ij}$ of the $SO(2m)$ may be written in terms of the Pauli
matrices $\sigma_{i}.$ We have

\begin{equation}
\sigma_{ij}=\frac{1}{4}i[\gamma_{i},\gamma_{j}]
\end{equation}
with

\begin{equation}
\begin{array}{rcl}
\gamma_1&=&\sigma_1\otimes \sigma_3\otimes\ldots\otimes \sigma_3 \\
\gamma_2&=&\sigma_2\otimes \sigma_3 \otimes\ldots\otimes \sigma_3 \\
\gamma_3&=&1\otimes\sigma_1\otimes \sigma_3 \otimes\ldots\otimes \sigma_3 \\
\gamma_4&=&1\otimes\sigma_2\otimes \sigma_3 \otimes\ldots\otimes \sigma_3 \\
&\vdots& \\
\gamma_{2m}&=&1\otimes\ldots\otimes 1 \otimes \sigma_2 \\
\end{array}
\end{equation}
Here the states $\ket{+}$ and $\ket{-}$ are the spin up and spin
down states for the Pauli matrices.

Now recall that the braid group $B_n$ is generated by $n-1$
elementary exchanges $\tau_1,\ldots,\tau_n$ subject to the relations

\begin{eqnarray}
\label{braidgroup}
\tau_i\tau_j&=&\tau_j\tau_i ~~~~~ (|i-j|\ge 2) \nonumber \\
\tau_i\tau_{i+1}\tau_i&=&\tau_{i+1}\tau_{i}\tau_{i+1}
\end{eqnarray}
Here, $\tau_i$ represents an exchange of quasihole $i$ and quasihole $i+1.$ 
The action of the braid group on the $n$-quasihole space is embedded
in the action of $SO(2n)\times U(1),$ as follows:

\begin{equation}
\tau_i\equiv e^{i\frac{\pi}{4}} e^{i\frac{\pi}{2}\sigma_{i,i+1}}
\end{equation} 
The $SO(2m)$ generators $\sigma_{i,i+1}$ which appear in this equation
are given by

\begin{equation}
\begin{array}{rcl}
\sigma_{1,2}&=&\frac{1}{2}\,\sigma_3\otimes 1\otimes\ldots\otimes 1 \\
\sigma_{2,3}&=&\frac{1}{2}\,\sigma_2\otimes \sigma_2 \otimes
1\otimes\ldots\otimes 1 \\
\sigma_{3,4}&=&\frac{1}{2}\,1\otimes\sigma_3\otimes 1\otimes\ldots\otimes 1 \\
\sigma_{4,5}&=&\frac{1}{2}\,1\otimes\sigma_2\otimes \sigma_2 \otimes
1\otimes\ldots\otimes 1,{\rm ~etc.} 
\end{array}
\end{equation}
So we see that, for odd $i,$ $\tau_i$ acts only on the $i^{\rm th}$ tensor
factor, whereas for even $i,$ $\tau_{i}$ acts only on the $(i-1)^{\rm
th}$ and $i^{\rm th}$ tensor factors. Moreover, the $2\times 2$-matrix
by which describes the action for even $i$ and the $4\times 4$-matrix
which describes it for odd $i$ do not vary with $i.$ Explicitly, they
are given by

\begin{equation}
\label{nwmats}
\tau_{2i+1}\equiv\left(
\begin{array}{cc}
 1 & 0 \\ 
 0 & i
\end{array} \right),~~~~  
\tau_{2i}\equiv \frac{1}{2}\left(
\begin{array}{cccc}
1+i& 0 & 0 & -1+i \\
 0 & 1+i & 1-i & 0 \\
 0 & 1-i & 1+i & 0 \\
 -1+i & 0 & 0 & 1+i 
\end{array} \right)   
\end{equation}

\section{The quantum group picture}
\label{qgsec}

In this extensive section, we give a description of the braiding for a
system of $n$ particles with a hidden quantum group symmetry. We
expect that the braiding properties of a quantum Hall state with $n$
quasiholes is conveniently described in terms of such a system. In the
first subsection, we motivate such a description and mention some
general features. In the remaining subsections, we give a fairly
detailed description to the quantum group $U_{q}(sl(2))$ and its
representation theory for $q$ a root of unity, which culminates in an
explicit description of the braid group representations that are
associated to this quantum group. We are well aware of the fact that
most of the material treated in this section is not new, except
possibly for the identities (\ref{x6jsyms}) for the $6j$-symbols,
however the input came from quite a variety of sources and putting it
together in comprehensive way seemed a nontrivial and even useful
thing to do. It should serve as a quick introduction to braiding in systems
with hidden quantum group symmetry, which is why we have tried to keep
the treatment as self-contained as possible.

\subsection{Using a quantum group rather than the full CFT}
\label{gaugesec}

One may always choose to describe a quantum system in terms of its
explicit wave functions but we know that it can be extremely
profitable to exploit its operator algebra, in particular its
symmetries. These allow one to extract many of the physical features
without reference to the explicit realisation in terms of
wave functions. Quite similarly one could in the present context remark
that there is an aspect of the description of the Read-Rezayi states
that is less than satisfactory: one has to use the full machinery of a
(conformal) field theory to calculate wave functions or even just
braiding properties for a finite number of quasiholes and
electrons. There are many questions one may want to answer for which
this seems like overkill: for example one would hope to be able to
describe the braiding of finitely many particles by means of a theory
with only finitely many degrees of freedom. Indeed, there is such an
alternative description and we pursue it here. It is well known that
conformal field theories possess a hidden quantum group symmetry (see
section \ref{qgcftsec} for details and references). What we propose is
to describe the electrons and quasiholes of a quantum Hall state that
would usually be described by a certain CFT as localised particles that
carry representations of the quantum group that is associated with
this CFT. Such a description has several advantages.
\begin{itemize}
\item
It avoids the introduction of a field theory to describe
a system with only a finite number of particle degrees of freedom.
\item
It provides a conceptual understanding of a phenomenon which emerges
in the usual CFT description. This is the fact that, while a state
with a low number of indistinguishable quasiholes can be described
with a one component wave function, a system with a higher number of
these quasiholes may need a several component wave function. Clearly,
it should only be possible to distinguish between the components of
the wave function by making a measurement that involves several holes
(otherwise the holes would not be indistinguishable). Hence, there
should be operators in the many hole Hilbert space that distinguish
states that cannot be distinguished by operators that act only on the
state of one of the particles. The quantum group picture provides
these in a natural way. They are the operators that correspond to the
global (i.e. invariant) quantum group charges of groups of
quasiholes. Even though all individual quasiholes have the same
quantum group charge i.e. belong to the same representation, a group
of $n$ such holes can occur in different representations leading to
distinguishable $n$-hole states. As a simple example, suppose that the
quasiholes carried the two dimensional representation of $SU(2)$ (or
equivalently, of the quantum group $U(sl(2))$). In that case a two
quasihole state could be either in the singlet or in the triplet
representation and the singlet states could be distinguished from the
triplet states by measuring the global charge.
\item
The quantum group picture allows for a very elegant description of the
braiding properties of the $n$-quasihole states; all braiding
properties are encoded into a single algebraic object: the quantum
group's $R$-matrix (cf. section \ref{rmatsec}). Starting from the
$R$-matrix, braiding calculations can be done in a purely algebraic
way and often a detailed picture of the braid group representation
that governs the exchanges of particles can be constructed.  In a CFT
description, the information contained in the $R$-matrix of the
quantum group would be much less manifest. In fact, to extract it from
this description of the system, one would have to calculate the
braiding and fusion matrices starting from the conformal blocks of the
CFT, which is usually quite hard.
\end{itemize}

Of course the description we propose also has its disadvantages when
compared to the CFT description. For instance, it seems much harder to
describe dynamical aspects of the quantum Hall states in this
picture. Still, we like to emphasise that the quantum group picture we
propose is a useful complementary way of thinking about nonabelian
quantum Hall states.

\subsection{$U_{q}(sl(2))$ for pedestrians}
\label{uqsl2sec}

In this section we have collected some basic information about quantum
groups in general and the quantum group $U_{q}(sl(2))$ in
particular. For those well versed in math there are excellent text
books, for example \cite{chapress}, covering the full story.

\subsubsection{The algebra and its unitary representations}

Quantum groups are actually not groups, but algebras. To be a quantum
group, an algebra has to have several structures associated with
it. As an algebra, it is already a vector space with a multiplication,
but next to this, there should also be a comultiplication, counit,
antipode and $*$-structure, (which makes the algebra a
Hopf-$*$-algebra), an $R$-matrix (which makes it a quasitriangular
Hopf-$*$-algebra) and optionally a coassociator (which would make it a
quasitriangular quasi-Hopf-*-algebra or something with a longer
name). We briefly describe these structures and their use in our context
in the following sections, always using the quantum group
$U_{q}(sl(2))$ as an example. In this section, we will describe the
algebra $U_{q}(sl(2))$ and its irreducible representations.

$U_{q}(sl(2))$ can be defined as the algebra generated by a
unit $1$ and the three elements $H,L^{+}$ and $L^{-}.$ These satisfy
the relations

\begin{eqnarray}
[H,L^{\pm}] &=&\pm 2 L^{\pm} \nonumber \\
~[L^{+},L^{-}] &=&\frac{q^{H/2}-q^{-H/2}}{q^{1/2}-q^{-1/2}}
\end{eqnarray}
where $q$ may be set to any non-zero complex value.
One may check that these relations reduce to those of the Lie algebra
$sl(2)$ when $q$ goes to one. Hence, $U_{1}(sl(2))$ is just the
universal enveloping algebra $U(sl(2))$ of $sl(2)$ and we say that
$U_{q}(sl(2))$ is a $q$-deformation of $U(sl(2))$.

If $q$ is taken as a formal variable and also for all $q \in \CC,$ $q$
not a root of unity, the representation theory of $U_{q}(sl(2))$ is
very similar to that of $U(sl(2)).$ For each non-negative $j \in
\frac{1}{2}\ZZ$ there is an irreducible highest weight representation
of dimension $2j+1.$ We will denote this representation by
$\pi^{\Lambda},$ where $\Lambda=d-1=2j$ is the highest weight. The
modules $V^{\Lambda}$ of these representations have an orthonormal
basis that consists of kets $\ket{j,m},$ with $m=-j,-j+1,\ldots,j$ and
the generating elements $H,L^{+}$ and $L^{-}$ act on this basis as
follows

\begin{eqnarray}
\label{repfrms}
H\ket{j,m}&=&2m\ket{j,m} \nonumber\\
L^{\pm}\ket{j,m}&=&\sqrt{\qnr{j\mp m}\qnr{j\pm m+1}}\;\;\ket{j,m\pm 1}
\end{eqnarray}
Here, the $q$-number $\qnr{m}$ is defined as

\begin{equation}
\qnr{m}=\frac{q^{m/2}-q^{-m/2}}{q^{1/2}-q^{-1/2}}
\end{equation}
These $q$-numbers enter the formulae for the representations through
the commutation relation of $L^{+}$ and $L^{-},$ the right hand side
of which can be written $\qnr{H}.$
The $q$-number $\qnr{m}$ approaches $m$ when $q$ goes to one and hence
we see that the representations given above reduce to the usual
$U(sl(2))$ representations for $q=1.$ Furthermore, when $q \in \RR$ or
$|q|=1,$ the $q$-number $\qnr{x}$ is a real number when $x$ is a real
number. 

We will call a representation of $U_{q}(sl(2))$
unitary when $(L^{+})^{\dag}=L^{-},~H^{\dag}=H.$ More formally, we
may define an antilinear algebra antihomomorphism $*$ on $U_{q}(sl(2))$ by

\begin{equation}
\label{stardef}
*(L^{\pm})=L^{\mp},~~~~*(H)=H
\end{equation}
and a representation $\rho$ of $U_{q}(sl(2))$ is then called unitary,
or a $*$-representation, exactly if
 
\begin{equation}
\forall x \in U_{q}(sl(2)):~~~\rho(*(x))=\rho(x)^{\dag} .
\end{equation}
In the representations defined above, we always have
$\pi^{\Lambda}(H)^{\dag}=\pi^{\Lambda}(H)$ and
$\pi^{\Lambda}(L^{\pm})^{\dag}=\bar{\pi}^{\Lambda}(L^{\mp}),$ where the
bar denotes complex conjugation of the elements.  Hence, these
representations are unitary when the matrix elements of $L^{\pm}$ are
real, i.e. if the square root in (\ref{repfrms}) is real for all
admissible values of $m$.  This will certainly be the case if $q$ is
real and positive and also if $q=e^{i\phi}$ with $\phi\in
\RR,~|\phi|\le\frac{2\pi}{2j+1}$. Thus we see that, for real $q$, all
the representations above are $*$-representations, while for $q=e^{i\phi}$,
the representations $\pi^{2j}$ with $|\phi|\le\frac{2\pi}{2j+1}$ are
$*$-representations.

The $q$-numbers satisfy many identities which are useful in
representation theoretic calculations. Two examples of such
identities, which hold for all $q \in \CC$ are

\begin{eqnarray}
\label{qnrids}
q^{n/2}\qnr{m}+q^{-m/2}\qnr{n}&=&\qnr{m+n} \nonumber\\
\qnr{n+m}\qnr{n-m}&=&\qnr{n}^2-\qnr{m}^2.
\end{eqnarray}

\subsubsection{Representations at $q=e^{2 i \pi/(k+2)}$}

\begin{figure}[h,t,b]
\label{indecfig}
\begin{picture}(450,50)(-10,0)

\multiput(0,20)(40,0){11}{\circle*{5}}

\put(-5,0){\scriptsize \mbox{$\ket{l}$}}
\put(105,0){\scriptsize \mbox{$(L^{-})^{k+1}\ket{h}$}}
\put(265,0){\scriptsize \mbox{$(L^{+})^{k+1}\ket{l}$}}
\put(395,0){\scriptsize \mbox{$\ket{h}$}}

\multiput(32,22)(40,0){2}{\vector(-1,0){24}}
\multiput(152,22)(40,0){7}{\vector(-1,0){24}}
\multiput(8,18)(40,0){7}{\vector(1,0){24}}
\multiput(328,18)(40,0){2}{\vector(1,0){24}}

\end{picture}

\caption{\footnotesize Diagram of an indecomposable representation as
defined by (\ref{repfrms}). The dots represent the basis states
$\ket{j,m},$ in particular, we have written $\ket{h}$ for the highest
weight state and $\ket{l}$ for the lowest weight state.  The arrows
$\rightarrow$ and $\leftarrow$ indicate the action of $L^{+}$ and
$L^{-}$ resp.}
\end{figure}
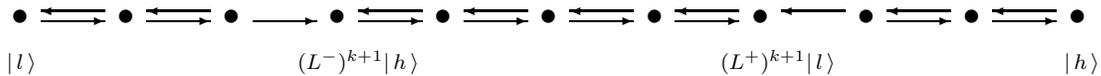

When $q$ is set equal to a root of unity, the properties of most of
the representations defined by (\ref{repfrms}) change quite
drastically. Specifically, at $q=e^{2 i \pi/(k+2)}$, the
representations $\pi^{2j}$ with $j>\frac{k+1}{2}$ will no longer be
irreducible. This can be traced back to the fact that For $q=e^{2\pi
i/(k+2)}$, the $q$-numbers satisfy the extra identity

\begin{equation}
\qnr{m+k+2}=-\qnr{m} 
\end{equation}
and more specifically

\begin{equation}
\qnr{k+2}=0 
\end{equation}
Because of this, one has $(L^{+})^{k+2}=(L^{-})^{k+2}=0$ in all the
representations defined by (\ref{repfrms}). Of course, in the
representations with $j<\frac{k+2}{2}$, this was already the case and
for these representations, nothing essential changes. In particular,
they are still irreducible. However, in the representations with $j\ge
\frac{k+2}{2}$, there will now be extra highest and lowest weight
states, which are annihilated by $L^{+}$ resp. $L^{-}$. For example,
the state $(L^{+})^{k+1}\ket{j,-j}$ in the module $V^{2j}$ of the
representation $\pi^{2j},~(j\le\frac{k+2}{2})$ will now be an extra
highest weight state, since $(L^{+})^{k+2}=0$ in this
representation. The descendants of this highest weight state (that is,
the states which can be obtained from it by applying powers of
$L^{-}$) now span an invariant subspace $W$ of $V^{2j}$, so $\pi^{2j}$
is no longer irreducible. Figure \ref{indecfig} illustrates this
situation in a simple case.

Although the module $\pi^{2j}$ is now reducible, it can not be written
as a direct sum of irreducibles. One says that it is
{\em indecomposable}. This indecomposability is directly related to the
fact that $\pi^{2j}$ is not a $*$-representation. For a
$*$-representation, the orthogonal complement of an invariant
submodule of the representation module is itself invariant and this
guarantees that any finite dimensional representation has an
orthogonal decomposition into irreducibles. The fact that $\pi^{2j}$
does not have a decomposition into irreducibles shows that it is not
just non unitary, but even non unitarisable. That is, it is impossible
to choose an inner product such that $\pi^{2j}$ is unitary with
respect to it.

Summarising, for $q=e^{2 i\pi/(k+2)}$, we are left with only $k+2$
irreducibles out of the infinitude that we would usually get from
(\ref{repfrms}). \footnote {Note that these irreps are by no means all
the irreps at $q=e^{i 2\pi/(k+2)};$ In fact, there are more irreps of
dimensions $1,\ldots,k+1$ (how many more depends on the precise
definition of $U_{q}(sl(2)),$ see e.g. \cite{chapress,keller}) and
there is a family of inequivalent representations of dimension $k+2,$
parametrised by a complex number $z.$ However, these representations
will not concern us here.} These are the unitary representations
$\pi^{2j}$ with $j<\frac{k+2}{2}$. The other representations defined
by (\ref{repfrms}) are no longer irreducible. They have become
indecomposable, and therefore they are non unitarisable.

\subsubsection{Tensor products of representations; the coproduct}
\label{tensprodsec}
 
The Hilbert space of a system of $N$ particles that carry quantum
group representations $\pi^{\Lambda_1}, \ldots,\pi^{\Lambda_{N}}$ is
just the tensor product $V^{\Lambda_1}, \ldots,V^{\Lambda_{N}}$ of the
modules of these representations. Thus, we want to have
representations of the quantum group on such tensor product
spaces. The decomposition of these tensor product representations into
irreducibles will then give the fusion rules for the particles in the
theory. Clearly, it is desirable that the tensor product decomposition
should be orthogonal, since otherwise the quantum group charges of
fusion products would not be well-defined. Orthogonality of the
decomposition can certainly be realised if the tensor product
representation is unitarisable. Therefore, we will spend some time
determining which tensor product representations of $U_{q}(sl(2))$ are
unitarisable.

For any quantum group $\mathcal{A}$, tensor product representations
are formed with the help of a so called coproduct. This coproduct is
an algebra homomorphism $\Delta$ from $\mathcal{A}$ to
$\mathcal{A}\otimes \mathcal{A}.$ The coproduct of $U_{q}(sl(2))$ is
given on the generators and the unit $1$ by

\begin{eqnarray}
\Delta(H)&=&1 \otimes H + H \otimes 1 \nonumber \\
\Delta(L^{\pm})&=&L^{\pm}\otimes q^{H/4}+q^{-H/4}\otimes L^{\pm}
\nonumber \\
\Delta(1)&=&1\otimes 1.
\end{eqnarray}
Given two representations $\pi^{\Lambda},\pi^{\Lambda'}$ of the
quantum group, we may form the tensor product representation by
the formula

\begin{equation}
\label{tensdef}
(\pi^{\Lambda}\otimes\pi^{\Lambda'})(x):=
(\pi^{\Lambda}\otimes\pi^{\Lambda'})(\Delta(x)). 
\end{equation} 
Here, $x$ is an arbitrary element of $\mathcal{A}$ and one may check
that, for $\mathcal{A}=U_{q}(sl(2))$ and $q=1,$ we recover the
ordinary tensor product of $U(sl(2))$ representations. the fact that
$\Delta$ is an algebra homomorphism ensures that the tensor product we
have just defined is indeed a representation of the algebra. Moreover,
this tensor product of representations is associative, because
$\Delta$ is coassociative, that is, $\Delta$ satisfies

\begin{equation}
\label{copcoas}
(1\otimes\Delta)\Delta=(\Delta\otimes 1)\Delta
\end{equation} 
A tensor product of $*$-representations will be itself a
$*$-representation with respect to the standard inner product on the
tensor product space if star and coproduct commute, that is, if one has

\begin{equation}
(*\otimes*)\Delta(x)=\Delta(*(x)).
\end{equation}
If this does not hold, it may still be possible to choose an inner
product other than the standard inner product with respect to which
the tensor product representation is a $*$-representation.  For
$U_{q}(sl(2))$, star and coproduct commute only when $q$ is real and
positive.  Thus, for $q\in \RR_{+}$, the tensor product of two unitary
representations is unitary and decomposes orthogonally with respect to
the standard inner product. When $q$ is not real, star and coproduct
do not commute and hence, the tensor product of two unitary
representations is not necessarily unitary with respect to the
standard inner product. Hence, tensor product decompositions should be
expected not to be orthogonal with respect to this inner product, but
they may still be orthogonal with respect to a different inner product
(see further on).
\footnote{Note that we do have $\sigma\circ (*\otimes
*)\Delta(x)=\Delta(*(x))$ for $|q|=1$. Here, $\sigma$ is the operator
which flips the factors of the tensor product. Thus, one might be
tempted to take $\sigma\circ (*\otimes *)$ as the $*$-structure on
$\mathcal{A}\otimes\mathcal{A}$ when $|q|=1$. This is done for example
in \cite{maschom}. In our context, this is not the right course to
pursue, since tensor product representations will still not be
$*$-representations with the new $*$ on
$\mathcal{A}\otimes\mathcal{A}$.}

Tensor product decompositions and even Clebsch-Gordan coefficients for
tensor products of unitary representations of $U_{q}(sl(2))$ may be
calculated similarly as for $U(sl(2)).$ The highest weight state
$\ket{jj}$ of each the irreducible representations in the tensor
product may be found by solving the equations $L^{+}\ket{jj}=0$ and
$(L^{-})^{2j}\ket{jj}=0.$ The other states are produced by the action
of $L^{-}$ on the highest weight states. In the calculations, the
following formula for the coproduct of $(L^{-})^{n}$ is a great help:

\begin{equation}
\label{cghelpfrm}
\Delta((L^{-})^{n})=(\Delta(L^{-}))^{n}=\sum_{m=0}^{n}
\qbin{n}{m}(L^{-1})^{m}q^{-(n-m)H/4}\otimes (L^{-})^{n-m}q^{m/4}
\end{equation} 
The $q$-binomial $\qbin{n}{m}$ in this formula is defined by

\begin{eqnarray}
\qbin{n}{m}&:=&\frac{\qnr{n}!}{\qnr{m}! \qnr{n-m}!} \\
\qnr{n}!&:=&\prod_{m=1}^{n}\qnr{m}
\end{eqnarray} 
When $q$ is not a root of unity, the tensor product representation
$\pi^{\Lambda}\otimes\pi^{\Lambda'}$ has the same decomposition into
irreps as for $q=1,$ i.e.

\begin{equation}
\label{tpdec}
\pi^{\Lambda}\hat{\otimes}\pi^{\Lambda'}=
\bigoplus_{\Lambda''=|\Lambda-\Lambda'|
}^{\Lambda+\Lambda'} \pi^{\Lambda''},
\end{equation}
where $\Lambda''$ increases in steps of $2.$ 

Explicit Clebsch-Gordan coefficients may be calculated for any tensor
product of irreducibles, using for example (\ref{cghelpfrm}). One
writes

\begin{equation}
\label{cgstate}
\ket{j,m}=\sum_{m_1,m_2}
\cgc{j_1}{j_2}{j}{m_1}{m_2}{m}_q\ket{j_1,m_1}
\ket{j_2,m_2}
\end{equation}
for the vector with $H$-eigenvalue $2m$ in the irrep $\pi^{2j}$ in the
decomposition of the tensor product $\pi^{2j_1}\otimes\pi^{2j_2}.$ The
above formula is only meant to introduce the notation for the
$q$-Clebsch-Gordan coefficients. Several general formulae for these
coefficients are proved in \cite{kirillov} and \cite{vaksman} and
collected in \cite{kirres}. We will not give these explicit
(complicated) formulae here, but we do give the coefficients for the
case $j_2=\frac{1}{2}$, as an illustration and because this case is of
interest to us later. For $j>0$, one has

\begin{eqnarray}
\label{cg2}
\ket{j+\shalf,j+\shalf-p}&=& 
q^{p/4}\sqrt{\frac{\qnr{2j+1-p}}{\qnr{2j+1}}}
\;\;\ket{j,j-p}\ket{\shalf,\shalf} +
\nonumber \\
~&~&q^{(p-2j-1)/4}\sqrt{\frac{\qnr{p}}{\qnr{2j+1}}}
\;\;\ket{j,j-p+1}\ket{\shalf,-\shalf} 
\nonumber \\
\ket{j-\shalf,j-\shalf-p}&=& 
q^{(p-2j)/4}\sqrt{\frac{\qnr{p+1}}{\qnr{2j+1}}}
\;\;\ket{j,j-p-1}\ket{\shalf,\shalf} -
\nonumber \\
~&~&q^{(p+1)/4}\sqrt{\frac{\qnr{2j-p}}{\qnr{2j+1}}}
\;\;\ket{j,j-p}\ket{\shalf,-\shalf}
\end{eqnarray}
The decomposition for $j=0$ should be obvious (it is the same as for
$U(sl(2))$). In making a decomposition such as the one above, one has
the freedom to multiply all the states in each summand irrep by a
constant phase factor.  Here, the phases are chosen in such a way
that, when $q$ goes to one, the coefficients reduce to the usual
Clebsch-Gordan coefficients for $U(sl(2)).$ One may check (for example
using (\ref{qnrids})) that, when $q$ is a real number, the tensor
product vectors on the right hand side are orthonormal, as they should
be, because the tensor product representation is unitary in this
case. However, as expected, if $q$ is not real, then the vectors above
are no longer normalised or orthogonal with respect to the standard
inner product on the tensor product space. This can be remedied by
choosing the inner product on the tensor product space which is {\em
defined} by the fact that it makes the states above orthonormal. With
respect to this $q$-deformed inner product, the tensor product
representation is a unitary representation of $U_{q}(sl(2))$ whenever
all the irreps that appear in its decomposition are unitary. This
happens when $q\in \RR$ (in which case the new inner product coincides
with the old one) and also when $|q|=e^{i\phi}$ with
$\phi\in\RR,~|\phi| \le \frac{2\pi}{2j+2}$.  For general tensor
products of two irreps, we have a similar situation. Here, too, we may
define an inner product by requiring that the different states given
by (\ref{cgstate}) are orthonormal and this inner product will make
the tensor product representation into a $*$-representation when the
irreps into which it decomposes are $*$-representations. {F}rom this
point, one may go on and define inner products on $N$-fold tensor
products of irreducibles by requiring that a complete set of states
obtained by iterative use of (\ref{cgstate}) is orthonormal. The set
of states that is declared orthonormal now depends on the order in
which one takes the tensor product of the irreducibles, so
$(V^{\Lambda_1}\otimes V^{\Lambda_2})\otimes V^{\Lambda_3}$ and
$V^{\Lambda_1}\otimes (V^{\Lambda_2}\otimes V^{\Lambda_3})$ may in
principle end up with a different inner product. We will say more
about this issue in sections \ref{sixjsec} and \ref{coassec}.

The orthogonality of the tensor product decomposition
which holds when $q$ is real is reflected in the following identity
for the $q$-Clebsch-Gordan coefficients:

\begin{equation}
\label{cgorth}
\sum_{m_1,m_2} \cgc{j_1}{j_2}{j}{m_1}{m_2}{m}_q
\cgc{j_1}{j_2}{j'}{m_1}{m_2}{m'}_{q}=\delta_{j,j'}\delta_{m,m'}
\end{equation}
Note that, although tensor product decomposition is orthogonal with
respect to the standard inner product only when $q$ is real and
positive, this equation holds by analytic continuation for all $q$
where the summands are not singular.

Another useful identity (taken from (\cite{kirres})), which relates
the coefficients for the tensor product $\pi^{2j_1}\otimes\pi^{2j_2}$
with those for the opposite tensor product is:

\begin{equation}
\label{cgopfrm}
\cgc{j_1}{j_2}{j_3}{m_1}{m_2}{m_3}_q=
(-1)^{j_1+j_2-j}
\cgc{j_2}{j_1}{j_3}{m_2}{m_1}{m_3}_{q^{-1}}
\end{equation}
In particular, this allows one to write down the Clebsch-Gordan
coefficients for $\pi^{1}\otimes\pi^{2j}$ using (\ref{cg2}).

\subsubsection{The root of unity case: truncated tensor products}
\label{ttpsec}

In the previous section, we described the tensor product of
representations of $U_{q}(sl(2))$ for the case that $q$ is not a root
of unity. In that case, many of the usual properties of tensor
products at $q=1$ could be recovered. For example, the tensor product
of two irreps could be decomposed into a direct sum of irreps
(see (\ref{tpdec})). Also, this decomposition is orthogonal with
respect to a suitable deformation of the standard inner product.
 
When $q$ is a root of unity, say $q=e^{i2\pi/(k+2)}$, the situation is
quite different. In this case, tensor products of two irreps will not
split into a direct sum of irreps, but will contain indecomposable
summands. This is not in itself surprising, because the
representations $\pi^{\Lambda}$ with $\Lambda>k+1$ that would occur in
the usual decomposition (\ref{tpdec}) become indecomposable for
$q=e^{i2\pi/(k+2)}$. However, what really happens is a bit more
complicated.  As an example, let us look at the decomposition of the
tensor product of the spin $\frac{1}{2}$ and the spin $\frac{k+1}{2}$
module.  As usual, the tensor product space may be decomposed into
eigenspaces of the operator $H$. These eigenspaces will be
one dimensional for the extremal eigenvalues $H=k+1$ and $H=-(k+1)$
and two dimensional for the other eigenvalues. If $q$ were not a root
of unity, then we would have two highest weight states in the tensor
product module. The $H=k+1$ state
$\ket{\frac{k+1}{2},\frac{k+1}{2}}\ket{\frac{1}{2},\frac{1}{2}}$ and
the $H=k-1$ state $\ket{\frac{k-1}{2},\frac{k-1}{2}}$ given in
(\ref{cg2}). At $q=e^{i2\pi/(k+2)}$, the coefficients of this second
state diverge, but if we multiply the state by $\qnr{k+2}$, then this
no longer happens and we still have two good highest weight
states. However, we have a third candidate highest weight state, which
is the $H=k-1$ state one gets when one lets $(L^{+})^{k+1}$ act on the
lowest weight state
$\ket{\frac{k+1}{2},-\frac{k+1}{2}}\ket{\frac{1}{2},-\frac{1}{2}}$
(remember $(L^{+})^{k+2}$ gives zero for this value of $q$). This new
highest weight state is just proportional to the state
$\ket{\frac{k+1}{2},\frac{k-1}{2}}$ given in (\ref{cg2}). Comparing
this state with the other highest weight state at $H=k-1$, we see that
although they would be linearly independent for any arbitrary $q$, they are
actually proportional to each other for $q=e^{2\pi i/(k+2)}$. It
follows that the irreducible spin $\frac{k-1}{2}$-module has become a
submodule of the module generated by the highest weight state at
$H=k+1$. Also, since we have only one highest weight state in the
$H=k-1$ eigenspace and since this space is two-dimensional, there must
also be a non-highest weight state in this module. The two dimensional
$H$-eigenspaces of the tensor product module will then be spanned by
a descendant of the highest weight state at $H=k-1$ and a descendant
of the non-highest weight state at $H=k-1$. We see thus that, at
$q=e^{2\pi i/(k+2)}$, the modules $\pi^{k+2}$ and
$\pi^{k}$ have disappeared from the decomposition of $\pi^{1}\otimes
\pi^{k+1}$ and in stead there is one indecomposable module, which has
the module $\pi^{k}$ as an irreducible submodule. 

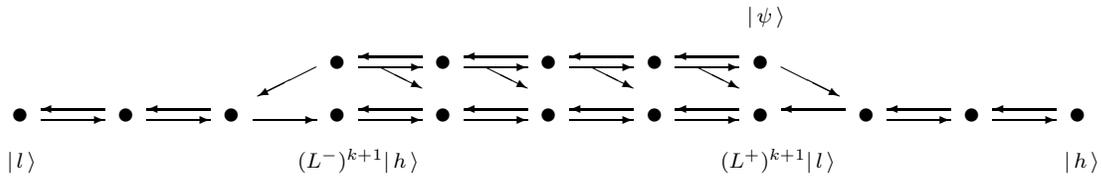
\begin{figure}[h,t,b]
\label{indectensfig}
\begin{picture}(450,65)(-10,0)

\multiput(0,20)(40,0){11}{\circle*{5}}
\multiput(120,40)(40,0){5}{\circle*{5}}

\put(-5,0){\scriptsize \mbox{$\ket{l}$}}
\put(105,0){\scriptsize \mbox{$(L^{-})^{k+1}\ket{h}$}}
\put(265,0){\scriptsize \mbox{$(L^{+})^{k+1}\ket{l}$}}
\put(395,0){\scriptsize \mbox{$\ket{h}$}}
\put(275,55){\scriptsize \mbox{$\ket{\psi}$}}

\multiput(32,22)(40,0){2}{\vector(-1,0){24}}
\multiput(152,22)(40,0){7}{\vector(-1,0){24}}
\multiput(8,18)(40,0){7}{\vector(1,0){24}}
\multiput(328,18)(40,0){2}{\vector(1,0){24}}

\multiput(152,42)(40,0){4}{\vector(-1,0){24}}
\put(112,38){\vector(-2,-1){22}}
\put(288,38){\vector(2,-1){22}}
\multiput(128,38)(40,0){4}{\vector(1,0){24}}
\multiput(136,38)(40,0){4}{\vector(2,-1){16}}

\end{picture}

\caption{\footnotesize Diagram of an indecomposable representation as
it would occur in the tensor product of two $U_{q}(sl(2))$-irreps at
$q=e^{2\pi i/(k+2)}$. The dots represent the basis states in the
module, the arrows $\rightarrow$ and $\leftarrow$ indicate the action
of $L^{+}$ and $L^{-}$ resp. The split arrows are meant to indicate
that the descendants of the state $\ket{\psi}$ are mapped onto linear
combinations of descendants of $\ket{\psi}$ and $(L^{+})^{k+1}\ket{l}$}
\end{figure}

This general picture extends to all tensor products of irreps; in
general, all the modules $\pi^{\Lambda}$ with $\Lambda>k$ and all the
corresponding modules $\pi^{2k-\Lambda}$ will disappear from the
decomposition \ref{tpdec} and in stead, there will be indecomposable
modules with the modules $\pi^{2k-\Lambda}$ as irreducible
submodule. The structure of these indecomposable modules is analogous
to the structure of the module we described above and is illustrated
in figure \ref{indectensfig}. For more detail on tensor product
decomposition when $q$ is a root of unity, one can consult for example
\cite{passal,keller,chapress}.

Clearly, the indecomposable representations which occur in the tensor
products are non unitarisable; this follows from the
indecomposability, but one can also see easily that any ``inner
product'' that would make these representations unitary would give the
states in the irreducible submodule zero norm. In fact, one can see
this happen explicitly for the limit of $q\rightarrow e^{2\pi
i/(k+2)}$ of the $q$-deformed inner product we defined at the end of
the previous section. 

In the present context, the states in the indecomposable representations
are considered non physical. Thus, we need to define a new
``tensor product'' $\ttp$ which is the the old tensor product with the
indecomposable modules projected out. However, the tensor product
defined this way would not be associative, since we would have for
example $(\pi^{1}\ttp\pi^{k})\ttp\pi^{k+1}=2\pi^{k+1}$ and
$\pi^{1}\ttp(\pi^{k}\ttp\pi^{k+1})=0$ for odd $k.$ (For even $k,$ we
get similar problems). Also, the fusion rule $\pi^{\Lambda}\otimes
\pi^{k+1}=0~ (\Lambda ~{\rm even})$ is clearly unphysical; after
adding a particle in the representation $\pi^{k+1}$ we would be left
without a Hilbert space ! These problems can be solved both at once by
projecting out not just the indecomposable modules, but also any
modules of type $\pi^{k+1}$ that may occur. The resulting tensor
product is called the {\em truncated tensor product.} Similarly as in
section (\ref{tensprodsec}), one may define an inner product on any
truncated tensor product of irreducible $U_{q}(sl(2))$-modules, which
makes the tensor product decomposition orthogonal. Moreover, since the
irreps into which a truncated tensor product factors are all
$*$-representations, this inner product makes the truncated tensor
product representation into a $*$-representation. The truncated tensor
product decomposition at $q=e^{i2\pi/(k+2)}$ is given by the following
formula, which is identical to the formula (\ref{wzwfus}) for the
fusion rules of $\widehat{sl(2)}_{k}$ chiral primaries:

\begin{equation}
\label{trunctens}
\pi^{\Lambda}\ttp\pi^{\Lambda'}=
\bigoplus_{\Lambda''=|\Lambda-\Lambda'|
}^{{\rm min}\{\Lambda+\Lambda',2k-\Lambda-\Lambda'\}}\pi^{\Lambda''}.
\end{equation}
From this formula, one may check easily that the truncated tensor
product is indeed associative, that is, the tensor product modules
$(\pi^{\Lambda_1}\ttp\pi^{\Lambda_2})\ttp\pi^{\Lambda_3}$ and
$\pi^{\Lambda_1}\ttp(\pi^{\Lambda_2}\ttp\pi^{\Lambda_3})$ are
isomorphic. Note however that these two modules are different
subspaces of the ordinary tensor product, so we might say that the
truncated tensor product is associative at the level of
$U_{q}(sl(2))$-modules, but not associative at the level of
states. This might seem like a problem at first sight, because we want
to have a unique three-particle Hilbert space, but this problem
disappears if we can find a canonical $U_{q}(sl(2))$-isomorphism
between the two three-particle spaces which preserves the inner
product.  We will say more about this in sections (\ref{sixjsec}) and
(\ref{coassec}).

As an illustration, let us take a closer look at the truncated tensor
product of the $2$-dimensional irrep $\pi^{1}$ with the unitary irreps
$\pi^{0},\pi^{1},\ldots,\pi^{k}$. For this case, the truncated tensor
product decomposition is given by

\begin{eqnarray}
\pi^{0}\ttp\pi^{1}&=&\pi^{1} \nonumber\\
\pi^{\Lambda}\ttp\pi^{1}&=&\pi^{\Lambda+1}\oplus\pi^{\Lambda-1}
~~~~(\Lambda\in\{1,\ldots,k-1\})
\nonumber \\
\pi^{k}\ttp\pi^{1}&=&\pi^{k-1}
\end{eqnarray}
As one can see, the only difference with the ordinary tensor product
occurs in the last line. The decomposition on the level of states can
be read off from (\ref{cg2}). Using this formula, we can also give an
example of the non-associativity at the level of states that we were
talking about: At $k=1$ (or $q=e^{2\pi i/3}$), the truncated tensor products
$V_1=(\pi^{1}\ttp\pi^{1})\ttp\pi^{1}$ and $V_2=\pi^{1}\ttp(\pi^1\ttp\pi^1)$
are both isomorphic to $\pi^1$ as $U_{q}(sl(2))$-modules, but any
state in $V_1$ may be written as 

\begin{equation}
\label{nonassex}
\frac{1}{\sqrt{\qnr{2}}}\left(
q^{-1/4} \ket{\shalf,-\shalf}\ket{\shalf,\shalf} -
q^{1/4}\ket{\shalf,\shalf}\ket{\shalf,-\shalf}\right)
\left(
\alpha_1\ket{\shalf,\shalf}+\alpha_2\ket{\shalf,-\shalf}
\right),
\end{equation}
while any state in $V_2$ may be written as

\begin{equation}
\frac{1}{\sqrt{\qnr{2}}}
\left(
\beta_1\ket{\shalf,\shalf}+\beta_2\ket{\shalf,-\shalf}
\right)
\left(
q^{-1/4}\ket{\shalf,-\shalf}\ket{\shalf,\shalf} -
q^{1/4}\ket{\shalf,\shalf}\ket{\shalf,-\shalf}\right).
\end{equation}
From this, we see that a vector in $V_1$ can only equal a vector in
$V_2$ if it is zero. Hence, $V_1$ and $V_2$ are different subspaces of
$\pi^1\otimes\pi^1\otimes\pi^1$.

Before ending this section, let us write down two useful identities
for truncated tensor decomposition which are related to the external
automorphism of $\widehat{sl(2)}_{k}$ that we discussed in relation to
the field identifications (\ref{fieldident}). If we define

\begin{equation}
\label{hatdef}
\hat{\Lambda}:=k-\Lambda,
\end{equation}
then we have

\begin{eqnarray}
\label{ttpids}
\hat{\Lambda}\ttp\Lambda'&=&
\bigoplus_{\Lambda''= |\Lambda-\Lambda'|}^{{\rm
min}\{\Lambda+\Lambda',2k-\Lambda-\Lambda'\}}\hat{\Lambda}'' 
\nonumber \\
\hat{\Lambda}\ttp\hat{\Lambda}'&=&
\bigoplus_{\Lambda''= |\Lambda-\Lambda'|}^{{\rm
min}\{\Lambda+\Lambda',2k-\Lambda-\Lambda'\}}\Lambda''
\end{eqnarray}
Here, we have written $\Lambda$ in stead of $\pi^{\Lambda}$ to avoid
notational overload. These identities tell us that the truncated fusion
rules of $U_{q}(sl(2))$ do not allow us to make a distinction between
a particle that carries the representation $\Lambda$ and a
particle that carries the representation $\hat{\Lambda}.$

\subsubsection{Counit, antipode and quantum trace}
 
Every quantum group $\mathcal{A}$ is required to have a one
dimensional representation $\epsilon,$ called the counit, which
represents the vacuum (or $\mathcal{A}$-neutral) sector of the
theory. The counit has to satisfy

\begin{equation}
\label{counit}
(\epsilon\otimes {\rm id})\Delta=({\rm id}\otimes\epsilon)(\Delta)={\rm id}
\end{equation}
where ${\rm id}$ is the identity map on $\mathcal{A}.$ It follows that
$\epsilon\otimes\pi=\pi\otimes\epsilon=\pi$ for any representation
$\pi$ of the quantum group, so the vacuum (or an $\mathcal{A}$-neutral
particle) has the fusion properties that one would expect. For
$U_q(sl(2)),$ we have already seen the counit; it is just the
representation $\pi^{0}.$ Explicitly, we have

\begin{equation}
\epsilon(1)=1,~~~~\epsilon(H)=0,~~~~\epsilon(L^{\pm})=0.
\end{equation}
The counit of a quantum group is an analogue of the trivial
representation of a group and hence, we will say that a state
$\ket{s}$ in a representation $\pi$ of the quantum group $\mathcal{A}$
transforms trivially if $\pi(a)\ket{s}=\epsilon(a)\ket{s}$ for all
$a\in \mathcal{A}.$ 

Next to the counit, $\mathcal{A}$ also required to have a linear
algebra antihomomorphism $S:\mathcal{A}\rightarrow\mathcal{A}$ which
satisfies 

\begin{equation}
\label{antipode}
\mu(S\otimes {\rm id})\Delta(a)=\mu({\rm id}\otimes
S)(\Delta)(a)=\epsilon(a) 1
\end{equation}
where, $\mu:\mathcal{A}\otimes\mathcal{A}\rightarrow\mathcal{A}$ is
just the multiplication of $\mathcal{A}$. $S$ is called the antipode
and $S(a)$ should be thought of as a quantum group analogue of the
inverse of $a$. If we are given a representation $\pi$ of
$\mathcal{A}$, then the antipode makes it possible to define the
representation $\bar{\pi}$ conjugate to $\pi$ by the formula

\begin{equation}
\bar{\pi}(a)=(\pi(S(a)))^{\rm t}
\end{equation}
The fact that $S$ is an antihomomorphism then makes sure that
$\bar{\pi}$ is a representation of $\mathcal{A}$, while the properties
(\ref{antipode}) ensure that the tensor product representations
$\pi\otimes\bar{\pi}$ and $\bar{\pi}\otimes\pi$ will contain the
trivial representation $\epsilon$ in their
decomposition. Thus, a particle which carries the representation
$\pi$ and its antiparticle, which carries the representation
$\bar{\pi}$ may indeed annihilate.

The antipode also allows us to define the action of
$\mathcal{A}$ on the space of linear operators on an
$\mathcal{A}$-module $V$ by

\begin{equation}
\label{opaction}
(a\cdot\hat{O})\ket{v}=a_{1}^{k}\hat{O}S(a_{2}^{k})\ket{v}
\end{equation}
Here, $\ket{v}$ is any state in the module $V$ and the $a_{1}^{k}$ and
$a_{2}^{k}$ are the left and right components of the coproduct of $a,$
i.e.

\begin{equation}
\Delta(a)=\sum_{k}a_{1}^{k}\otimes a_{2}^{k}
\end{equation}
Again using the fact that $S$ is an antihomomorphism, one can see that
(\ref{opaction}) defines a representation of $\mathcal{A},$
while using the property (\ref{antipode}), one can see that
$\mathcal{A}$ acts trivially on operators that commute with the action
of $\mathcal{A}$ on $V$. 

For $U_{q}(sl(2)),$ the antipode is given by

\begin{equation}
S(1)=1,~~~~S(H)=-H,~~~~S(L^{\pm})=-q^{\frac{\mp 1}{4}}L^{\pm}.
\end{equation}
The representations of $U_{q}(sl(2))$ are clearly all isomorphic to
their conjugates (one may see this for example from the decomposition
(\ref{tpdec}) without even using the explicit form of the antipode)
and the action of $U_{q}(sl(2))$ on an operator $\hat{O}$ is given by

\begin{eqnarray}
\label{opssl2q}
H\cdot\hat{O}&=&[H,\hat{O}] \nonumber\\
L^{\pm}\cdot\hat{O}&=&L^{\pm}\hat{O}q^{-H/4}-q^{-(H\pm 1)/4}\hat{O}L^{\pm},
\end{eqnarray}
which reduces to the usual commutator for $q\rightarrow 1$.

One can define a kind of trace on operators, which has the property
that it transforms trivially under $U_{q}(sl(2))$ when the operator is
transformed. For $q=1,$ the ordinary trace has this property, since
${\rm Tr}([a,\hat{O}])=0=\epsilon(a){\rm Tr}(\hat{O})$ for all $a\in
sl(2)$ and for arbitrary $\hat{O}.$ However, for $q\neq 1,$ we have to
use a modified trace to get this property. This trace is usually
called the quantum trace and we will denote it ${\rm Tr}_{q}$. Of
course, the quantum trace is supposed to preserve some nice properties
of the ordinary trace. Most importantly, the trace of a tensor product
of operators should be the product of the traces of the tensor factors. A
quantum trace with this property can be defined for a large class of
quantum groups (see cf. \cite{chapress}). For $U_{q}(sl(2))$, it is
given by

\begin{equation}
{\rm Tr}_{q}(\hat{O})={\rm Tr}(q^{H/2}\hat{O})
\end{equation}
One may verify readily that ${\rm Tr}_{q}(a\cdot
\hat{O})=\epsilon(a){\rm Tr}_{q}(\hat{O})$. The fact that ${\rm
Tr}_{q}(\hat{O}_1\otimes\hat{O}_2)={\rm Tr}_{q}(\hat{O}_{1}){\rm
Tr}_{q}(\hat{O}_{2})$ follows from the comultiplication
$\Delta(q^{H/2})=q^{H/2}\otimes q^{H/2}.$ 

Using the quantum trace, one may now define the quantum dimension
${\rm dim}_{q}(\pi)$ of a representation of the quantum group as the
quantum trace of the unit operator in this representation. For the
representations $\pi^{\Lambda}$ of $U_{q}(sl(2))$, this yields ${\rm
dim}_{q}(\pi^{\Lambda})=\qnr{\Lambda+1}=\qnr{{\rm dim}(\pi^{\Lambda})}.$
In particular, the quantum dimension of $\pi^{k+1}$ is zero. The
quantum dimensions of all the indecomposable modules of dimension
$2k+4$ that appeared in the (untruncated) tensor products of the
$\pi^{\Lambda}$ are also zero, since these modules were a
(non-direct) sum of two modules of dimensions $k+2-d$ and $k+2+d$
and we have $\qnr{k+2-d}+\qnr{k+2+d}=\qnr{d}+\qnr{-d}=0.$ Since the
quantum dimensions of the modules $\pi^{1},\dots,\pi^{k}$ are non
zero, we see that we might also have defined the truncated tensor
product of two modules in this set as the ordinary tensor product with
the modules of quantum dimension zero projected out. With this
definition, the truncated tensor product is automatically associative
and the module $\pi^{k+1}$ does not need separate treatment.

Quantum traces may also be used to construct knot invariants (see for
example \cite{gras},\cite{chapress} and references therein). For
$U_{q}(sl(2)),$ one of the knot invariants which can be constructed
this way is the famous Jones polynomial. 

\subsubsection{Braiding: the $R$-matrix}
\label{rmatsec}

Suppose that we have two particles that carry the $U_{q}(sl(2))$
representation $\pi^{\Lambda}$. The total internal state of the system
can then be represented by a state $\ket{s}$ in the (possibly
truncated) tensor product $\pi^{\Lambda}\ttp\pi^{\Lambda}.$ Now if we
exchange the particles, how does the state of the system change?  In
the $q=1$ case, we can describe the exchange simply by exchanging the
tensor factors in the state $\ket{s}$. Let us call this exchange of
the factors in the tensor product $\sigma.$ An exchange of two
adjacent particles in a system of $N$ identical particles may then be
described by the action of $\sigma$ on the corresponding factors of
the $N$-fold tensor product that describes the system. For example, in
a $4$-particle system, the exchange of the second and third particles
is affected by the operation $1\otimes\sigma\otimes 1.$ Such exchanges
generate a representation of the permutation group $S_{N}$. Also, they
commute with the action of the algebra at $q=1.$ this can be seen from
the formulae for the coproduct at $q=1$; the coproduct of any element
of the algebra is invariant under the exchange of the tensor
factors. We say that the coproduct is cocommutative. Because the
algebra action and the permutation group action commute, the
$N$-particle state space of the system can be decomposed into
representations of $U(sl(2))\times S_{N}.$

When $q\neq 1,$ the situation is different. The coproduct is now no
longer cocommutative and hence, exchanging the order in the tensor
product no longer commutes with the $U_{q}(sl(2))$ action. However,
this can be remedied with the use of a universal $R$-matrix. This is
an invertible element of $U_{q}(sl(2))\otimes U_{q}(sl(2))$ which acts
before $\sigma$ in the appropriate representation. For example, in a
system of three particles, all of which carry the representation $\pi$
of $U_{q}(sl(2)),$ the exchange of the first and second particles will
now be affected by $(\sigma\circ\pi\otimes\pi(R))\otimes 1.$ Note
that, in any tensor product of two representations, the universal
$R$-matrix does indeed act as a matrix, but the matrix in question
depends on the representations. The $R$-matrix is required to have the
following properties:

\begin{eqnarray}
\label{rprops}
\Delta^{\rm op}R&=&R\Delta \nonumber \\
(\Delta\otimes 1)R&=&R_{13}R_{23} \nonumber \\
(1\otimes \Delta)R&=&R_{13}R_{12}
\end{eqnarray} 
Here, $\Delta^{\rm op}$ is the comultiplication, followed by an exchange
of the tensor factors in $\mathcal{A}\otimes\mathcal{A}$ and
$R_{ij}$ is an abbreviation for the action of $R$ on the factors $i$
and $j$ of $\mathcal{A}^{\otimes 3},$ so for example
$R_{12}=R\otimes 1.$ The first of the properties above ensures that
the exchanges affected by $\sigma R$ in the tensor product
representations commute with the action of the quantum group, as we
wanted. The second and third property make sure that braiding of two
particles around a third one and then fusing them together gives the
same result as fusing the two particles first and then braiding the
result around the third one. Using the above properties of the
universal $R$-matrix, one may also prove the following equality, which
is called the quantum Yang-Baxter equation.

\begin{equation}
R_{12}R_{13}R_{23}=R_{23}R_{13}R_{12}
\end{equation}
This equation implies that in any representation, we have 

\begin{equation}
\label{qybe}
(\sigma R \otimes 1)(1\otimes\sigma R)
(\sigma R\otimes 1) =
(1\otimes\sigma R)(\sigma R \otimes1)
(1\otimes\sigma R).
\end{equation}
It follows that, for a system of $N$ identical particles that carry a
representation of the quantum group, the exchanges of adjacent
particles, as performed using $\sigma R,$ satisfy the relations
(\ref{braidgroup}). Hence, since $R$ is invertible, they generate a
representation of the braid group on $N$ strands, $B_N.$ Since the
exchanges commute with the action of the quantum group $\mathcal{A}$,
the system thus carries a representation of $\mathcal{A}\times B_{N}.$

When the particles do not all carry the same quantum group
representation (and are hence not identical), then the $R$-matrix no
longer gives us a representation of the braid group on the Hilbert
space of the system, simply because the exchanges now act between
different vector spaces; the flip operator $\sigma$ sends
$V^{\Lambda_1}\otimes V^{\Lambda_2}$ into $V^{\Lambda_2}\otimes
V^{\Lambda_1}$. This is not a problem, because exchanges of
non-identical particles are not symmetries of the system. What we do
still get from the $R$-matrix is a representation of a so called
coloured braid group, which consists of the braids for which the final
position of any particle is the original position of a particle of the
same kind (or ``colour''). These coloured braidings will still commute
with the quantum group action. One should note that the colouring
restriction leaves plenty of room for non-trivial and even nonabelian
monodromies between distinguishable particles. All the braiding
transformations in coloured braid groups may still be generated by
elementary exchanges of adjacent particles, although some of these
will no longer have any physical meaning and should be called
half-monodromies rather than braidings. We will be a bit sloppy about
this in the rest of this article, but we assume that this will not
cause confusion.

Before going on, let us note that one could multiply the $R$-matrix by
a constant phase factor. The matrices $(\sigma R)_{12},(\sigma
R)_{23},\ldots,(\sigma R)_{n-1,n},$ which model the exchange of
adjacent particles would then still generate a representation of the
braid group and commute with the quantum group. The second and third
condition in (\ref{rprops}), which express compatibility of fusion and
braiding, may also be satisfied if we give the coproduct $\Delta$ the
same phase factor as the $R$-matrix. The tensor product
``representation'' defined in (\ref{tensdef}) will then in general no
longer be a representation of the quantum group, but it will still be
a projective representation, which is really what we should expect in
a quantum mechanical setting. Thus, we may say that the universal
$R$-matrix determines the braiding of the particles only up to a
constant phase factor in each (clockwise) exchange (obviously, the
counterclockwise exchanges get the inverse phase factor).

The universal $R$-matrix which fulfills the requirements
(\ref{rprops}) for $U_{q(sl(2))}$ is given by

\begin{equation}
\label{univrmat}
R=q^{\frac{H\otimes H}{4}}\sum_{n=0}^{\infty}
\frac{(1-q^{-1})^{n}}{\qnr{n}!} q^{n(1-n)/4}
(q^{nH/4}(L^{+})^{n})\otimes (q^{-nH/4}(L^{-})^{n}),
\end{equation}
We see that, when $q$ approaches one, only the $n=0$ term in
(\ref{univrmat}) contributes and we get $R=1\otimes 1,$ as
expected. 

The action of the universal $R$-matrix on the module
$V^{\Lambda_1}\otimes V^{\Lambda}$ of the
tensor product representation $\pi^{\Lambda_1}\otimes\pi^{\Lambda}$
is given by

\begin{equation}
\label{rmatelts}
\begin{array}{rcl}
R\ket{j_1,m_1}\ket{j_2,m_2}&=&
\sum_{n\geq 0}\sqrt{\mbox{\scriptsize $\qbin{j_1-m_1}{n}\qbin{j_2+m_2}{n}$}
\frac{\qnr{j_1+m_1+n}!\qnr{j_2-m_2+n}!}{\qnr{j_1+m_1}!\qnr{j_2-m_2}!}}
q^{\frac{n(1-n)}{4}}\nonumber \\
~&~& \times
q^{\frac{1}{2}(m_2 n-m_1 n+2m_1m_2)}(1-q^{-1})^{n}
\ket{j_1,m_1+n}\ket{j_2,m_2-n}
\end{array}
\end{equation}
where the sum extends over all $n$ for which the kets on the right hand
side are well defined. Using this formula, one may easily find the
exchange matrix $\sigma R$ in any tensor product module. For example,
in the tensor product $\pi^{1}\otimes\pi^{1}$ of two two-dimensional
modules, we have

\begin{equation}
\label{xcmat}
\sigma R^{1,1}:=q^{-1/4}\left(
\begin{array}{cccc}
q^{1/2}&0&0&0\\
0&0&1&0\\
0&1&q^{1/2}-q^{-1/2}&0\\
0&0&0&q^{1/2}
\end{array}
\right).
\end{equation}
Note that, if $q\neq 1$, this is not a unitary matrix, which is not
good if it is supposed to represent a symmetry transformation on a
physical system.  Still, we could hope to make $\sigma R$ unitary by
choosing a suitable inner product on the module $V^{1}\otimes V^{1}.$
This will certainly not succeed unless $|q|=1$. To see this, note that
the eigenvalues of $R^{1,1}$ are $q^{1/4}$ (with multiplicity $3$) and
$-q^{-3/4}$ and these will only have norm $1$ (as required for the
eigenvalues of a unitary transformation) if $q$ does. However, if
$|q|=1,$ then $R^{1,1}$ is indeed unitary with respect to the
$q$-deformed inner product which we defined below formula (\ref{cg2})
and which makes the tensor product decomposition orthogonal. Of
course, the reason that this works is the fact that the action of $R$
commutes with the quantum group action, which means that the
eigenspaces of $R$ are submodules of the tensor product module. We
will say more about inner products and unitarity in section
\ref{sixjsec}.

In the above, we have always implicitly assumed that $q$ was not a
root of unity. If it is, then the whole story is more complicated,
because of the truncation of the tensor product which has to be taken
into account for these values of $q$. The $R$-matrix (\ref{univrmat})
still describes the braiding of two particles
\footnote {Note that this $R$-matrix is not well defined if
$q=e^{i2\pi/(k+2)}$, since the $q$-factorial $\qnr{n}!$ which appears
in the $(L^{+})^{n})\otimes (L^{-})^{n})$ term becomes zero for $n \ge
k+2.$ This problem is usually resolved by adding the relations
$(L^{+})^{k+2}=(L^{-})^{k+2}=0$ to the algebra for this value of $q.$
To us, this subtlety is not very important, since these relations
already hold in the unitary representations we are interested in},
but if we go to three or more particles, then we can get problems. For
example, three particles in the representation $\pi^{\Lambda}$ may be
described by a state in the truncated tensor product space
$(V^{\Lambda}\ttp V^{\Lambda})\ttp V^{\Lambda}$ and we can exchange
the two leftmost particles by means of $\sigma
(\pi^{\Lambda}\ttp\pi^{\Lambda})(R)\otimes 1$, which gives us a state
in $(V^{\Lambda}\ttp V^{\Lambda})\ttp V^{\Lambda}$, as it
should. However, if we want to exchange the two rightmost particles,
then we can leave the space $(V^{\Lambda}\ttp V^{\Lambda})\ttp
V^{\Lambda}$ if we just apply $1\otimes\sigma
(\pi^{\Lambda}\ttp\pi^{\Lambda})(R)$. One may see this explicitly in
the example we gave in formula (\ref{nonassex}); exchanging the last
two particles in this state, we get a state which can clearly not be
written in the same form and hence does not belong to $(V^{1}\ttp
V^{1})\ttp V^{1}$. If we use the other bracketing of the truncated
tensor product (i.e. $V^{\Lambda}\ttp (V^{\Lambda}\ttp V^{\Lambda})$),
then we can exchange the last two particles in the expected way, but
then the problem occurs in the exchange of the first two. In this way,
we can always expect problems when we try to exchange two particles
over a bracket. Thus, we will not get a representation of the braid
group on the truncated tensor product, unless we modify the way in
which we exchange particles. We will explain the modification that is
needed in some detail in section \ref{coassec}

\subsubsection{$q$-$6j$-symbols and their properties}
\label{sixjsec}

If we take a tensor product of three $U_{q}(sl(2))$ modules
$\pi^{\Lambda_1},\pi^{\Lambda_2}$ and $\pi^{\Lambda_3},$ then there are
two different ways to decompose this tensor product into
irreducibles. We may either first decompose the product
$\pi^{\Lambda_1}\otimes\pi^{\Lambda_2}$ and then the resulting modules
$\pi^{\Lambda'}\otimes \pi^{\Lambda_3}$, or we may first decompose the
product $\pi^{\Lambda_2}\otimes\pi^{\Lambda_3}$ and then the resulting
modules $\pi^{\Lambda_1}\otimes \pi^{\Lambda''}.$ These two procedures
yield two different natural bases for the vector space
$V^{\Lambda_1}\otimes V^{\Lambda_2}\otimes V^{\Lambda_3}.$ In each
case, the basis vectors are labelled by their $H$-eigenvalue, the
label of their overall fusion channel and the label of their
intermediate fusion channel (which is the representation into which
$\pi^{\Lambda_1}$ and $\pi^{\Lambda_2}$ fuse in the first case and the
representation into which $\pi^{\Lambda_2}$ and $\pi^{\Lambda_3}$ fuse
in the second case). Let us call the vectors in the first basis
$e^{j_1,j_2,j_3}_{j_{12},j,m}$ and the vectors in the second basis
$f^{j_1,j_2,j_3}_{j_{23},j',m'}.$ Here, $j_1,j_2$ and $j_3$ correspond
to $\Lambda_1,\Lambda_2$ and $\Lambda_3,$ $m$ and $m'$ give the
$H$-eigenvalues, $j$ and $j'$ give the overall fusion channels and
$j_{12}$ and $j_{23}$ represent the intermediate fusion channels. 
The vectors $e^{j_1,j_2,j_3}_{j_{12},j,m}$ and
$f^{j_1,j_2,j_3}_{j_{23},j',m'}$ may be written in terms of the
standard (product) basis for the tensor product by means of the
Clebsch-Gordan coefficients. We have

{\small
\begin{eqnarray}
\label{efcgexpr}
e^{j_1,j_2,j_3}_{j_{12},j,m}&=&
\sum_{m_1,m_2,m_3}
\cgc{j_1}{j_2}{j_{12}}{m_1}{m_2}{m_{12}}_q
\cgc{j_{12}}{j_3}{j}{m_{12}}{m_3}{m}_q
\ket{j_1,m_1}\ket{j_2,m_2}\ket{j_3,m_3}
\nonumber \\
f^{j_1,j_2,j_3}_{j_{23},j,m}&=&
\sum_{m_1,m_2,m_3}
\cgc{j_2}{j_3}{j_{23}}{m_2}{m_3}{m_{23}}_q
\cgc{j_1}{j_{23}}{j}{m_1}{m_{23}}{m}_q
\ket{j_1,m_1}\ket{j_2,m_2}\ket{j_3,m_3}
\end{eqnarray}}
where $m_{12}=m_1+m_2$ and $m_{23}=m_2+m_3$.
The vectors in the $e$-basis may also be expressed in terms of the
$f$-basis vectors and this expression takes the following form:

\begin{equation}
e^{j_1,j_2,j_3}_{j_{12},j,m}=
\sum_{j_{23},j',m'}\delta_{jj'}\delta_{mm'}
\sixj{j_1}{j_2}{j_3}{j}{j_{12}}{j_{23}}
f^{j_1,j_2,j_3}_{j_{23},j',m'}
\end{equation}
The coefficients represented by the curly brackets are now called the
$6j$-symbols of $U_{q}(sl(2)).$ By definition, these $q$-$6j$-symbols
must be equal to the $6j$-symbols for $SU(2)$ when $q$ equals
one. The $6j$-symbol in the formula above will clearly be zero unless the
representation $j_{12}$ occurs in the tensor product of the
representations $j_1$ and $j_2,$ the representation $j$ occurs in the
tensor product of the representations $j_{12}$ and $j_3,$ etcetera. It
follows that the $6j$-symbol will be zero unless its
arguments satisfy the following requirements:

\begin{eqnarray}
\label{argconds}
|j_1-j_2|\leq j_{12} \leq j_1+j_2,&~~~&j_1+j_2+j_{12}\in \ZZ
\nonumber \\
|j_2-j_3|\leq j_{23} \leq j_2+j_3,&~~~&j_2+j_3+j_{23}\in \ZZ
\nonumber \\
|j_{12}-j_3|\leq j \leq j_{12}+j_3,&~~~&j_{12}+j_3+j\in \ZZ
\nonumber \\
|j_1-j_{23}|\leq j \leq j_1+j_{23},&~~~&j_1+j_{23}+j\in \ZZ
\end{eqnarray}
If these requirements are met, then the $6j$-symbol may be written in
terms of Clebsch-Gordan coefficients; using (\ref{efcgexpr}) and the
relations (\ref{cgorth}), one easily finds that

{\small
\begin{equation}
\sixj{j_1}{j_2}{j_3}{j}{j_{12}}{j_{23}}=
\frac{{\displaystyle \sum_{m_2,m_3}}
\cgc{j_1}{j_2}{j_{12}}{m_1}{m_2}{m_{12}}_{q}
\cgc{j_{12}}{j_3}{j}{m_{12}}{m_3}{m}_{q}
\cgc{j_2}{j_3}{j_{23}}{m_1}{m_2}{m_{23}}_{q}}
{\cgc{j_1}{j_{23}}{j}{m_1}{m_{23}}{m}_{q}}
\end{equation}}
From this formula, one may obtain explicit formulae for the
$6j$-symbols. We will not do the (long) computations here, but just
give one of the possible explicit answers, as given in
\cite{kirres}(see also \cite{kaklim}).

\begin{equation}
\label{sixjsymbs}
\begin{array}{l}
\sixj{j_1}{j_2}{j_3}{j}{j_{12}}{j_{23}} \rule[-7mm]{0mm}{5mm}= \\[2mm]
~~~~~
\sqrt{\qnr{2j_{12}+1}\qnr{2j_{23}+1}}\,
\Delta(j_1,j_2,j_{12})\Delta(j_{12},j_3,j)
\Delta(j_2,j_3,j_{23})\Delta(j_1,j_{23},j)  \\[2mm]
~~~~~
\times \sum_{z} \left\{
\frac{(-1)^{z}\qnr{z+1}!}{\qnr{z-j_1-j_2-j_{12}}!
\qnr{z-j_{12}-j_3-j}!\qnr{z-j_2-j_3-j_{23}}!\qnr{z-j_1-j_{23}-j}!}
\right.
 \\[2mm]
~~~~~~~~~~~~~~~~~
\times \left. 
\frac{1}{\qnr{j_1+j_2+j_3+j-z}!\qnr{j_1+j_{12}+j_3+j_{23}-z}!
\qnr{j_2+j_{12}+j+j_{23}-z}!}
\right\}
\end{array} 
\end{equation}
where

\begin{equation}
\Delta(a,b,c):=
\sqrt{\frac{\qnr{-a+b+c}!\qnr{a-b+c}!\qnr{a+b-c}!}{\qnr{a+b+c+1}!}}
\end{equation}
The sum in (\ref{sixjsymbs}) is taken over all $z$ for which all the
$q$-factorials in the summands are well-defined.  

The $q$-$6j$-symbols are invariant under many symmetries (described in
\cite{kirres,kaklim}) which are analogues of the symmetries of the
$6j$-symbols of $SU(2)$ (see \cite{varshalovich}). For us, the most
important of these are the so called classical symmetries. These
symmetries can be treated slightly more elegantly if one works with
the $q$-Racah coefficients in stead of the $q$-$6j$-symbols. The Racah
coefficients are just the $6j$-symbols with a different normalisation;
they are given by the formula for the $6j$-symbols above with the
first square root factor left out. Invariance under the classical
symmetries means that the Racah coefficients remain unchanged under
permutations of the columns and under exchanging the upper and lower
entry in two columns simultaneously. In effect, this means that we
have the following identities for the $6j$-symbols

\begin{equation}
\label{classyms}
\begin{array}{ccccc}
\sixj{j_1}{j_2}{j_3}{j}{j_{12}}{j_{23}}&=&
\sixj{j_2}{j_1}{j}{j_3}{j_{12}}{j_{23}}&=&
\sqrt{\frac{\qnr{2j_{12}+1}\qnr{2j_{23}+1}}
{\qnr{2j_{2}+1}\qnr{2j+1}}}
\sixj{j_1}{j_{12}}{j_3}{j_{23}}{j_2}{j}\rule[-7mm]{0mm}{5mm} \\
\sixj{j_1}{j_2}{j_3}{j}{j_{12}}{j_{23}}&=&
\sixj{j_1}{j}{j_3}{j_2}{j_{23}}{j_{12}}&~&~
\end{array}
\end{equation}
and all the identities generated by these. The other symmetries of the
$6j$-symbols are analogues of the Regge and reflection symmetries.

When $q\in \RR_{+}$, the bases for the three-fold tensor product given
in (\ref{efcgexpr}) are orthonormal and hence the basis transformation
between these bases is unitary. As a consequence, the $6j$-symbols
satisfy the following orthogonality relation (see
cf. \cite{kirres})

\begin{equation}
\label{orth6j}
\sum_{j_{12}}\sixj{j_1}{j_2}{j_3}{j_4}{j_{12}}{j_{23}}
\sixj{j_1}{j_2}{j_3}{j_4}{j_{12}}{j^{'}_{23}}=\delta_{j_{23}j^{'}_{23}}
\end{equation}
Here, we have used the fact that the $6j$-symbols are real for $q\in
\RR_{+}$.  The equation above indeed tells us that the matrix for the
coordinate transformation between the $e$ and $f$ bases for the charge
$j$ subspace of $V^{j_1}\otimes V^{j_2}\otimes V^{j_3}$ is
real-orthogonal. When $q$ is not a positive real number, the above
relation for the $6j$-symbols remains valid by analytic continuation,
as long as the summands are not singular, but it does not tell us that
the matrix for the coordinate transformation we mentioned is
orthogonal unless all the $6j$-symbols that appear are real.  For
$|q|=1$, these $6j$-symbols will certainly be real as long as
$|\arg(q)|$ is small enough to make sure that all the $q$-numbers that
appear in these $6j$-symbols are positive. This will be the
case(cf. formula(\ref{sixjsymbs})) when

\begin{equation}
|\arg(q)|<\min_{j_{12}} \{
\frac{2 \pi}{j_1+j_2+j_{12}+1},\frac{2 \pi}{j_2+j_3+j_{23}+1},\frac{2
\pi}{j_{12}+j_3+j_4+1},\frac{2 \pi}{j_1+j_{23}+j_4+1} \},
\end{equation}
where the minimum is over all $j_{12}$ that appear in (\ref{orth6j}).
Hence we see that also for $|q|=1$, $|\arg(q)|$ small enough,
the matrix of the coordinate transformation from the $e$ to the $f$
basis of the charge $j$ subspace of the space $V^{2j_1}\otimes
V^{2j_2}\otimes V^{2j_3}$ is real-orthogonal.

It follows by iterative use of this fact that, for $|q|=1$, the
$q$-deformed inner product on a given $N$-fold tensor product of
irreducible $U_{q}(sl(2))$-modules that we defined in section
\ref{tensprodsec} can be made independent of the choice of the order
in which the tensoring of the irreducibles is performed by taking
$|\arg(q)|$ small enough. For the case $N=3$, the different sets of
vectors that are declared orthonormal are just the $e$ and the
$f$-vectors and the fact that the matrix which relates these is
unitary shows that declaring one set to be orthonormal is equivalent
to declaring the other set to be orthonormal. On the other hand, given
any fixed value of $|\arg(q)|$, it will not be difficult to construct
tensor product representations in which the inner product does depend
on the order of the tensoring. In fact, we can expect this to happen
as soon as the decomposition of the tensor product module contains
non-unitary irreps.

\subsubsection{Truncated $6j$-symbols; the coassociator}
\label{coassec}

When $q$ is a root of unity ($q=e^{\pi i/(k+2)}$), we can define
truncated $6j$-symbols, related to the truncated tensor product. For
these to be non-zero, the conditions (\ref{argconds}) have to be
changed in such a way that they require that $j_{12}$ be not just in
the tensor product, but even in the truncated tensor product of $j_1$
and $j_2,$ etcetera. This means that the upper bounds
$j_1+j_2,\ldots,j_1+j_{23}$ in (\ref{argconds}) are sharpened to ${\rm
min}\{j_1+j_2,k-j_1-j_2\}, \dots, {\rm
min}\{,j_1+j_{23},k-j_1-j_{23}\}.$ When the arguments satisfy these
sharpened conditions, the truncated $6j$-symbols are still given by
the formula (\ref{sixjsymbs}). The truncated $6j$-symbols defined in
this way give a canonical isomorphism between the truncated tensor
product modules $(V^{2j_1}\ttp V^{2j_2})\ttp V^{2j_3}$ and
$V^{2j_1}\ttp (V^{2j_2}\ttp V^{2j_3})$. Moreover, this isomorphism is
preserves the $q$-deformed inner products on the truncated tensor
product spaces. To see this, note first of all that the truncated
$6j$-symbols are real (this follows easily from
(\ref{sixjsymbs})). Also, it is known that the truncated $6j$-symbols
satisfy an analogue of the orthogonality relations (\ref{orth6j}). We
have

\begin{equation}
\label{truncorth6j}
\sum_{j_{12}}\sixj{j_1}{j_2}{j_3}{j_4}{j_{12}}{j_{23}}
\sixj{j_1}{j_2}{j_3}{j_4}{j_{12}}{j^{'}_{23}}=\delta_{j_{23}j^{'}_{23}},
\end{equation}
where the sum is now restricted to the $j_{12}$ that are allowed by
the truncated tensor product. It follows that the matrix of the
mapping between $(V^{2j_1}\ttp V^{2j_2})\ttp V^{2j_3}$ and
$V^{2j_1}\ttp (V^{2j_2}\ttp V^{2j_3})$ is real-orthogonal and hence
that the mapping preserves the inner product.  The proof of the
relations (\ref{truncorth6j}) uses the usual orthogonality relations
and the formula (ref{krsyms}) in section \ref{truncsymsec}.

Using the isomorphism given by the truncated $6j$-symbols, we can
identify the spaces $(V^{2j_1}\ttp V^{2j_2})\ttp V^{2j_3}$ and
$V^{2j_1}\ttp (V^{2j_2}\ttp V^{2j_3})$, so that we have a well-defined
three-particle Hilbert space. The isomorphism may also be used to
define braiding transformations on truncated tensor products. Recall
from the end of section \ref{rmatsec} that we could use the $R$-matrix
to define braiding of two particles, but that there were difficulties
if we wanted to braid particles ``over a single bracket'' in a
multi-particle Hilbert space. These difficulties can now be resolved
using the mappings given by the truncated $6j$-symbols. For example,
if we want to exchange the two rightmost particles in the
representation $(\pi^{2j}\ttp\pi^{2j})\ttp\pi^{2j}$, then we can first
use the $6j$-symbols to map the representation space onto that for
$\pi^{2j}\ttp(\pi^{2j}\ttp\pi^{2j})$, then use the $R$-matrix to
exchange the particles and finally use the inverse of the mapping
given by the $6j$-symbols to get back to the representation space of
$(\pi^{2j}\ttp\pi^{2j})\ttp\pi^{2j}$. Similarly, any braiding in a
multiple truncated tensor product may now be achieved by using the
$6j$-symbols to move the brackets around before and after the actual
braiding. 

\begin{figure}[h,t,b]
\label{assocfig}
\begin{picture}(450,150)(0,0)
\put(20,75){\scriptsize\mbox{$((V^{2j_1}\otimes V^{2j_2})\otimes
V^{2j_3})\otimes V^{2j_4}$}}
\put(110,85){\vector(1,1){30}}
\put(110,65){\vector(1,-1){30}}
\put(150,125){\scriptsize\mbox{$(V^{2j_1}\otimes V^{2j_2})\otimes
(V^{2j_3}\otimes V^{2j_4})$}}
\put(280,115){\vector(1,-1){30}}
\put(280,75){\scriptsize\mbox{$V^{2j_1}\otimes (V^{2j_2}\otimes
(V^{2j_3}\otimes V^{2j_4}))$}}
\put(70,25){\scriptsize\mbox{$(V^{2j_1}\otimes (V^{2j_2}\otimes
V^{2j_3}))\otimes V^{2j_4}$}}
\put(195,25){\vector(1,0){25}}
\put(230,25){\scriptsize\mbox{$V^{2j_1}\otimes ((V^{2j_2}\otimes
V^{2j_3})\otimes V^{2j_4})$}}
\put(280,35){\vector(1,1){30}}
\end{picture}
\caption{\footnotesize This diagram shows two ways of going from one
bracketing of a fourfold tensor product to another. The arrows denote
the canonical isomorphisms given by the coassociator (or the truncated
$6j$-symbols). The diagram will commute if the condition
(\ref{assoccond}) on the coassociator is satisfied}
\end{figure}
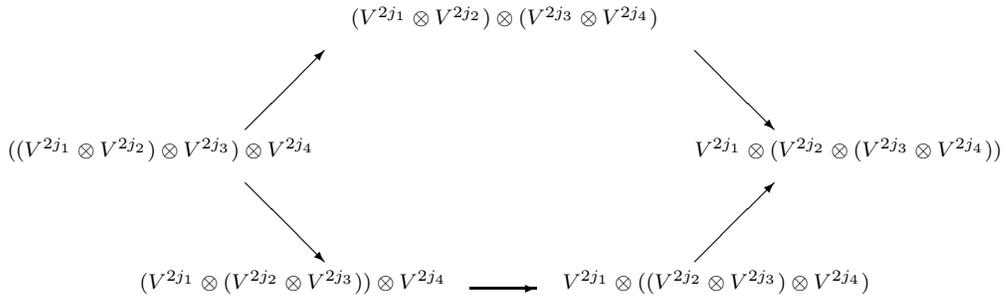

The whole procedure of truncating the tensor product so that it is no
longer associative and then defining braiding by identification
mappings may be elegantly formalised and brought to the level of the
algebra, at the cost of making the connection with non-truncated
$U_{q}(sl(2))$ somewhat less apparent. This has been done in
\cite{maschom} and the resulting structure is called a weak
quasitriangular quasi-Hopf algebra, or a weak quasi quantum group. Let
us call this $\mathcal{A}$. Some important features of the resulting
picture are the following. The coproduct is modified in such a way
that it has the truncation built in. As a result, one no longer has
$\Delta(1)=1\otimes 1$ and one also loses coassociativity. A so-called
coassociator is introduced to compensate for this loss. This
coassociator is an element
$\phi=\sum_{k}\phi^{1}_{k}\otimes\phi^{2}_{k}\otimes\phi^{3}_{k}$ of
$\mathcal{A}^{\otimes 3}$ which is not invertible in $\mathcal{A}$, but
which has a quasi-inverse called $\phi^{-1}$ which is the inverse in
all the representations that one is interested in and which has the
following important property for all $a\in \mathcal{A}$:

\begin{equation}
\phi (\Delta\otimes 1)\Delta(a) = (1 \otimes \Delta)\Delta(a)\phi
\end{equation}
This ensures that the representations
$(\pi^{2j_1}\ttp\pi^{2j_2})\ttp\pi^{2j_3}$ and $\pi^{2j_1}\ttp
(\pi^{2j_2}\ttp\pi^{2j_3})$ are isomorphic, with the isomorphism given
by $\pi^{2j_1}\ttp\pi^{2j_2}\ttp\pi^{2j_3}(\phi)$. Of course, this
isomorphism is just the one given by the truncated $6j$-symbols.
Clearly, one would like to be able to go from one bracketing of a
multiple tensor product to another, using $\phi$, in such a way that
it does not matter which individual steps are taken on the way. This
will be the case if the diagram in figure \ref{assocfig} commutes

To make this diagram commute, we need to impose the following
condition on the coassociator \cite{drinfquasi}:

\begin{equation}
\label{assoccond}
(1\otimes 1\otimes\Delta)(\phi)(\Delta\otimes 1\otimes 1)(\phi)=
(1\otimes \phi)(1\otimes\Delta\otimes 1)(\phi)(\phi\otimes 1)
\end{equation}
In terms of $6j$-symbols, this condition becomes

\begin{equation}
\label{penteq}
\begin{array}{l}
\sixj{j_{12}}{j_3}{j_4}{j}{j_{123}}{j_{34}}
\sixj{j_1}{j_2}{j_{34}}{j}{j_{12}}{j_{234}}\rule[-7mm]{0mm}{5mm}= \\
~~~~~~~
\sum_{j_{23}}\sixj{j_1}{j_2}{j_3}{j_{123}}{j_{12}}{j_{23}}
\sixj{j_1}{j_{23}}{j_4}{j}{j_{123}}{j_{234}}
\sixj{j_2}{j_3}{j_4}{j_{234}}{j_{23}}{j_{34}}
\end{array}
\end{equation}
This condition will clearly be satisfied for non-truncated
$6j$-symbols, since the sides of the equation just correspond to two
ways of doing the same coordinate transformation in that case. For
non-truncated $6j$-symbols the coordinate transformations change to
mappings that really do something, but the equation above still
holds.

When there is a non-trivial coassociator, the last two conditions in
(\ref{rprops}), which guaranteed the compatibility of fusion and
braiding, change to

\begin{eqnarray}
\label{rcoasprops}
(\Delta\otimes 1)(R)=\phi_{312}R_{13}\phi_{132}^{-1}R_{23}\phi_{123}
\nonumber \\
(1 \otimes\Delta)(R)=\phi^{-1}_{231}R_{13}\phi_{231}R_{12}\phi^{-1}_{123}
\end{eqnarray}
and these in turn imply the following quasi-Yang-Baxter equation
\cite{drinfquasi}, which is the analogue of (\ref{qybe}):

\begin{equation}
\label{quasiybe}
R_{12}\phi_{132}R_{13}\phi_{132}^{-1}R_{23}\phi_{123}=
\phi_{321}R_{23}\phi_{231}^{-1}R_{13}\phi_{213}R_{12}
\end{equation}
This relation ensures that the recipe that we gave for performing braidings
does indeed give a representation of the braid group.

\subsubsection{Symmetries of the truncated $6j$-symbols}
\label{truncsymsec}

The truncated $6j$-symbols still satisfy the classical, Regge and
reflection symmetries of the $6j$-symbols, but as it turns out, we
also have two sets of extra symmetries, which do not have an analogue at
$q=1.$ 

The first of these two sets was mentioned already in \cite{kirres} and
it is important in the proof of the truncated orthogonality relations
(\ref{truncorth6j}). If we define $\overline{j}:=k+1-j,$ then the
symmetries in this set can all be generated from the untruncated
symmetries (such as the classical symmetries (\ref{classyms})) and the
identity

\begin{equation}
\sixj{\overline{j_1}}{j_2}{j_3}{j}{j_{12}}{j_{23}}=
(-1)^{j_2+j_{23}-j-j_{12}+2j_1+1}
\sixj{\overline{j_1}}{j_2}{j_3}{j}{j_{12}}{j_{23}}
\end{equation}
In particular, we get from this that

\begin{equation}
\label{krsyms}
\sixj{{j_1}}{j_2}{j_3}{j}{j_{12}}{j_{23}}=
i (-1)^{j_2+j_3-j_1-j+2j_{12}+1}
\sixj{j_1}{j_2}{j_3}{j}{\overline{j_{12}}}{j_{23}}
\end{equation}
To prove the truncated orthogonality relations (\ref{truncorth6j}), we
may now start from the the untruncated orthogonality relations
(\ref{orth6j}). We may split the sum in (\ref{orth6j}) into three
parts as in

\begin{equation}
\begin{array}{l}
\sum_{j_{12}=\max\{|j_1-j_2|,|j_3-j_4|\}}^{\min\{j_1+j_2,j_3+j_4\}}
\rule[-7mm]{0mm}{5mm}= \\
\sum_{j_{12}=\max\{|j_1-j_2|,|j_3-j_4|\}}^{\min\{
j_1+j_2,j_3+j_4,k-j_1-j_2,k-j_3-j_4\}}+
\sum_{\min\{j_1+j_2,j_3+j_4,k-j_1-j_2,k-j_3-j_4\}+1}^{k-j_1-j_2}+
\sum_{k-j_1-j_2+1}^{\min\{j_1+j_2,j_3+j_4\}}
\end{array}
\end{equation}
Now if $\min \{ j_1+j_2,j_3+j_4,k-j_1-j_2,k-j_3-j_4\}$ equals
$j_1+j_2$ or $j_3+j_4,$ then all the $6j$-symbols in the last two
summations are zero, because their arguments don't satisfy the
conditions (\ref{argconds}). If $\min \{
j_1+j_2,j_3+j_4,k-j_1-j_2,k-j_3-j_4\}$ equals $k-j_1-j_2$, then the
second summation on the right hand side is empty and the third is zero
because the $j_{12}$ and $\overline{j_{12}}$ terms cancel each other
using (\ref{krsyms}) (if there is a middle term in the summation then
this also vanishes using (\ref{krsyms})). Finally, if $\min \{
j_1+j_2,j_3+j_4,k-j_1-j_2,k-j_3-j_4\}$ equals $k-j_3-j_4$, then one
can use the explicit formula (\ref{sixjsymbs}) for the $6j$-symbols to
show that all the terms of the middle summation vanish, while
(\ref{krsyms}) still makes sure that the last summation vanishes
because of pairwise cancellation of terms. In any case, the summation
on the left, which is the summation in (\ref{orth6j}), equals the
first summation on the right, which is the summation in
(\ref{truncorth6j}) and this shows the validity of the truncated
orthogonality relations.

We will now present the second set of symmetries of the truncated
$6j$-symbols, These symmetries are related to the identities
(\ref{ttpids}) for the truncated fusion rules and the seem to have
gone unnoticed until now. They are generated by the following
identities

\begin{equation}
\label{x6jsyms}
\sixj{j_1}{j_2}{j_3}{j}{j_{12}}{j_{23}}=
(-1)^{k+j_1+j_3+j_{12}+j_{23}}
\sixj{\hat{j}_1}{j_2}{\hat{j}_3}{j}{\hat{j}_{12}}{\hat{j}_{23}}=  
(-1)^{k+j_{12}+j_3+j}
\sixj{\hat{j}_1}{\hat{j}_2}{j_3}{j}{j_{12}}{\hat{j}_{23}}.
\end{equation}
Here, we have defined $\hat{j}:=\frac{k}{2}-j,$ in accordance with the
definition of (\ref{hatdef}) of $\hat{\Lambda}$. Of course all the
identities related to these by the classical, Regge and reflection
symmetries are also symmetries. The above identities may be proved in
the following way. First notice that the replacements of spins are
made in a way that is consistent with the truncated tensor product
decomposition. Hence, the arguments of the $6j$-symbol on the left
satisfy the truncated version of the conditions (\ref{argconds})
exactly if the arguments in the other two $6j$-symbols do. This means
we can fill in formula (\ref{sixjsymbs}) in all three cases. To show
that the results are equal, one needs an identity which holds for
$q$-factorials at $q=e^{2\pi i/(k+2)}$. We have

\begin{equation}
\label{qfacident}
\qnr{k+1-a}!=\frac{\qnr{k+1}!}{\qnr{a}!}
\end{equation}
Using this identity, it is easy to show that the $\Delta$-factors are
equal for all three $6j$-symbols in (\ref{x6jsyms}). For the middle
$6j$-symbol in (\ref{x6jsyms}), we can now see that the sum over $z$
in (\ref{sixjsymbs}) is equal to that for the untransformed
$6j$-symbol by making the substitution $z\rightarrow
z+k-(j_1+j_3+j_{12}+j_{23})$ and using the $q$-factorial identity
above twice. The proof for the rightmost $6j$-symbol in
(\ref{x6jsyms}) is similar, but uses the substitution $z\rightarrow
-z+k+j_3+j+j_{12}$.

\subsubsection{Braiding and $6j$-symbols}
\label{braidsixjsec}

In this section, we will give a systematic description of the braid
group representations that are associated with (truncated) tensor
products of $U_{q}(sl(2))$ representations. First of all, let us look
at the braiding in a tensor product of two irreps $\pi^{\Lambda_1}$
and $\pi^{\Lambda_2}$. We can decompose this tensor product into irreps
as in equation (\ref{tpdec}) or (\ref{trunctens}). {F}rom these
formulae, we see that any irrep can occur at most once in this
decomposition; we say that the tensor product decomposition is
multiplicity-free. It follows from this, using Schur's lemma, that
any map from the tensor product module $V^{\Lambda_1}\otimes
V^{\Lambda_2}$ to the tensor product module $V^{\Lambda_2}\otimes
V^{\Lambda_1}$ that commutes with the quantum group action on these
modules, is a constant on each of the irreducible summands of
$V^{\Lambda_1}\otimes V^{\Lambda_2}$ The exchange matrix $\sigma R
\equiv \sigma (\pi^{\Lambda_1}\otimes \pi^{\Lambda_2})(R)$ is such a
map. Hence, we can choose bases for $V^{\Lambda_1}\otimes
V^{\Lambda_2}$ and $V^{\Lambda_2}\otimes V^{\Lambda_1}$ such that the
action of $\sigma R$ is described by a diagonal matrix with respect to
these bases. Of course, the basis vectors in each case are just the
basis vectors $\ket{\frac{\Lambda}{2},m}$ of each irreducible summand
$\pi^{\Lambda}$ and the action of $\sigma R$ on these will depend
on $\Lambda_1,\Lambda_2$ and $\Lambda$ and not on $m$. Explicitly, one has

\begin{equation}
\label{braidconst}
\sigma (\pi^{\Lambda_1}\otimes\pi^{\Lambda_2})(R)|_{V^{\Lambda}}=
(-1)^{\frac{\Lambda_1}{2}+\frac{\Lambda_2}{2}-\frac{\Lambda}{2}}q^{\frac{1}{2}
(c_{\Lambda}-c_{\Lambda_1}-c_{\Lambda_2})}
\end{equation}
where $c_{\Lambda_{i}}=\frac{\Lambda_i}{2}(\frac{\Lambda_i}{2}+1)$ is
the value of the undeformed Casimir for the representation
$\pi^{\Lambda_i}$. This can be derived from the formula
(\ref{rmatelts}) for the elements of the $R$-matrix, using the
formulae for the Clebsch-Gordan coefficients given in
\cite{kirres}. For the case $\Lambda_2=2,$ one may also check it from
(\ref{cg2}), using (\ref{cgopfrm}). Note that the action of $\sigma R$
given above is effected by a unitary matrix exactly if $q$ is a root
of unity. Therefore, if we use the $q$-deformed inner product on the
tensor product space that makes the basis described above orthonormal,
then $\sigma R$ is a unitary operation, as we already remarked in a
special case below formula \nolinebreak (\ref{xcmat}).

Now let us look at a tensor product of $n$ quantum group irreps
$\pi^{\Lambda_1},\ldots,\pi^{\Lambda_n}$. In such a tensor product,
there are a number of natural bases which reflect the structure of the
tensor product. More precisely, there is on such basis for each way in
which the tensor product can be built up by adding subsequent
factors. We have already described the situation for three tensor
factors in the section \ref{sixjsec}. In this case, there were two of
these natural bases and the transformation that related these was
described by the $6j$-symbols.  In the case of $n$ factors, we will
choose to work with the natural basis one gets by adding subsequent
tensor factors on the right, i.e. the basis induced by the following
``bracketing'' of the tensor product:

\begin{equation}
\pi^{\Lambda_1}\otimes\pi^{\Lambda_2}\otimes\ldots\otimes\pi^{\Lambda_n}=
(\ldots(\pi^{\Lambda_1}\otimes\pi^{\Lambda_2})\otimes\pi^{\Lambda_3})
\ldots \otimes\pi^{\Lambda_{n-1}})\otimes\pi^{\Lambda_n})
\end{equation}
The elements of this basis can be labelled by their overall $H$
eigenvalue $m$, their overall fusion channel $j_n$ and and $n-1$
intermediate fusion channels $j_1,\ldots,j_{n-1}$. We may thus write
these basis elements as $e^{\mathcal{J}_1,
\ldots,\mathcal{J}_n}_{j_1,\ldots,j_n,m}$, where we have defined
$\mathcal{J}_i=\frac{\Lambda_i}{2}.$ Here, we have $j_1=\mathcal{J}_1$
and $j_i$ is one of the summands in $j_{i-1}\ttp \mathcal{J}_i$ for
$i>1.$ If there is no cause for confusion, we will suppress the upper
indices and write $e_{j_1,\ldots,j_n,m}$. It is easy to show that the
set of $e_{j_1,\ldots,j_n,m}$ for which all the $j$'s are equal forms
a basis for an irrep of $U_{q}(sl(2))$ of type $\pi^{2j_n},$ i.e. the
action on this set corresponds to the action given in formula
(\ref{repfrms}). Hence, it follows that the tensor product
representation becomes a $*$-representation if we take the inner
product which makes the $e_{j_1,\ldots,j_n,m}$ orthonormal and if each
of the possible $\pi^{2j_n}$ is itself a $*$-representation. Note that
if we are working with a truncated tensor product, then there will be
a different truncated tensor product space for each bracketing,
because of the non-associativity of this tensor product. The bases we
have described here then provide canonical bases for the different
subspaces of the ordinary tensor product that one gets from the
different bracketings.

The basis of $e_{j_1,\ldots,j_n,m}$ is very suited to a description of
the braiding. Suppose we want to exchange particles $i$ and $i+1,$
i.e. we want to calculate the action of the exchange $\tau_i$ on
$e_{j_1,\ldots,j_n,m}$. We can do this in three steps:

\begin{enumerate}
\item
Move particle $i$ completely to the left, using right-over-left
exchanges. Since the representations
$\pi^{\Lambda_1},\ldots,\pi^{\Lambda_{i-1}}$ fuse together to the
representation $\pi^{2j_{i-1}}$ and since the fusion of this
$\pi^{2j_{i-1}}$ with $\pi^{\Lambda_i}$ gives $\pi^{2j_i}$, this
operation gives us just a constant factor. We have

\begin{equation}
\label{step1}
e^{\mathcal{J}_1, \ldots,\mathcal{J}_n}_{j_1,\ldots,j_n,m}\rightarrow
(-1)^{j_i-\mathcal{J}_i-j_{i-1}}
q^{\frac{1}{2}(c_{j_{i-1}}+c_{\mathcal{J}_i}-c_{j_i})}
f^{\mathcal{J}_i,\mathcal{J}_1, \ldots,\mathcal{J}_n}_{j_1,\ldots,j_n,m}
\end{equation}
Here we have defined $c_j=j(j+1),$ in accordance with the definition
of $c_{\Lambda}$ above. Also, the vector
$f^{\mathcal{J}_i,\mathcal{J}_1,
\ldots,\mathcal{J}_n}_{j_1,\ldots,j_n,m}$ is an element of the natural
basis for the tensor product that one gets by first tensoring together
$\pi^{\Lambda_1},\ldots,\pi^{\Lambda_{i-1}}$, adding successive factors
on the right, then tensoring on $\pi^{\Lambda_i}$ from the left and
finally tensoring on the remaining factors from the right.  To get the
result (\ref{step1}), one uses (\ref{braidconst}) and, repeatedly,
(\ref{rprops}) or, for truncated tensor products, (\ref{rcoasprops}).

\item
Now change the bracketing, using the $6j$-symbols, so that we end up
in a basis in which the representations $\pi^{\Lambda_i}$ and
$\pi^{\Lambda_{1}},\ldots,\pi^{\Lambda_{i-1}}$ no longer fuse to a
fixed representation, but the representations
$\pi^{\Lambda_{1}},\ldots,\pi^{\Lambda_{i-1}}$ and
$\pi^{\Lambda_{i+1}}$ do. The new basis is the natural basis for the
tensor product which one gets by first tensoring together
$\pi^{\Lambda_1},\ldots,\pi^{\Lambda_{i-1}}$, adding successive factors
on the right, then tensoring on $\pi^{\Lambda_{i+1}}$ from the right,
then tensoring on $\pi^{\Lambda_i}$ from the left and finally
tensoring on the remaining factors from the right.  In the new basis,
the label $j_{i}$ (which gave the overall quantum group charge of
particles $1$ to $i$) is replaced by a new label $j',$ which gives the
overall quantum group charge of particles $1,2,\ldots,i-1,i+1$. All
the other labels are as before. If we denote the elements of the new
basis by $g^{\mathcal{J}_i,\mathcal{J}_1,
\ldots,\mathcal{J}_n}_{j_1,\ldots,j',\ldots,j_n,m}$, then the
$f$-basis can be written in terms of the $g$'s as

\begin{equation}
\label{step2}
f^{\mathcal{J}_i,\mathcal{J}_1,
\ldots,\mathcal{J}_n}_{j_1,\ldots,j_i,\ldots,j_n,m}=
\sum_{j'}\sixj{\mathcal{J}_i}{j_{i-1}}{\mathcal{J}_{i+1}}{j_{i+1}}{j_{i}}{j'}
g^{\mathcal{J}_i,\mathcal{J}_1, 
\ldots,\mathcal{J}_n}_{j_1,\ldots,j',\ldots,j_n,m}
\end{equation}
where we have used the fact that the representations carried by the
particles $1,\ldots,i-1$ fuse to $\pi^{2j_{i-1}}$ and that these
particles can thus be treated as one particle that carries the
representation $\pi^{2j_{i-1}}.$

\item
Now we move particle $i$ to the right, using left-over-right
exchanges, until it has reached the position to the right of particle
$i-1.$ At the end of this process, we have effectively only produced a
left-over-right exchange of the particles $i$ and $i+1$, as we
wanted. In the $g$ basis, the process of exchanging particle $i$ past
particles $1,\ldots,i-1$ and $i+1$ is described once again by a
simple phase factor (compare the first step of the calculation), since
the representations on particles $1,\ldots,i-1,i+1$ fuse to
$\pi^{2j'}$ and this fuses with $\pi^{2\mathcal{J}_{i}}$ into the fixed
fusion channel $\pi^{2j_{i+1}}.$ We get

\begin{equation}
\label{step3}
g^{\mathcal{J}_i,\mathcal{J}_1, 
\ldots,\mathcal{J}_n}_{j_1,\ldots,j',\ldots,j_n,m} \rightarrow
(-1)^{j'+\mathcal{J}_i-j_{i+1}}
q^{\frac{1}{2}(c_{j_{i+1}}-c_{j'}-c_{\mathcal{J}_i})}
e^{\mathcal{J}_1,\ldots,\mathcal{J}_{i+1},\mathcal{J}_{i}, 
\ldots,\mathcal{J}_n}_{j_1,\ldots,j',\ldots,j_n,m}
\end{equation}
where $e^{\mathcal{J}_1,\ldots,\mathcal{J}_{i+1},\mathcal{J}_{i}, 
\ldots,\mathcal{J}_n}_{j_1,\ldots,j',\ldots,j_n,m}$ is an element of
the basis we started with.
\end{enumerate}

We may now write down the action of the elementary exchange $\tau_i$
on the $e$-basis as the cumulative effect of these three steps. We
have

\begin{equation}
\label{braid6jeq}
\begin{array}{l}
\tau_i e^{\mathcal{J}_1, \ldots,\mathcal{J}_{i},\mathcal{J}_{i+1},
\ldots,\mathcal{J}_n}_{j_1,\ldots,j_i,\ldots,j_n,m}=  \\ ~~~~
\sum_{j'}(-1)^{j_i-j_{i-1}+j'-j_{i+1}}
q^{\frac{1}{2}(c_{j_{i-1}}-c_{j_i}+c_{j_{i+1}}-c_{j'})}
\sixj{\mathcal{J}_i}{j_{i-1}}{\mathcal{J}_{i+1}}{j_{i+1}}{j_{i}}{j'}
e^{\mathcal{J}_1,\ldots,\mathcal{J}_{i+1},\mathcal{J}_{i}, 
\ldots,\mathcal{J}_n}_{j_1,\ldots,j',\ldots,j_n,m}
\end{array}
\end{equation}
Using equation (\ref{truncorth6j}), one may check easily that the matrix
that describes this transformation is unitary if $q$ is a root
of unity, which is the case we are interested in. Hence, if we take the
inner product which makes the $e_{j_1,\ldots,j_n,m}$ orthonormal, then
the braid group representation which governs the exchanges of
particles with $U_{q}(sl(2))$-charges is unitary, as it should be. If
either $\mathcal{J}_{i}=\mathcal{J}_{i+1}$ or $j_{i-1}=j_{i+1},$ then
it follows from the classical symmetries (\ref{classyms}) that the
matrix for $\tau_i$ is also symmetric.

\subsubsection{Hidden quantum group symmetry}
\label{hidsymsec}

We will say that a quantum mechanical system has a hidden quantum
group symmetry if there is an action of a quantum group $\mathcal{A}$
on the Hilbert space of the theory which has the property that it
commutes with all the observables of the theory. For a system of
particles which carry $U_{q}(sl(2))$-representations, this means in
particular that the $H$-eigenvalues associated to the particles will
not be observable, while on the other hand, one can allow observables
which make it possible to determine the $U_{q}(sl(2))$-representation
associated to each of the particles. In other words, the total
``quantum spin'' of each particle would be measurable, but the
components of this quantum spin would not be measurable. The above
definition of hidden quantum group symmetry is just what we have
distilled from various sources in the literature that mention hidden
quantum group symmetries (see section \ref{qgcftsec} for
references). Note however that there does not seem to be a completely
standard definition of the concept of this concept. Let us say more
about what the above definition means within our context.  Suppose we
have a system of $n$ particles that carry representations
$\pi^{\Lambda_1},\ldots,\pi^{\Lambda_n}$ of a quantum group
$\mathcal{A}$. In that case the whole system will be in a state in the
tensor product space $V^{\Lambda_1}\otimes\ldots\otimes
V^{\Lambda_n}.$ If this tensor product may be decomposed into
irreducibles then the decomposition will take the form

\begin{equation}
\label{hiddecomp}
V^{\Lambda_1}\otimes\ldots\otimes V^{\Lambda_n}=
\bigoplus_{\Lambda} U^{\Lambda}_{\Lambda_1,\ldots,\Lambda_n}\otimes V^{\Lambda}
\end{equation}
Here, $U^{\Lambda}_{\Lambda_1,\ldots,\Lambda_n}$ is a vector space
whose dimension equals the multiplicity of the irrep $V^{\Lambda}$ of
$\mathcal{A}$ in the tensor product. When no confusion seems possible,
we will just write $U^{\Lambda}$. If the $\mathcal{A}$-symmetry of
this system is a hidden symmetry, then it follows that all the
observable operators act only on the spaces $U^{\Lambda}$ without
mixing these. That is, every observable $\hat{O}$ should take the form

\begin{equation}
\hat{O}=
\sum_{\Lambda} \hat{O}^{\Lambda}\otimes I_{V^{\Lambda}}
\end{equation}
where each $\hat{O}^{\Lambda}$ is an operator acting on $U^{\Lambda}$
and $I_{V^{\Lambda}}$ is the identity operator on $V^{\Lambda}$.
Since all the observables have this structure, the state of the system
can be uniquely characterised by a list of vectors, one for each of
the spaces $U^{\Lambda}$. Usually, the overall quantum group charge(s) of
the system will be well-defined. In other words, the state of the
system will be described by a vector in one of the summands in the
decomposition (\ref{hiddecomp}). In fact, there may be superselection
rules which prevent superposition of states from different summands in
(\ref{hiddecomp}). If the system as a whole is in the quantum group
representation $\pi^{\Lambda}$, then the state of the system may be
described by a vector in the space $U^{\Lambda}.$ Now note that the
braid group representation on the tensor product
$V^{\Lambda_1}\otimes\ldots\otimes V^{\Lambda_n}$ which comes from the
action of the $R$-matrix of $\mathcal{A}$ induces an action of the
braid group on each of the $U^{\Lambda}$. This follows from the fact
that the action of the braid group elements, like the action of any
observable, commutes with the action of $\mathcal{A}$. Any operator
that represents a braid group element will thus be of the general form
given above for observables. Thus, if one wants to describe only the
monodromy or braid group representation that governs the statistics of
the system at a fixed number of particles with given overall quantum
group charge $\Lambda$, one can restrict oneself to the space
$U^{\Lambda}$. It should be clear from the previous section what form
such a representation would take for a system of $n$ particles with a
hidden $U_{q}(sl(2))$ symmetry. In this case we have the canonical
basis of the $e^{\mathcal{J}_1,
\ldots,\mathcal{J}_n}_{j_1,\ldots,j_n,m}$ for the tensor product of
the $n$ representations of $U_{q}(sl(2))$. Of these, we need only
retain the ones whose overall charge $j_n$ is equal to the fixed total
charge of the system, say $j_n=j.$ The remaining vectors may then be
written as tensor product vectors:

\begin{equation}
e^{\mathcal{J}_1,
\ldots,\mathcal{J}_n}_{j_1,\ldots,j,m}=
e^{\mathcal{J}_1,
\ldots,\mathcal{J}_n}_{j_1,\ldots,j}\otimes\ket{j,m}
\end{equation}
where $e^{\mathcal{J}_1, \ldots,\mathcal{J}_n}_{j_1,\ldots,j}$ now
denotes a vector in the space $U^{j}$. The braid group
representation on $U^{j}$ may now be read off immediately from the
formula (\ref{braid6jeq}) which gave the braiding for the full tensor
product of $U_{q}(sl(2))$-representations. The matrix elements between
the $e^{\mathcal{J}_1, \ldots,\mathcal{J}_n}_{j_1,\ldots,j_n,m}$ which
are given in this formula can be used in unchanged form for the
vectors $e^{\mathcal{J}_1, \ldots,\mathcal{J}_n}_{j_1,\ldots,j}$,
since they already did not depend on $m$ and did not mix different
$j_n$. A similar treatment of braid group representations for systems
with hidden quantum group symmetry is possible in any situation in
which $6j$-symbols may be defined for the quantum group
representations involved. This is the case if the tensor products of
these representations have a multiplicity free decomposition into
irreducibles.

\subsubsection{Braiding of identical particles and fusion diagrams}
\label{braiddiagsec}

In the previous subsections, we have described the braiding for a
system of $n$ particles with a hidden $U_{q}(sl(2))$-symmetry and we
have indicated that a similar description should hold for many systems
that involve other quantum groups. Let us now look at the special case
in which the particles are identical. This case is of interest for the
description of the braiding of identical quasiholes in the
RR-states. When the particles are identical, they all carry the same
quantum group representation $\pi^{2\mathcal{J}}$ and hence the upper indices
on the elements $e^{\mathcal{J}_1,
\ldots,\mathcal{J}_n}_{j_1,\ldots,j}$ of the canonical basis for the
space $U^j$ are all equal to $\mathcal{J}$. Fixing $\mathcal{J}$, we
may thus forget about the upper indices and write just
$e_{j_1,\ldots,j_n}$. As in the previous section, we also fix
$j_n=j$. Now the $n$-tuple $(j_1,\ldots,j_n)$ may be seen as a path of
length $n$ through the space of representation labels of
$U_{q}(sl(2))$, which starts at the trivial representation $j_0=0$ and
ends at $j$. Of course, not all paths through the space of
representation labels will correspond to an element of the canonical
basis. A path will represent a basis vector precisely if the
representation at position $m$ of the path may always be found in the
tensor product of the representation at position $m-1$ with the
representation $\pi^{2\mathcal{J}}$. This in turn means precisely that the
path lies on the fusion (or Bratteli) diagram for the representation
$\pi^{2\mathcal{J}}$. Thus, the paths of length $n$ on the Fusion diagram of
the representation $\pi^{2\mathcal{J}}$ may be taken as a basis for the braid
group representation that describes exchanges in a system of $n$
particles that carry the quantum group representation
$\pi^{2\mathcal{J}}$. {F}rom equation (\ref{braid6jeq}), one may now easily
read off that the braid group generator $\tau_m$ will only mix paths
that are identical everywhere except at position $m$. 

As an example let us look at the case of $n$ particles in the
$2$-dimensional representation of $U_{q}(sl(2))$. The fusion diagram
for this representation is just the diagram drawn in figure
\ref{sl2fusdiag}.  Let $p=(p^{(1)},\ldots,p^{(n)})=
((\Lambda_{p}^{(1)},1),\ldots,(\Lambda_{p}^{(n)},n))$ be a path on
this diagram which starts at the point
$p^{(1)}=(\Lambda_{p}^{(1)},1)=(0,0)$ and ends at the point
$p^{(n)}=(\Lambda_{p}^{(n)},n)=(\Lambda,n).$ Then there is either no
path which differs from $p$ only at its $m^{\rm th}$ vertex or there
is exactly one such path. If there is such a path, we will call it
$\sigma_m(p).$ Let us write down the action of the exchange $\tau_{m}$
on a path $p.$ We start with the cases in which $p$ does not have a
partner path. Using equation (\ref{braid6jeq}), we see that such paths
just get a phase factor. There are four cases:

\begin{eqnarray}
\Lambda_{p}^{(m-1)}=2j<\Lambda_{p}^{(m)}<\Lambda_{p}^{(m+1)}~
&\Rightarrow&
\tau_m p =  q^{1/4}\, p 
\nonumber \\
\Lambda_{p}^{(m-1)}=2j>\Lambda_{p}^{(m)}>\Lambda_{p}^{(m+1)}~
&\Rightarrow&
\tau_m p =  q^{1/4}\, p 
\nonumber \\
\Lambda_{p}^{(m-1)}=\Lambda_{p}^{(m+1)}=0,~\Lambda_{p}^{(m)}=1~
&\Rightarrow&
\tau_m p = -q^{-3/4}\, p 
\nonumber \\
\Lambda_{p}^{(m-1)}=\Lambda_{p}^{(m+1)}=k,~\Lambda_{p}^{(m)}=k-1~
&\Rightarrow&
\tau_m p = -q^{-3/4}\, p 
\end{eqnarray}
These equations may be summarised by saying that the path $p$ gets a
factor of $q^{1/4}$ if it does not change direction at its $m^{\rm th}$
vertex, whereas it gets a factor $-q^{-3/4}$ if it does change
direction (which can only happen at the boundary of the diagram).  In
obtaining the equations, we used the following values for the
$6j$-symbols involved:

\begin{eqnarray}
\sixj{\frac{1}{2}}{j}{\frac{1}{2}}{j+1}{j+\frac{1}{2}}{j+\frac{1}{2}} &=&
\sixj{\frac{1}{2}}{j}{\frac{1}{2}}{j-1}{j-\frac{1}{2}}{j-\frac{1}{2}} =
\sixj{\frac{1}{2}}{0}{\frac{1}{2}}{0}{\frac{1}{2}}{\frac{1}{2}} = 1 
\nonumber\\
\sixj{\frac{1}{2}}{\frac{k}{2}}{\frac{1}{2}}{\frac{k}{2}}
{\frac{k-1}{2}}{\frac{k-1}{2}} &=& -1
\end{eqnarray}
We are now left  with the
case in which $p$ does have a partner path $\sigma(p).$ In this case,
we will certainly have $\Lambda_{p}^{(m-1)}=\Lambda_{p}^{(m+1)}=
\Lambda_{\sigma(p)}^{(m-1)}=\Lambda_{\sigma(p)}^{(m+1)}=2j$ and,
exchanging $p$ with $\sigma(p)$ if needed, we can also make sure that
$\Lambda_{p}^{(m)}>\Lambda_{\sigma(p)}$, so that
$\Lambda_{p}^{(m)}=2j+1,~\Lambda_{\sigma(p)}^{(m)}=2j-1$. The relevant
$6j$-symbols for this case are given by

\begin{eqnarray}
\sixj{\frac{1}{2}}{j}{\frac{1}{2}}{j}{j+\frac{1}{2}}{j+\frac{1}{2}} &=&
\frac{1}{\qnr{2j+1}}
\nonumber\\
\sixj{\frac{1}{2}}{j}{\frac{1}{2}}{j}{j-\frac{1}{2}}{j-\frac{1}{2}} &=&
\frac{-1}{\qnr{2j+1}}
\nonumber\\
\sixj{\frac{1}{2}}{j}{\frac{1}{2}}{j}{j+\frac{1}{2}}{j-\frac{1}{2}} &=&
\sixj{\frac{1}{2}}{j}{\frac{1}{2}}{j}{j-\frac{1}{2}}{j+\frac{1}{2}}=
-\frac{\sqrt{\qnr{2j+2}\qnr{2j}}}{\qnr{2j+1}}
\end{eqnarray}
and combining this with the phase factors in (\ref{braid6jeq}), we see
that, in the linear space with basis $\{p,\sigma(p)\}$, the exchange
$\tau_m$ is represented by the matrix

\begin{equation}
\label{sl2taummat}
\tau_m \equiv \frac{q^{-1/4}}{\qnr{d}}\left(
\begin{array}{cc}
-q^{-d/2} & -\sqrt{\qnr{d+1}\qnr{d-1}} \\
-\sqrt{\qnr{d+1}\qnr{d-1}}& q^{d/2}
\end{array} \right)
\end{equation}
where we have defined $d:=2j+1.$ This matrix for $\tau_m$ is obviously
symmetric. It is also unitary, as can be easily seen, using the fact
that $\qnr{d+1}\qnr{d-1}$ equals $\qnr{d}^2-1.$ We will denote the
braid group representation on the paths which start from $(0,0)$ and
end at $(\Lambda,n)$ by $\rho^{\Lambda}_{n}$ and the corresponding
modules by $U_{n}^{\Lambda}$. An induction argument taken from
\cite{wenzl} shows that the $\rho^{\Lambda}_{n}$ are all irreducible
and that they are non-isomorphic for different $\Lambda$: The
representation $\rho^{1}_{1}$ of the trivial group $B_0$ is
irreducible because it is one dimensional and $\Lambda=1$ is the only
possibility at $n=1$. Now suppose that, for all $\Lambda$ and all
$n<m$, all $\rho^{\Lambda}_{n}$ are irreducible and non-isomorphic for
different $\Lambda$. Then the representations $\rho^{\Lambda}_{m}$ are
irreducible and mutually non-isomorphic for all $\Lambda$. For let $S$
be a non-zero $B_{m-1}$-invariant submodule of $U^{\Lambda}_{m}$ and
suppose for convenience that $\Lambda$ does not equal $0$ or $k$. Then
$S$ has a unique decomposition into $B_{m-2}$-invariant submodules
which is clearly given by $U^{\Lambda}_{n}=U^{\Lambda-1}_{m-1}\oplus
U^{\Lambda+1}_{m-1}$ (just forget the last step in the paths). Because
the $\rho^{\Lambda}_{m-1}$ are non-isomorphic for different values of
$\Lambda$ (by the induction hypothesis), it follows immediately that
the $\rho^{\Lambda}_{m}$ are also non-isomorphic for different values
of $\Lambda$; their modules have different decompositions into
irreducible $B_{m-2}$-modules. Moreover, since
$\rho^{\Lambda-1}_{m-1}$ and $\rho^{\Lambda+1}_{m-1}$ are irreducible
and non-isomorphic, it follows that the only possible proper
$B_{m-1}$-invariant submodules of $U^{\Lambda}_{m}$ are
$U^{\Lambda-1}_{m-1}$ and $U^{\Lambda+1}_{m-1}$.  However, these will
clearly be mixed by the exchange $\tau_{m-1}$, so that
$U^{\Lambda}_{m}$ has no proper $B_{m-1}$-invariant subspaces. Hence
$S=U^{\Lambda}_{m}$ and $\rho^{\Lambda}_{m}$ is irreducible. Of course
if $\Lambda$ equals $0$ or $k$ then the argument becomes even simpler
and we need not repeat it.

\section{Conformal field theory and quantum groups}
\label{qgcftsec}

In this section, we review the correspondence between conformal field
theory and quantum groups. In section \ref{qgcftgen}, we give a short
general description of this correspondence, illustrated with the
example of $U_{q}(sl(2))$ versus the $\widehat{sl(2)}_{k}$
WZW-theory. In the next section, we go on to describe the quantum
group $U_{q}(sl(m))$ and its relation to the $\widehat{sl(m)}_{k}$
WZW-theory. In section \ref{heckesec}, we describe representations of
the braid group $B_n$ which factor over the Hecke algebra $H_{n,q}$.
These are important in the description of the braiding of a system of
$n$ particles with hidden $U_{q}(sl(m))$-symmetry. In section
\ref{bossec}, we describe a quantum group for the chiral
boson. Finally, in section \ref{qgparaf}, we indicate quantum groups
which correspond to the parafermion theory that is used in the
description of the Read-Rezayi states.

\subsection{The CFT-QG relation}
\label{qgcftgen}

The relation between quantum groups and conformal field theories has
been much studied over the years and it is believed that every
conformal field theory has associated to it some quantum group (or
generalisation thereof) with the following properties:

\begin{itemize}
\item
Each chiral primary field of the CFT (or equivalently: each
irreducible representation of the chiral algebra) corresponds to an
irreducible representation of the quantum group.
\item
The fusion algebra of the CFT is identical to the representation ring
of the quantum group, i.e. fusion of chiral primaries corresponds to
taking the tensor product of quantum group irreps.
\item
The braiding of the chiral primary fields in conformal blocks
corresponds to the braiding in the tensor product of
quantum group representations, as described by means of an
$R$-matrix and, if needed, a coassociator.  
\end{itemize}
The points above can be illustrated by the case of the
$\widehat{sl(2)}_k$ WZW-theory, whose associated quantum group is
$U_{q}(sl(2))$ at $q=e^{\frac{2\pi i}{k+2}}$. For this value of $q$,
the unitary irreducible representations $\pi^{\Lambda}$ of
$U_{q}(sl(2))$ that have positive quantum dimension are indeed in one
to one correspondence with the affine primary fields $G^{\Lambda}$ of
the WZW-theory. Moreover, comparing equations (\ref{wzwfus}) and
(\ref{trunctens}), we see that the fusion rules for the WZW-fields are
the same as the decomposition rules for tensor products of
$U_{q}(sl(2))$-representations. We described the braid group
representations associated to the fundamental representation of
$U_{q}(sl(2))$ in section \ref{braiddiagsec}. The braiding of the
corresponding conformal blocks of the WZW-theory was calculated by
Tsuchiya and Kanie \cite{tsuka87,tsuka88} and this braiding is indeed
the same as that described in section \ref{braiddiagsec}, up to a
renormalisation of the blocks. In connection with this, the
$q$-$6j$-symbols may be identified with the fusion matrix of the
$\widehat{sl(2)}_{k}$ conformal field theory as defined by Moore and
Seiberg \cite{ms88,ms89cqcft}. The pentagon equation for this fusion
matrix then corresponds to the equation (\ref{penteq}) (see also
figure \ref{assocfig}) and the hexagon equation is just the
quasi-Yang-Baxter equation (\ref{quasiybe}), written in terms of
$6j$-symbols by means of (\ref{braid6jeq}).

Note that it is essential in the above, that the truncated tensor product
of $U_{q}(sl(2))$-representations is used, rather than the ordinary
one. In other words, we may say that it is essential that one uses a
weak quasi-quantum group rather than an ordinary quantum group. This
is not a very special situation; the fusion rules of many CFTs cannot
be reproduced by those of ordinary quantum groups (or quantum groups
with an ordinary tensor product). On the other hand, there is
mathematical work \cite{majid, schomerus} in which it is shown that,
given a CFT, one may always find weak quasi quantum groups that will
reproduce its fusion and braiding properties. This does not mean that
it is known for all conformal field theories how the quantum group
generators can be represented in terms of operators in the conformal
field theory. In fact, no general construction for these operators
seems to be known, although several proposals have been made for CFTs
that have a Coulomb gas description
\cite{gomsier90,gomsier91,bmccp91}. Through this work, much is known
about the quantum groups for the WZW-models. In particular, it is well
known that for any semisimple Lie algebra $g,$ the $\hat{g}_{k}$
WZW-model and the quantum group $U_{q}(g)$ at $q=e^{2\pi
i/(k+\hat{g})}$ are related in the way we have described above (here
$\hat{g}$ is the dual Coxeter number of $g$). In the following, we
shall be especially interested in the case $g=sl(m),$ because of the
close relation between the parafermion theory that describes the
RR-states and the $\widehat{sl(2)}_k$ and $\widehat{sl(k)}_2$
WZW-theories.

Before we go on, let us cite a few general references on the relation
between CFTs and quantum groups.  Books that include information on
this are for example \cite{gras,fuchs} and a review article is
\cite{ags90}.  An early description of the correspondence between
$U_{q}(sl(2))$ and the $\widehat{sl(2)}_k$ WZW-theory can be found in
\cite{ags89}.

\subsection{$U_{q}(sl(m))$ and the $\widehat{sl(m)}_{k}$ WZW-theory}
\label{uqslmsec}

In this subsection, we recall some facts about the
quantum group $U_{q}(sl(m))$ that is associated with the
$\widehat{sl(m)}_{k}$ WZW-theory. $U_{q}(sl(m))$ is a $q$-deformation of
the universal enveloping algebra $U(sl(m))$ of $sl(m)$. Such a
$q$-deformation can be constructed for any simple Lie algebra $g$. If we
denote the simple roots of $g$ by $\alpha_{i}$, then we can
associate to each of these three generators
$H_{i},L^{+}_{i},L^{-}_{i}$ and these will generate $U_{q}(g)$ as an algebra,
subject to the relations

\begin{equation}
\begin{array}{l}
[H_{i},H_{j}]=0 \\ {} [H_{i},L^{\pm}_{j}]=\pm
2 \frac{(\alpha_{i},\alpha_{j})}{(\alpha_{i},\alpha_{j})}L^{\pm}_{j}
\rule[-5mm]{0mm}{5mm}\\
{} [L^{+}_{i},L^{-}_{j}]=\delta_{ij}\qnr{H_{i}}
\rule[-5mm]{0mm}{5mm} \\ {}
[L^{\pm}_{i},L^{\pm}_{j}]=\left\{
\begin{array}{l}
0~~{\rm if}~ A_{ij}=0~{\rm or}~i=j \\
{\displaystyle \sum_{s=0}^{1-A_{ij}}} (-1)^s q^{(\alpha_i,\alpha_i)s(s+A_{ij}-1)/2}
\mbox{\small $\qbin{\textstyle 1-A_{ij}}{\textstyle s}$}
(L^{\pm}_{i})^{1-A_{ij}-s}L^{\pm}_{j}(L^{\pm}_{i})^{s}~~{\rm
otherwise}
\end{array}\right.
\end{array}
\end{equation}
Here, $A$ is the Cartan matrix of $g$.  When $q=1$, these relations
reduce to the relations for the Chevalley-Serre basis of $U(g)$ and
when $g=sl(2),$ they reduce to the relations we gave in section
\ref{uqsl2sec}. If $q$ is not a root of unity, the irreducible
representation of $U_{q}(g)$ are labelled by dominant integral weights
of $g$ and one may give formulae for the action of the generators
which are similar to those given in (\ref{repfrms}). When $q$ is a
root of unity, one finds again that all these representations
remain well-defined, but many are no longer irreducible and in
particular there are indecomposable representations.

The coproduct $\Delta$, counit $\epsilon$ and antipode
$S$ are given by

\begin{eqnarray}
\Delta(H_{i})&=& 1\otimes H_{i} + H_{i}\otimes 1 \nonumber\\
\Delta(L^{\pm}_{i})&=&L^{\pm}_{i}\otimes
q^{H_{i}/4}+q^{-H_{i}/4}\otimes L^{\pm}_{i} \nonumber\\
\epsilon(1)&=&1,~~~\epsilon(L^{\pm}_{i})=\epsilon(H_{i})=0
\nonumber\\
S(H_{i})&=&-H_{i},~~~S(L^{\pm}_{i})=-q^{\rho/2}L^{\pm}_{i}q^{-\rho/2}
\end{eqnarray}
Here, $\rho$ is the Weyl-vector of $g$, which is equal to half the sum
of the positive roots, or equivalently, to the sum of the fundamental
weights. One may check easily that this comultiplication, counit and
antipode satisfy the conditions given in section \ref{uqsl2sec}. As
usual, one can define the tensor product of representations through
the formula (\ref{tensdef}) and as in the case of $U_{q}(sl(2))$, this
tensor product will usually not be fully decomposable if $q$ is a root
of unity. However, it is once more possible to define a truncated
tensor product which involves only a finite set of unitary irreducible
representations which is fully decomposable. If $q=e^{2\pi
i/(k+\hat{g})}$, where $\hat{g}$ is the dual Coxeter number of $g$,
then the irreducible representations involved are each labelled by a
dominant integral weight $\Lambda$ such that $(\Lambda,\theta)\le k$,
where $\theta$ is the highest root of $g$. Hence, they are in one to
one correspondence with the affine primary fields of the $\hat{g}_{k}$
WZW-theory. Moreover, as in the case of $sl(2)$, the decomposition
rules of the truncated tensor product are identical to the fusion
rules of the WZW-primaries. One may also define a quantum trace and a
corresponding quantum dimension and one may then go from the ordinary
to the truncated tensor product by projecting out modules of zero
quantum dimension.

An $R$-matrix that satisfies the requirements (\ref{rprops}) with the
given comultiplication is also known (see for example chapter eight of
\cite {chapress} for details and references), but it is in general not
so easy to obtain the exchange matrices in any given tensor product of
representations from this $R$-matrix. The reason for this is that, to
calculate the action of the $R$-matrix on a tensor product of
representations, one needs formulae for the action of the elements of
$U_{q}(g)$ associated to the roots of $g$ on both representations in
the tensor product. Although it is quite easy to obtain formulae
similar to (\ref{repfrms}) for the action of the raising and lowering
operators $L^{\pm}_{\alpha_{i}}$ associated with the simple roots
$\alpha_{i}$,, the same does not go for the action of the raising and
lowering operators that correspond to non-simple roots. Nevertheless,
the exchange matrices have been calculated in special cases, one of
which is important to us. This is the case of the tensor product of
the fundamental $m$-dimensional representation of $U_{q}(sl(m))$ with
itself (see \cite{chapress} for a detailed calculation). Let us denote
this fundamental representation by $\pi_{e_1}$, where $e_1$ denotes
the highest weight of the representation as in section \ref{slkcossec}.
We may then write

{\small
\begin{equation}
\label{slkxcmat}
\sigma (\pi_{e_1}\otimes\pi_{e_1})(R)=
q^{\frac{1}{2m}}\left(
q^{1/2}\sum_{i=1}^{m} E_{ii}\otimes E_{ii}+
\sum_{i\neq j}E_{ij}\otimes E_{ji} +
(q^{1/2}-q^{-1/2})\sum_{i<j}E_{jj}\otimes E_{ii}
\right)
\end{equation}
}
where $E_{ij}$ denotes the matrix whose $(i,j)$ entry is one and whose
other entries are zero. One may check easily that this formula gives
back the matrix (\ref{xcmat}) in the case of $sl(2)$. 

\begin{figure}[h,t,b]

\begin{picture}(300,120)(-40,-20)

\multiput(8,8)(60,0){4}{\vector(1,1){14}}
\multiput(38,38)(60,0){4}{\vector(1,1){14}}
\multiput(68,68)(60,0){3}{\vector(1,1){14}}
\multiput(38,22)(60,0){4}{\vector(1,-1){14}}
\multiput(68,52)(60,0){3}{\vector(1,-1){14}}
\multiput(98,82)(60,0){3}{\vector(1,-1){14}}

\put(0,2){\circle*{5}}
\put(27,22){\yodit{4pt}{1}{0}}
\put(57,50){\yodit{4pt}{2}{0}}
\put(90,85){\thrryodi{4pt}{1}{1}{1}}

\put(55,-6){\yodit{4pt}{1}{1}}
\put(85,23){\yodit{4pt}{2}{1}}
\put(117,52){\thrryodi{4pt}{2}{1}{1}}
\put(147,85){\thrryodi{4pt}{3}{1}{1}}

\put(115,-6){\yodit{4pt}{2}{2}}
\put(147,23){\thrryodi{4pt}{2}{2}{1}}
\put(175,52){\thrryodi{4pt}{3}{2}{1}}
\put(207,85){\thrryodi{4pt}{3}{3}{1}}

\put(176,-3){\thrryodi{4pt}{2}{2}{2}}
\put(207,23){\thrryodi{4pt}{3}{2}{2}}
\put(237,52){\thrryodi{4pt}{3}{3}{2}}

\put(236,-3){\thrryodi{4pt}{4}{2}{2}}
\end{picture}

\caption{\footnotesize The Bratteli diagram for the fundamental
representation of $U_{q}(sl(3))$ at $q=e^{2 i \pi/5}$. This is in fact
the same diagram as that shown in figure \ref{slkfusdiag}, but this
time each site in the diagram is uniquely labelled by a Young diagram
only. The diagrams in figure \ref{slkfusdiag} may be recovered by
removing columns of $3$ boxes.}
\label{youngfig}
\end{figure}
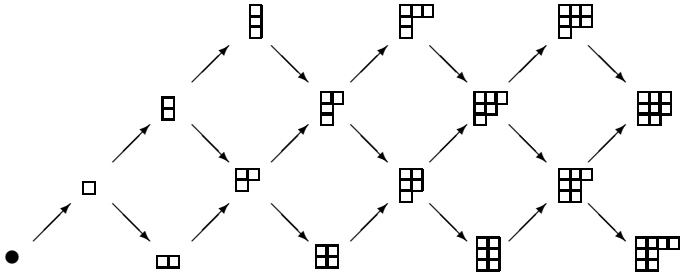

In the following, we want to describe the braid group representation
that is associated with an $n$-fold truncated tensor product
$\pi_{e_1}^{\hat{\otimes} n}$. Since tensor products that involve
$\pi_{e_1}$ are multiplicity free, we can apply the methods described
in sections \ref{braidsixjsec} and onwards. That is, we may define
$6j$-symbols for the tensor products involved and describe the
braiding by formula (\ref{braid6jeq}) and finally graphically, in
terms of paths on the fusion diagram of the representation
$\pi_{e_1}.$ In fact, we can already say quite a lot about the
braiding just from the fusion diagram, without a detailed knowledge of
the $6j$-symbols. So let us describe this fusion diagram. For
$q=e^{2\pi i/(k+\hat{g})}$, each vertex of the fusion diagram may be
labelled the number of fundamental representations that have been
tensored up to that point, together with a dominant integral weight of
$sl(m)$ which satisfies the requirement $(\Lambda,\theta)\le k$. We
may equivalently represent this weight by its Young diagram and if we
do this, then the requirement that $(\Lambda,\theta)\le k$ translates
to the restriction that the diagrams should not have more than $k$
columns. Fusion diagrams of this kind have already been drawn in
figures \ref{sl2fusdiag} and \ref{slkfusdiag}. Instead of using the
particle number and the Young diagram for the overall
$U_{q}(sl(m))$-charge, one may also use just a Young diagram to
represent each vertex. This Young diagram is then the diagram which
reduces to the Young diagram for $U_{q}(sl(m))$-charge if columns of
$m$ boxes are removed and whose number of boxes is equal to the number
of representations tensored up to that vertex. As an example, we show
a diagram for $U_{q}(sl(3))$ in figure \ref{youngfig}.

The connections between the different vertices are of course
determined by the fusion rules for the fundamental
representation. These can be elegantly described in terms of Young
diagrams. The truncated tensor product of the fundamental
representation with the representation that has Young diagram $Y$
decomposes into the sum of the representations whose Young diagrams
have at most $m$ columns and may be formed by adding one box to $Y$
and removing any columns of length $m$ that result. If one keeps
the columns of length $m$ then one obtains the Young diagrams which
label the vertices of the Bratteli diagram. The restriction on the number of
columns then applies only to the number of columns of length less than
$m$. 

The representation of $B_{n}$ that describes exchanges for a system of
$n$ particles with $U_{q}(sl(m))$ symmetry may now be described in
terms of these Bratteli diagrams. In fact, given the overall
$U_{q}(sl(m))$-charge of the system, we may find the Young diagram $Y$
with $n$ boxes which gives this overall charge and then the Braid
group representation space is just the space of paths on the diagram
which start at the empty diagram and end at $Y$. Moreover, each of the
exchanges $\tau_i$ will only mix paths that are the same everywhere
except possibly at the $i^{\rm th}$ vertex. Note that such paths occur
at most in pairs, since there are no more than two orders in which one
can place the two boxes that are added in going from vertex $i-1$ to
vertex $i+1$ (if the boxes are added in the same row, for example,
then there is only one admissible order and thus only one path). The
paths which are mixed transform into each other by means of unitary
matrices and since the diagrams all become periodic after a while, one
needs only to find a finite number of such matrices (see also section
\ref{tensbraidsec} for this). To find these matrices exactly, one
should calculate the $6j$-symbols of $U_{q}(sl(m))$. However, we will
not do this here, but in stead take a short cut by using the fact that
all the Braid group representations we need are related to
representations of the Hecke algebra $H_{n,q}$, whose representation
theory has been well studied.

\subsection{The Hecke algebra $H_{n,q}$}
\label{heckesec}

In this paragraph, we give a short description of an algebra which
plays an important role in our understanding of the braiding of
$U_{q}(sl(m))$-representations: the Hecke algebra $H_{n,q}$. We will
also describe the irreps of this algebra that are relevant to
us. 

$H_{n,q}$ may be defined as the complex algebra with generators
$1,g_1,g_2,\ldots,g_{n-1},$ subject to the relations

\begin{eqnarray}
\label{halgebra}
g_ig_j&=&g_jg_i ~~~~~~ (|i-j|\ge 2) \nonumber \\
g_ig_{i+1}g_i&=&g_{i+1}g_{i}g_{i+1} \nonumber \\
g_{i}^{2}&=&(q-1)g^{i}+q
\end{eqnarray}
From these relations, we see that $H_{n,q}$ is a $q$-deformation of
the group algebra $\CC S_{n}$ of the symmetric group, to which it
reduces at $q=1$. The reason that the Hecke algebra comes into play
in the braiding of $U_{q}(sl(2))$ representations is that the exchange
matrix for the fundamental representation of $U_{q}(sl(m))$, given in
(\ref{slkxcmat}), satisfies the following extra relation next to the
braiding relation given in (\ref{qybe}):

\begin{equation}
\label{rhecke}
(\sigma
R^{e_1,e_1})^{2}=(q^{\frac{m-1}{2m}}-q^{\frac{-m-1}{2m}})\sigma R^{e_1,e_1}+
q^{-1/m}(1\otimes 1)
\end{equation}
As a consequence of this relation, the braid group representation that
can be constructed from the $R$-matrix also gives a representation of
the Hecke algebra $H_{n,q}.$ This representation is given by the
prescription

\begin{equation}
\label{gtor}
g_i\mapsto q^{\frac{m+1}{2m}}(R^{e_1,e_1})_{i,i+1}
\end{equation}
and one may easily verify that (\ref{rprops}) and (\ref{rhecke})
guarantee that the relations (\ref{halgebra}) for the $g_i$ are
satisfied. 

The representations of the Hecke algebra which are induced by
$U_{q}(sl(m))$ in this way all factor over a quotient of the Hecke
algebra, the so called $m$-row quotient. This is because the exchange
matrix (\ref{slkxcmat}) satisfies even further relations apart from the
ones already given. For example, the $2$-row quotient of the Hecke
algebra (which is also called the Temperley-Lieb-Jones algebra) can be
defined by adding the following relations to those given in
(\ref{halgebra}):

\begin{equation}
1+g_i+g_{i+1}+g_{i}g_{i+1}+g_{i+1}g_{i}+g_{i}g_{i+1}g_{i}=0
\end{equation}
and one may check that the matrix $q^{3/4}\,R^{e_1,e_1}$ for
$U_{q}(sl(2))$ satisfies the corresponding equation. Note that the
situation we are describing already occurs in the $q=1$ case. In that
case, we are describing representations of $\CC S_{n}$ in which the
$m+1$-row antisymmetriser vanishes (as it should do for exchanges in a
tensor product of $m$-dimensional spaces). The Young tableaus for
these representations thus have at most $m$ rows. The equation above
indeed just says that the $3$-row antisymmetriser vanishes, if one
identifies the generators $\sigma_i$ of the permutation group with the
elements $-g_i$ of the Hecke algebra at $q=1$. We have kept this minus
sign for better compatibility with the literature on Hecke algebra
representations (for example \cite{wenzl}). The relations that need to
be added to the Hecke algebra relations to obtain the general $m$-row
quotient may be written similarly as above; they are just the
equations that say that the $m+1$-row antisymmetrisers vanish, with each
$\sigma_i$ replaced by $-g_i$.

The representation theory of the Hecke algebra is analogous to that of
the group algebra of the symmetric group as long as $q$ is not a root
of unity, but when $q$ is a root of unity (and $q\neq 1$), there are
complications, similar to those that arise in the representation
theory of $U_{q}(sl(m))$. In particular, the representation ring of
$H_{n,q}$ is no longer semisimple for these values of $q$. The same is
true for the representation ring of the $m$-row quotients. However,
one may restrict oneself to representations that factor over a certain
subquotient of the $m$-row quotient and these representations do form
a semisimple ring.  We will now give a quick description of the
irreducible representations of $H_{n,q}$ at $q=e^{2\pi i/(k+2)}$, that
factor over this semisimple quotient.  These representations (among
many others) have been constructed by Wenzl \cite{wenzl} and,
independently, by Ocneanu \cite{ocneanu}. Another relevant early
reference is \cite{hoefsmit}.  They are $q$-deformations of Young's
orthogonal representations of the symmetric group (see for example
\cite{boerner}). Each one of the representations we are interested in
is characterised by a Young diagram $Y$ that has at most $m$ rows and
at most $k$ columns of length less than $m$. The module of the
representation characterised by $Y$ is the module generated by all
paths on the Bratteli diagram of the fundamental representation of
$U_{q}(sl(m))$ that start at the empty diagram and end at $Y$. Thus,
we do indeed get the same representation modules that we described in
section \ref{uqslmsec}. Also, the action of the elementary exchanges
is of the kind we described in section \ref{uqslmsec}; $g_i$ does not
mix paths that differ in any place other than at their $i^{\rm th}$
vertex. Using the work of Wenzl and Ocneanu, we may now write down the
matrices through which $g_i$ acts on the spaces of paths that do
differ only at this vertex. Let us denote the Young diagram at the
$i^{\rm th}$ vertex of the path $p$ as $Y_{p}^{(i)}$, so that
$p=(Y_{p}^{(1)},Y_{p}^{(2)},\ldots,Y_{p}^{(n)}=Y)$. Then two paths $p$
and $p'$ can only be mapped into each other by $g_{i}$ if one has
$Y_{p}^{(n)}=Y_{p'}^{(n)}$ for all $n\neq i$. As we have already
remarked, the spaces of paths that are mapped into each other by
$g_{i}$ are at most two dimensional, since the last two boxes that are
added in going from the Young diagram at vertex $i-1$ to that at
vertex $i+1$ can be added in at most two different orders. Suppose
that there are indeed two admissible orders. These two orders then
correspond to two paths $p$ and $p'$ that form a basis for the space
that we are interested in. One may define the distance between the two
boxes involved as the number of hops from box to box that one has to
make if one walks from the first to the second box along the right
hand side of the Young diagram $Y_{p}^{(i+1)}$. One may also define a
signed version of this distance, which we will denote
$d_{p,i}$. Suppose that the first box is added in row $r_1$ and column
$c_1$ of the Young diagram $Y_{p}^{(i)}$ and that the second is added
in row $r_2$ and column $c_2$ of the Young diagram $Y_{p}^{(i+1)},$
then this signed distance is given by

\begin{equation}
d_{p,i}=r_2-r_1+c_1-c_2
\end{equation}
Clearly, this is equal to the ordinary distance if the first box is
located higher up and more to the right than the second. In the
opposite case the formula above gives minus the distance. Using this
signed distance, we may now write the action of the exchange $g_i$ on
the path $p$ as

\begin{equation}
g_i p = -\frac{q^{(1-d_{p,i})/2}}{\qnr{d_{p,i}}}\,p + {\rm
sign}(d_{p,i})\frac{\sqrt{\qnr{d_{p,i}+1}\qnr{d_{p,i}-1}}}
{\qnr{d_{p,i}}}\,p'
\end{equation}
Hence if $d_{p,i}>0$ (which may be achieved by exchanging $p$ and $p'$
if necessary), then the matrix for the action of the exchange $g_i$ on
the basis $\{p,p'\}$ is given by

\begin{equation}
\label{heckexcmat}
g_i \equiv -\frac{q^{1/2}}{\qnr{d_{p,i}}}\left(
\begin{array}{cc}
q^{-d_{p,i}/2} & \sqrt{\qnr{d_{p,i}+1}\qnr{d_{p,i}-1}} \\
\sqrt{\qnr{d_{p,i}+1}\qnr{d_{p,i}-1}}& -q^{d_{p,i}/2}
\end{array} \right)
\end{equation}
and one may check easily that this matrix is symmetric and unitary.

The action of the exchange $g_i$ on a path $p$ that
does not have a partner path, i.e. for which there is no other path
$p'\neq p$ such that $Y_{p}^{(n)}=Y_{p'}^{(n)}$ for all $n\neq i$, is
just multiplication by a phase factor. This phase factor equals $q$ if
$d_{p,i}=-1$, which means that the two boxes were added in the same
row, and it equals $-1$ if $d_{p,i}=1$, which means the boxes were
added to the same column, or if $d_{p,i}=k+1$, which happens  
at the ``edges'' of the fusion diagrams.

Clearly, the representations we have just described are the right ones
for the description of the braiding of particles with a hidden
$U_{q}(sl(m))$-symmetry. For a system of $n$ particles with overall
quantum group charge $\Lambda$, we need the representation labelled by
the Young diagram that consists of $n$ boxes and which reduces to the
Young diagram of the charge $\Lambda$ on removal of all columns with
$m$ boxes. With this correspondence, one may indeed check that the
formulae given in this section are the same as the ones we gave for
$U_{q}(sl(2))$ in section \ref{braiddiagsec}, up to the phase factor
in (\ref{gtor}). Using the explicit form of the exchange matrices, it
is not difficult to prove some nice mathematical properties of the
representations above. For example, Wenzl has proved \cite{wenzl} that
they are irreducible and that representations labelled by different
Young diagrams are non-isomorphic. The argument used in the proof
is essentially the same as the one we described at the end of section
\ref{braiddiagsec} for the case of $U_{q}(sl(2))$.

\subsection{A quantum group for the chiral boson}
\label{bossec}

In this section, we will indicate how a quantum group may reproduce
the fusion and braiding of the CFT that describes a chiral boson on a
circle of radius $\sqrt{m}$, where $m \in \ZZ$. This CFT has $m$
chiral primary fields, which are the vertex operators
$\nu_{l}=e^{il\phi/\sqrt{m}}$, for $m\in\{0,\ldots,k-1\}$ The vertex operator
$\nu_{l}$ has conformal weight $\frac{l^2}{2m}$ and the fusion is very
simple, one has

\begin{equation}
\label{bosfus}
\nu_{l_1}\times\nu_{l_2}=\nu_{l_1+l_2 ~{\rm mod}~m}
\end{equation}
All conformal blocks of primaries may be calculated explicitly, giving

\begin{equation}
\label{boscor}
\left<\nu_{l_1}(z_1),\ldots,\nu_{l_n}(z_n)\right>=
\prod_{i<j}(z_i-z_j)^{l_{i}l_{j}/m}
\end{equation}
It follows that the braiding of these blocks is abelian; the
half-monodromy which takes a $\nu_{l_1}$ around a $\nu_{l_2}$ gives
the block a factor of $e^{i l_1 l_2\pi/m}$.

We would like to reproduce the above data through a quantum group,
that is we would like to find a quantum group which has $m$
irreducible representations whose fusion rules and braiding are
identical to those of the boson vertex operators. It turns out that
this can not be achieved by a quantum group whose coassociator is
trivial. This can be most easily seen in the case $m=2$. In this case,
we would need a quantum group with two representations $\pi_0$ and
$\pi_1$ satisfying

\begin{eqnarray}
\pi_0\otimes\pi_0&=&\pi_0 \nonumber \\
\pi_0\otimes\pi_1&=&\pi_1 \nonumber \\
\pi_1\otimes\pi_1&=&\pi_0
\end{eqnarray}
Now if we look at the threefold tensor product
$\pi_1\otimes\pi_1\otimes\pi_1$, then we know on the one hand that
braiding the left $\pi_1$ over the other two must give two factors of
$e^{i\pi/2}$ yielding a total factor of $e^{i\pi}=-1$. On the other
hand, the braiding factor $bf$ is also given by the following formula
(which is only valid if the coassociator is trivial)

\begin{equation}
bf=\pi_{1}\otimes\pi_1\otimes\pi_1((1\otimes\Delta)(R))=\pi_1\otimes\pi_0(R)
\end{equation}
Here we have used the information that the two rightmost
representations must fuse to $\pi_0$. We are thus really just
exchanging the left $\pi_1$ over a $\pi_0$ and this should give a
factor of $e^{0 i\pi/m}=1$, which yields a contradiction.  To describe
the chiral boson, we should thus either use a quantum group with a
non-trivial coassociator or relax the demands on the correspondence
between quantum group and CFT. A good candidate for a
non-coassociative quantum group would be $U_{q}(sl(m))$ at
$q=e^{2i\pi/3}$ (or $k=1$). This weak quasi-quantum group does have
$m$ representations with the right fusion rules and the braiding for
the fundamental representation does reproduce that for the vertex
operator $\nu_1$, up to a trivial scalar factor in every
exchange. However, checking the correctness of the braiding for all
the other representations of $U_{q}(sl(m))$ seems to be rather
complicated and therefore we will choose for a different approach.

If we relax the demands on our quantum group such that more than one
quantum group representation may correspond to the same charge sector
in the CFT, then we can reproduce the braiding of the chiral boson CFT
by a very simple finite dimensional coassociative quantum group.  As a
Hopf algebra this quantum group is the group algebra of the cyclic
group $\ZZ_{2m}$. A convenient basis for this algebra is given by the
primitive idempotents $e_0, e_1,\dots, e_{2m-1}$ which project on the
isotypical components of the representations of $Z_{2m}$ in the group
algebra. In formulae: the $e_i$ are elements that satisfy

\begin{equation}
e_i e_j =\delta_{ij} e_i
\end{equation}
and the full set $\pi_0,\ldots, \pi_{2m-1}$ of irreps of $\ZZ_{2m}$ is given
by

\begin{equation}
\label{bosreps}
\pi_i(e_j)=\delta_{ij}
\end{equation}
The coproduct reads

\begin{equation}
\Delta(e_j)=\sum_i e_i \otimes e_{j-i~{\rm mod}~ 2m},
\end{equation}
and one may easily check that this leads to the fusion rules

\begin{equation}
\pi_i\otimes\pi_j=\pi_{i+j~{\rm mod}~ 2m}
\end{equation}
Counit and antipode are given by

\begin{equation}
\epsilon(e_i)=\pi_0(e_i)=\delta_{0i},~~~~~~S(e_i)=e_{2m-i}
\end{equation}
So far, we have just described the group algebra of $\ZZ_{2m}$ in
terms of the $e_i$. Now let us introduce an $R$-matrix. One may easily
check that the most general $R$-matrix which satisfies the requirements
(\ref{rprops}) is of the form

\begin{equation}
\label{bosrmat}
R=\sum_{i,j} q^{ij}e_i\otimes e_j
\end{equation}
where $q$ is an $m^{\rm th}$ root of unity. If we take $q=1,$ then
this is just the identity on $\CC\ZZ_{2m}\otimes\CC\ZZ_{2m}$ and
braiding is trivial. On the other hand, if $q\neq 1$ then we have
non-trivial braiding and the braiding factor we get when taking a
$\pi_i$ around a $\pi_j$ will be $q^{ij}.$

Let us call the quantum group we have just described $\CC\ZZ_{2m,q}$.
The correspondence between this simple quantum group and the chiral
boson CFT can now be made as follows: The chiral sector corresponding
to $\nu_{l}$ is represented by the two quantum group representations
$\pi_{l}$ and $\pi_{l+m}$. Of these, the $\pi_l$ with $0\le l<m$
represent the primary fields and their even conformal descendants,
while the $\pi_l$ with $m\le l<2m$ represent all the odd
descendants. The fusion rules of the quantum group are then consistent
with those of the CFT. In particular, it is impossible to distinguish
particles represented by the representations $\pi^{l}$ and $\pi^{l+m}$
by means of the tensor product decomposition rules for these
representations, just as it is impossible to distinguish the primary
field $\nu_{l}$ from one of its descendants by means of the CFTs fusion
rules.  The braiding is also correct, if we choose $q=e^{i\pi/m}$. The
braiding factors we get for two different representatives of the same
CFT-sector may now differ by a minus sign, but this is in fact just
what we want. To clarify this, let us look once more at the example we
gave for the case $m=2.$ For this case, the vacuum sector will now be
represented by the representation $\pi_0$ and also by the
representation $\pi_2,$ but if we exchange a $\pi_1$ with a $\pi_0$
then we get a factor of $1$, while if we exchange a $\pi_1$ with a
$\pi_2,$ we get factor of $-1$. We already know that if we have three
$\nu_1$ fields and we exchange the first over the last two, we will
get a factor of $-1$. This is due to the fact that the correlator
(\ref{boscor}) will have a single zero at any point where two
$\nu_1$-fields are brought together. If one would just place a $\nu_0$
at this point there would be no such zero and hence also no braiding
factor. Hence we can think of $\pi_0$ as representing the vacuum
sector, while we can think of $\pi_2$ as representing a charge-neutral
bound state of two $\nu_1$ fields.

\subsection{Quantum groups for the parafermions}
\label{qgparaf}

In this section, we shall describe quantum groups for the
$\ZZ_{k}$-parafermion conformal field theory that is used in the
description of the RR-states. Since there are two different coset
descriptions of this CFT (cf sections \ref{slkcossec} and
\ref{sl2cossec}), one can also expect to get two different quantum
group theoretic descriptions.

\subsubsection{The quantum group for $\widehat{sl(2)}_{k}/\widehat{U(1)}_{k}$}
\label{qgsl2sec}

Let us start with the coset $\widehat{sl(2)}_{k}/\widehat{U(1)}_{k}$. For this
coset, we have the factorisation formula (\ref{parafwzw}), which
describes a Virasoro primary field of the $\widehat{sl(2)}_{k}$ theory
as a product of a parafermion field an a vertex operator for a chiral
boson that lives on a circle of radius $\sqrt{2k}$. We already used
this formula to explain the conformal weights an fusion rules of the
parafermions and clearly, it can also be used to calculate
braidings. To see this, look at the following equality of correlators
which follows from (\ref{paraffus}):

\begin{equation}
\label{braidfaceq}
\left<G^{\Lambda_1}_{\lambda_1}(z_1),
\ldots,G^{\Lambda_n}_{\lambda_n}(z_n) \right>=
\left<\Phi^{\Lambda_1}_{\lambda_1}(z_1),
\ldots,\Phi^{\Lambda_n}_{\lambda_n}(z_n)\right>
\left<e^{i\lambda_1\phi}(z_1),
\ldots,e^{\lambda_n \phi}(z_n)\right>
\end{equation}
The braiding on the left hand side of this equation is just a braiding
of $\widehat{sl(2)}_{k}$-fields and the matrices which describe this
are known to be the same matrices that describe the braiding of
$U_{q}(sl(2))$ representations at $q=e^{2i\pi/(k+2)}$ (the labels
$\lambda_i$ do not play a role in the braiding). The braiding on the
right hand side will be described by matrices which are products of a
matrix for braiding the parafermions and a known scalar factor for braiding
the boson's vertex operators. Thus, we may obtain the braiding
matrices for the parafermion fields by just bringing the scalar factor
obtained from the bosonic correlator to the left.

The braiding matrices which are obtained from this recipe are the same
braiding matrices that one gets for the quantum group
$\mathcal{A}_{q_1,q_2}:=U_{q_1}(sl(2))\otimes \CC\ZZ_{4k,q_2}$,
where $q_1=e^{2i\pi/(k+2)}$ and $q_2=e^{-i\pi/{2k}}$. The irreducible
representations of this quantum group are tensor products of
$U_{q_1}(sl(2))$-irreps and $\CC\ZZ_{4k,q_2}$-irreps and hence
they are labelled by an $sl(2)$-weight $0\le\Lambda\le k$ and an
integer $0\le \lambda<4k$. We will write these representations as
$\pi^{\Lambda}_{\lambda}$. The representation
$\pi^{\Lambda}_{\lambda}$ will represent the
$\Phi^{\Lambda}_{\lambda~{\rm mod}~2k}$-sector of the parafermion
CFT. As in the case of the chiral boson we thus have more than one
quantum group representation that corresponds to the same sector of
the CFT. In fact, we now have four quantum group representations for
every sector of the CFT, since not only have we doubled the period of
the label $\lambda$ (as we did for the boson), but we have also not
taken the second of the field identifications (\ref{fieldident}) into
account. Looking at this identification, we see that the labels
$(\Lambda,\lambda)$ and $(k-\Lambda,\lambda-k),$ that should be
identified, usually stand for quantum group representations of
different dimensions ($\Lambda+1$ and $k-\Lambda+1$ respectively),
although their quantum dimensions are equal ($\qnr{\Lambda+1}=
\qnr{k+2-(\Lambda+1)} =\qnr{k-\Lambda+1}$).  Nevertheless, we believe
that the quantum group $\mathcal{A}_{q_1,q_2}$ will give a good
description of the CFTs braiding properties. To motivate this
statement, let us first look at the tensor product decomposition of
$\mathcal{A}_{q_1,q_2}$. This is given by

\begin{equation}
\pi^{\Lambda}_{\lambda}{\ttp}\pi^{\Lambda'}_{\lambda'}=
\bigoplus_{\Lambda''=|\Lambda-\Lambda'|
}^{{\rm min}\{\Lambda+\Lambda',2k-\Lambda-\Lambda'\}}
\pi^{\Lambda''}_{\lambda+\lambda'},
\end{equation}
which is the same as the fusion rules (\ref{paraffus}) for the
parafermions, except that the identifications (\ref{fieldident}) are
not incorporated. Nevertheless, using the formulae (\ref{ttpids}) for
the truncated tensor product of the $U_{q_1}(sl(2))$-representations,
on can see that it is impossible to distinguish particles in
representations that correspond to the same parafermion sector by
means of these fusion rules alone. In other words, it is consistent
with these fusion rules to declare that particles in the
representations $\pi^{\Lambda}_{\lambda},\pi^{\Lambda}_{\lambda+2k}$
and $\pi^{k-\Lambda}_{\lambda\pm k}$ are indistinguishable, just as it
is consistent with the fusion rules of a CFT to declare all
descendants of a field indistinguishable.

Now let us look at the braiding of
$\mathcal{A}_{q_1,q_2}$-representations. To describe this braiding, we
can use the bases that we introduced for $U_{q_1}(sl(2))$ in section
(\ref{braidsixjsec}), because the representations of
$\CC\ZZ_{4k,q_2}$ are one-dimensional. The matrices that describe
the braiding w.r.t. these bases will be the product of the matrices
for $U_{q_1}(sl(2))$ that we gave in (\ref{braid6jeq}) with the powers
of $q_2$ that we get from the $R$-matrix (\ref{bosrmat}) for
$\CC\ZZ_{4k,q_2}$. Using the symmetries (\ref{x6jsyms}) of the
truncated $6j$-symbols, one may then check that, when one changes the
representations which represent CFT-sectors in a way which is
consistent with the quantum group's fusion rules, the elements of the
braiding matrices will at most get minus signs. Again, this situation
is similar to the situation for the chiral boson that we discussed in
section \ref{bossec}.

We are now left with the difficulty of choosing the quantum group
representations which should represent the fields $\Phi^{0}_{2}$ and
$\Phi^{1}_{1}$, which are important for the description of electrons
and quasiholes in the $RR$-states. We will use the representations
$\pi^{0}_{2}$ and $\pi^{1}_{1}$ for this (rather than for example
$\pi^{k}_{k-2}$ and $\pi^{1}_{k+1}$). This choice keeps comparison to
the CFT-picture easy and it gives good results. Also, it gives results
which are consistent with those of the quantum group for the coset
$\widehat{sl(k)}_{1}\times\widehat{sl(k)}_{1}/\widehat{sl(k)}_{2}$,
for which there is a one-to-one correspondence between quantum group
representations and CFT-sectors. 

\subsubsection{The quantum group for
$\widehat{sl(k)}_{1}\times\widehat{sl(k)}_{1}/\widehat{sl(k)}_{2}$}
\label{qgslksec}

For the coset
$\widehat{sl(k)}_{1}\times\widehat{sl(k)}_{1}/\widehat{sl(k)}_{2}$, we
do not have a factorisation formula like (\ref{paraffus}) and
therefore we cannot find the braiding matrices for this coset by the
method we used for $\widehat{sl(2)}_{k}/\widehat{U(1)}_{k}$ in the previous section
(cf. formula (\ref{braidfaceq})). Still, the results of the previous
section and also the fusion rules and modular properties
\footnote{For modular properties of cosets, one can consult for
example \cite{gepner89},\cite{dsm}. The relationship between modular
and braiding properties of CFTs and quantum groups is clarified in
\cite{ags90}}
of $\widehat{sl(k)}_{1}\times\widehat{sl(k)}_{1}/\widehat{sl(k)}_{2}$
suggest a natural candidate for a quantum group related to this coset:
the quantum group $U_{q_3}(sl(k))\otimes U_{q_3}(sl(k))\otimes
U_{q_4}(sl(k))$ with $q_3=e^{i\pi/(k+1)}$ and
$q_4=e^{-i\pi/{k+2}}=(q_1)^{-1}$. The irreducible representations of
this quantum group are tensor products of those of the factors and
hence they are labelled by two $\widehat{sl(k)}_1$-weights and an
$\widehat{sl(k)}_2$-weight, just like the fields
$\Phi^{\mu_1,\mu_2}_{\mu}$ of section \ref{slkcossec}. Let us thus
denote the quantum group irreps as $\pi^{\mu_1,\mu_2}_{\mu}$. We are
now in the same situation that we encountered in the case of the coset
$\widehat{sl(2)}_{k}/\widehat{U(1)}_{k}$; we have several quantum group
representations per CFT-sector, because the quantum group does not
take the identifications (\ref{slkidents}) into account. However, in
this case, there is a subset of representations of the quantum group
which closes under fusion and which contains exactly one
representation for each CFT-sector. In fact, there are two such
subsets: the set of $\pi^{\mu_1,\mu_2}_{\mu}$ with $\mu_1=0$ and the
set of $\pi^{\mu_1,\mu_2}_{\mu}$ with $\mu_2=0$. If we restrict to one
of these sets (clearly, it does not matter which of the two we use),
then we are effectively forgetting about one of the
$U_{q_3}(sl(k))$-factors of the quantum group and hence we may say
that the quantum group for the coset
$\widehat{sl(k)}_{1}\times\widehat{sl(k)}_{1}/\widehat{sl(k)}_{2}$ is
$\mathcal{B}_{q_3,q_4}:=U_{q_3}(sl(k))\otimes U_{q_4}(sl(k))$. The
irreducible representations of this quantum group are labelled by an
$\widehat{sl(k)}_1$-weight $\mu_1$ and an $\widehat{sl(k)}_2$-weight
$\mu$. For the irreps that are relevant to the description of the
coset CFT, the $\widehat{sl(k)}_1$-weight is uniquely determined by
the $\widehat{sl(k)}_2$-weight through the branching rule
(\ref{slkbranch}), so that we may choose to label the relevant irreps
by just a single $\widehat{sl(k)}_2$-weight $\mu$. We may write these
representations $\pi_{\mu}$ and they are in one-to one correspondence
with the fields $\Phi_{\mu}$ we defined in section
(\ref{slkcossec}). Clearly, the fusion rules of the $\pi_{\mu}$ are
the same as those of the $\Phi_{\mu}$; they are identical to the
fusion rules for the corresponding $\widehat{sl(k)}_{2}$-fields or
$U_{q2}(sl(k))$-representations. We have not checked if all the
braiding representations we get from the quantum group
$U_{q_3}(sl(k))\otimes U_{q_4}(sl(k))$ are equivalent to those one
gets from the quantum group $\mathcal{A}_{q_1,q_2}$ of the previous
section, but we do know this for the representations that are related
to the quasiholes of the $RR$-states. In section \ref{qgsl2sec}, the
quasihole was represented by the irrep $\pi^{1}_{1}$ of
$\mathcal{A}_{q_1,q_2}$, whereas here, it must clearly be represented
by the irrep $\pi_{e_1}$ of $\mathcal{B}_{q_3,q_4}$ (for the notation
$e_1$, see section \ref{slkcossec}). Explicit calculation of braiding
matrices, using formulae (\ref{heckexcmat}), (\ref{gtor}) and
(\ref{bosrmat}) shows that the braid group representations related to
these irreps are indeed equivalent. Rather than writing out all these
calculations in detail here, we will make some remarks which make this
result very plausible. First of all, the braid group representations
we get from the tensor products $(\pi^{1}_{1})^{\otimes n}$ and
$\pi_{e_1}^{\otimes n}$ have the same fusion diagram associated to
them. This guarantees for example that the representations will have a
similar structure (see the discussion at the end of section
\ref{uqslmsec}) and in particular that their dimensions are
equal. Second, the eigenvalues of the matrices that represent the
fundamental exchanges may be easily found if we note that the
representation matrices of the canonical Hecke algebra generators
always have eigenvalues $-1$ and $q$ (this follows directly from the
last relation in (\ref{halgebra})). Thus, if we denote the eigenvalues
of the braiding for $\pi^{1}_{1}$ by $\alpha_1$ and $\alpha_2$, then we
have, using (\ref{gtor}) and (\ref{bosrmat}):

\begin{eqnarray}
\alpha_1&=& (q_1)^{-3/4} q_1 q_2 =e^{\frac{-i\pi}{k(k+2)}} \nonumber\\
\alpha_2&=& (q_1)^{-3/4}(-1) q_2 = -e^{\frac{-i(2k+1)\pi}{k(k+2)}}
\end{eqnarray}
On the other hand, the eigenvalues $\beta_1$ and $\beta_2$ for the braiding
associated with $\pi_{e_1}$ can be found using (\ref{gtor}) and this yields

\begin{eqnarray}
\beta_1 &=& (q_4)^{-\frac{k+1}{2k}}(-1)(q_2)^{-\frac{k+1}{2k}}q_2=
-e^{\frac{-i(2k+1)\pi}{k(k+2)}} = \alpha_2 \nonumber\\
\beta_2 &=& (q_4)^{-\frac{k+1}{2k}}(-1)(q_2)^{-\frac{k+1}{2k}}(-1)=
e^{\frac{-i\pi}{k(k+2)}} = \alpha_1
\end{eqnarray}
so that the eigenvalues of the braidings are equal, as they should be.

\section{Quantum group picture and braiding for the Read-Rezayi states}
\label{rrbraidsec}

In this section, we will describe the Read-Rezayi states as systems of
point particles with a hidden quantum group symmetry. We also give an
explicit description of braiding representations that are associated
with these states. All of this will be done in subsection
\ref{qgrrsec}. In section \ref{tensbraidsec}, we will give an
alternative description of the resulting braid group representations,
which does not make any explicit use of quantum groups. In this
description, it is also somewhat easier to change the number of
quasiholes in the system. Finally, in section \ref{naywilrepro},
we show how the results a of Nayak and Wilczek \cite{naywil} for the
Pfaffian state arise as a special case.

\subsection{RR-states and hidden quantum group symmetry}
\label{qgrrsec}

In section \ref{rrdefsec}, we described the RR-states as conformal
blocks in a CFT which was the tensor product of the parafermion CFT
and a CFT for a chiral boson.  In section \ref{qgparaf}, we derived a
relation between the parafermion CFT and the quantum groups
$\mathcal{A}_{q_1,q_2}=U_{q_1}(sl(2))\otimes\CC\ZZ_{4k,q_2}$ and
$\mathcal{B}_{q_3,q_4}=U_{q_3}(sl(k))\otimes U_{q_4}(sl(k))$. In
section \ref{bossec}, we gave a quantum group for the chiral
boson. Clearly, we can thus make a quantum group which will describe
fusion and braiding for the Read-Rezayi states by tensoring the
parafermion and boson quantum groups. However, since the extra boson
factor does not affect the fusion of the relevant representations and
only adds some scalar factors to the braiding matrices, we will choose
to work with $\mathcal{A}_{q_1,q_2}$ and $\mathcal{B}_{q_3,q_4}$ and
to add the scalar factors by hand.  Thus, we see the following picture
of the RR-states emerge.  The RR-system of electrons and quasiholes can
be seen as a system of point particles with hidden quantum group
symmetry (cf. section \ref{hidsymsec}). The electrons, which were
represented by the operator $\psi=\Phi^{0}_{2}=\Phi_{2e_1}$ in the
CFT-picture, are now point particles which carry the representation
$\pi^{0}_{2}$ of $\mathcal{A}_{q_1,q_2}$ or the representation
$\pi_{2e_1}$ of $\mathcal{B}_{q_3,q_4}$.  Similarly, the quasiholes,
which used to be represented by the field
$\sigma=\Phi^{1}_{1}=\Phi_{e_1}$, now carry the representation
$\pi^{1}_{1}$ of $\mathcal{A}_{q_1,q_2}$ or the representation
$\pi_{e_1}$ of $\mathcal{B}_{q_3,q_4}$. The state of an RR-system with
$N$ electrons and $n$ quasiholes may then be described as a vector
in the tensor product of $N$ $\pi^{0}_{2}$ (or $\pi_{2e_1}$) modules
and $n$ $\pi^{1}_{1}$ (or $\pi_{e_1}$) modules. However, not all of
the vectors in this tensor product correspond to physical
states. First of all, we have to restrict to the states in a truncated
tensor product with a given bracketing, as explained in sections
\ref{ttpsec} and \ref{coassec}. Second, there is a restriction that
comes from the fact that the conformal block in (\ref{rrwf}) vanishes
unless all the fields that appear in it fuse into the vacuum
sector. This now means that the system as a whole is in one of the
$\mathcal{A}_{q_1,q_2}$-representations $\pi^{0}_{0},\pi^{k}_{k}$ or
in the $\mathcal{B}_{q_3,q_4}$-representation $\pi_{0}$. Thus, the
physical states in the tensor product are those that lie in a
truncated tensor product and are in a global quantum group
representation that corresponds to the CFT's vacuum sector. The second
condition has the same consequence as the corresponding condition for
the CFT; one has to have $N+n$ equal to zero modulo $k,$ because
otherwise there are no states that fulfill this condition. The
condition can be interpreted as saying that the incorporation of more
electrons in an RR-system and the creation of quasiholes in such a
system are $\mathcal{A}$-charge preserving processes. It follows as in
the CFT-picture that quasiholes may only be created in multiples of
$k$ at a time (if the number of electrons is kept fixed).

Now let us look at the braiding of electrons and quasiholes. For
convenience, we will do this in terms of $\mathcal{A}_{q_1,q_2}$, but
the treatment in terms of $\mathcal{B}_{q_3,q_4}$ will give equivalent
results (see the discussion at the end of section \ref{qgslksec}).
Since the representation $\pi^{0}_{2}$ is one dimensional, the
braiding between electrons is abelian. This means any exchange of
electrons will just give a phase factor. To find this factor, one may
use the formulae (\ref{univrmat}) and (\ref{repfrms}) for the
universal $R$-matrix and for the representations of $U_{q_1}(sl(2))$,
the analogous formulae (\ref{bosrmat}) and (\ref{bosreps}) for the
universal $R$-matrix of $\CC\ZZ_{4k,q_2}$, and the explicit factors
from the boson vertex operators which appear in the expression
(\ref{rrwf}) for the wave function. All this together just gives a
factor of $-1,$ as is appropriate for fermions.  Similarly, one may
show that there is no nonabelian braiding between electrons and
quasiholes and that the braiding factor for electron-quasihole
exchanges is equal to one. Hence, as far as the braiding is concerned,
the electrons and quasiholes can be treated separately. Since only the
quasiholes have nonabelian braiding, we will from now on focus on these.

The braiding associated to a system of identical particles with hidden
quantum group symmetry can be elegantly described in terms of a basis
that is labelled by the paths on the fusion diagram of the quantum
group representation carried by the particles (we described this in
detail ins sections \ref{braidsixjsec} to \ref{braiddiagsec}).  The
quasiholes of the RR-states carry the
$\mathcal{A}_{q_1,q_2}$-representation $\pi^{1}_{1}$ and the fusion
diagram for this representation is the same as that for the
parafermionic field $\Phi^{1}_{1}$, which is in turn the same as the
fusion diagram for the spin-$\frac{1}{2}$-representation of
$U_{q_1}(sl(2))$. The braiding representations associated to this
$U_{q}(sl(2))$-representation were described in detail in section
\ref{braiddiagsec} (and they were also included in the material of
section \ref{heckesec}). The only difference between the braid
representations described there and the braiding for the RR-quasiholes
lie in a scalar factor for every exchange, which comes from the
$\CC\ZZ_{4k,q_2}$ part of $\mathcal{A}_{q_1,q_2}$ and from the
explicit factors in the wave function (\ref{rrwf}). Thus, a basis for
the space of states with $n$ quasiholes in fixed positions is
labelled by the paths on the fusion diagram of figure \ref{sl2fusdiag}
which start at the point $(0,0)$ and which end at the point
$(n,\Lambda)$, where $\Lambda=-N~{\rm mod}~k$, so that the
$\mathcal{A}_{q_1,q_2}$-charge of the whole system corresponds to the
vacuum sector of the CFT. We will call the braid group representation
on this space $\rho_{n}^{\Lambda}$. In this representation, the braid
group generator $\tau_m$ will only mix paths which differ from each
other only at the $m^{\rm th}$ node. Given any path $p$, there will be
at most one path $p'$ which differs from $p$ only at the $m^{\rm th}$
node and we will call this the partner path of $p$ at node $m$. If a
path does not have a partner path at node $m$, then the action of
$\tau_m$ on this path will be multiplication by a scalar factor. To
give this factor, let us first take $q=e^{2\pi i/(k+2)}$, so that we
have

\begin{equation}
q_1=q,~~~~q_2=q^{-\frac{1}{2}-\frac{1}{k}}. 
\end{equation}
The path will then get a factor of $-q^{-1+\frac{1-M}{2(kM+2)}}$ if it
changes direction at the $m^{\rm th}$ node and a factor of
$q^{\frac{1-M}{2(kM+2)}}$ otherwise.  When $M$ takes its lowest
physical value of $1$, these factors reduce to $-q^{-1}$ and $1$
respectively. If a path does have a partner path at its $m^{\rm th}$
node, then the path and its partner path will have the same
representations at nodes $m-1$ and $m+1$ and these representations
will have the same dimension. Let us call this dimension $d$. The
exchange $\tau_m$ will then act on the vector space generated by the
path and its partner path through the matrix

\begin{equation}
\rho_{n}^{\Lambda}(\tau_m) \equiv
\frac{-q^{-\frac{1}{2}+\frac{1-M}{2(kM+2)}}}{\qnr{d}}\left(
\begin{array}{cc}
q^{-d/2} & \sqrt{\qnr{d+1}\qnr{d-1}} \\
\sqrt{\qnr{d+1}\qnr{d-1}}& -q^{d/2}
\end{array} \right).
\end{equation}
Here we have ordered the basis so that the first of the basis vectors
corresponds to the path with the highest representation at node $m$ of
the diagram. The matrix above is symmetric and unitary and which
should be compared with (\ref{sl2taummat}). The braid group
representations $\rho_{n}^{\Lambda}$ are irreducible by the same
arguments as those we used for the braid group representations
associated with $U_{q}(sl(2))$. Information on their dimensions has
been gathered in section \ref{telsec}.

\subsection{Tensor product description of the braiding}
\label{tensbraidsec}

We will now set up an alternative description for the braid group
representations $\rho_{n}^{\Lambda}$ of the previous section. In this
description, the representation spaces are seen as subspaces of
$n$-fold tensor product spaces. This makes it somewhat closer in
spirit to the description Nayak and Wilczek have given of the braiding
for the Pfaffian state \cite{naywil}, a fact we will utilise in
section \ref{naywilrepro}. We also feel that the description of this
section is useful in itself, because it shows very clearly how
braidings in systems with an arbitrary number of quasiholes can be
performed by a recipe that depends very little on this number.

Let us start by defining $V_{k,l}$ to be the $k$-dimensional vector
space spanned by orthonormal vectors which represent the possible
$l^{\rm th}$ steps in a path on the fusion diagram of figure
\ref{sl2fusdiag}. We write:

\begin{equation}
V_{k,l}= \left\{ 
\begin{array}{ll}
{\rm Span}\{v_{0,1},v_{2,3},\ldots,v_{k-2,k-1},
v_{2,1},v_{4,3},\ldots,v_{k,k-1}\}& 
(k {\rm ~even~},l{\rm ~odd}) \\
{\rm Span}\{v_{1,2},v_{3,4},\ldots,v_{k-1,k},
v_{1,0},v_{3,2},\ldots,v_{k-1,k-2}\} & 
(k {\rm ~even~},l{\rm ~even}) \\
{\rm Span}\{v_{0,1},v_{2,3},\ldots,v_{k-1,k},
v_{2,1},v_{4,3},\ldots,v_{k-1,k-2}\} & 
(k {\rm ~odd~},l{\rm ~odd}) \\
{\rm Span}\{v_{1,2},v_{3,4},\ldots,v_{k-2,k-1},
v_{1,0},v_{3,2},\ldots,v_{k,k-1}\} & 
(k {\rm ~odd~},l{\rm ~even}) 
\end{array}\right.
\end{equation}
Here the indices on each basis vector represent the weights at the
starting points and end points of the piece of path represented by the
vector.  In order to simplify the description, we have also, for $l\le k$,
included some vectors which do not actually correspond to bits of path
in the fusion diagram of figure \ref{sl2fusdiag} (for example the vector
$v_{2,1}$ at $l=1$). Clearly, any continuous path of length $n$
through the fusion diagram may be represented by a canonical basis
vector of the ``domino'' form $v_{\Lambda_1,\Lambda_2} \otimes
v_{\Lambda_2,\Lambda_3} \otimes v_{\Lambda_3,\Lambda_4} \otimes \ldots
\otimes v_{\Lambda_{n-1},\Lambda_n}$ in the tensor product space
$V_{k,1} \otimes V_{k,2} \otimes \ldots \otimes V_{k,n}.$ The paths in
the representation space of $\rho^{\Lambda}_{n}$ can be isolated by
requiring $\Lambda_0=0$ and $\Lambda_{n}=\Lambda$.

We can now define a matrix representation $\Gamma_{k,n}$ of the
relations (\ref{braidgroup}) on the given tensor product space which
has a very simple form and which reduces to the braid group
representation $\rho^{\Lambda}_{n}$ when one restricts to the subspace
of the tensor product which corresponds to the paths in
$\rho^{\Lambda}_{n}.$ The action of the $\tau_i$ is defined as
follows. $\Gamma_{k,n}(\tau_i)$ is always of the form
$1\otimes\ldots\otimes 1\otimes R_{k,i}\otimes 1\otimes\ldots\otimes
1,$ where the matrix $R_{k,i}$ acts only on $V_{k,i}\otimes
V_{k,i+1}.$ On the basis vectors which are of domino form in the
$i^{\rm th}$ and $(i+1)^{\rm th}$ factor, we take

\begin{eqnarray}
\label{brmats}
R_{k,i} v_{\Lambda_i,\Lambda_i+1} \otimes v_{\Lambda_i+1,\Lambda_i+2}
&=& \alpha v_{\Lambda_i,\Lambda_i+1} \otimes
v_{\Lambda_i+1,\Lambda_i+2} 
\nonumber \\
R_{k,i}v_{\Lambda_i,\Lambda_i-1} \otimes v_{\Lambda_i-1,\Lambda_i-2}
&=& \alpha v_{\Lambda_i,\Lambda_i-1} \otimes
v_{\Lambda_i-1,\Lambda_i-2} 
\nonumber \\
R_{k,i}v_{\Lambda_i,\Lambda_i+1} \otimes v_{\Lambda_i+1,\Lambda_i} &=&
\frac{-\alpha q^{-\frac{\Lambda_i}{2}-1}}
{\qnr{\Lambda_i+1}}v_{\Lambda_i,\Lambda_i+1} \otimes
v_{\Lambda_i+1,\Lambda_i} - 
\nonumber \\ 
~&~&
\frac{\alpha q^{-\frac{1}{2}}
\sqrt{\qnr{\Lambda_i+2}\qnr{\Lambda_i}}}{\qnr{\Lambda_i+1}}
v_{\Lambda_i,\Lambda_i+1} \otimes v_{\Lambda_i+1,\Lambda_i} 
~~~(1 \le \Lambda_{i} \le k-1) \rule[-10mm]{0mm}{5mm} 
\nonumber \\ 
R_{k,i} v_{\Lambda_i,\Lambda_i-1} \otimes v_{\Lambda_i-1,\Lambda_i}
&=& \frac{\alpha q^{\frac{\Lambda_i}{2}}}{\qnr{\Lambda_i+1}}
v_{\Lambda_i,\Lambda_i-1} \otimes v_{\Lambda_i-1,\Lambda_i} -
\nonumber \\ 
~&~&
\frac{\alpha q^{-\frac{1}{2}}
\sqrt{\qnr{\Lambda_i+2}\qnr{\Lambda_i}}}{\qnr{\Lambda_i+1}}
v_{\Lambda_i,\Lambda_i+1} \otimes v_{\Lambda_i+1,\Lambda_i} 
~~~(1 \le \Lambda_{i} \le k-1) 
\nonumber \\
R_{k,i} v_{\Lambda_i,\Lambda_i+1} \otimes v_{\Lambda_i+1,\Lambda_i}
&=& -\alpha q^{-1} 
v_{\Lambda_i,\Lambda_i+1} \otimes v_{\Lambda_i+1,\Lambda_i}
~~~(\Lambda_{i}=0) 
\nonumber \\ 
R_{k,i} v_{\Lambda_i,\Lambda_i-1} \otimes v_{\Lambda_i-1,\Lambda_i}
&=& -\alpha q^{-1}
v_{\Lambda_i,\Lambda_i-1} \otimes v_{\Lambda_i-1,\Lambda_i}
~~~(\Lambda_{i}=k)
\end{eqnarray}
where we have defined 

\begin{equation}
\alpha=q^{\frac{1-M}{2(kM+2)}}
\end{equation}
This factor of course reduces to $1$ when $M=1$.  For basis
vectors $v$ which are not of the domino form in $V_{k,i}\otimes
V_{k,i+1},$ we define $R_{k,i}v=0.$ With this definition, the matrices
in $\Gamma_{k,n}$ satisfy the relations (\ref{braidgroup}), but they
are not invertible (of course they are invertible if we restrict to
the space generated by vectors of the domino form). Alternatively, one
may set $R_{k,i}v=v.$ In that case, the matrices in $\Gamma_{k,n}$ are
invertible, but they no longer satisfy the second relation in
(\ref{braidgroup}) (of course, they still do satisfy this relation on
the ``domino state space''). It is not difficult to see that, on the
set of vectors which corresponds to the paths of $\rho^{\Lambda}_{n},$
the representation defined here indeed reduces to
$\rho^{\Lambda}_{n}.$ The matrices $R_{k,i},$ for given $k,$ depend
only on the parity of $i,$ which means that knowledge of $R_{k,1}$ and
$R_{k,2}$ is enough to determine the representation $\Gamma_{k,n}$ and
hence also all the $\rho^{\Lambda}_{n}$ completely. This is quite
useful, because it gives us an easy way to go from a description of
$n$ particles to a description of $n+1$ particles; we just add another
tensor factor and use the same matrices $R_{k,1}$ and $R_{k,2}$ as
before to implement particle exchanges. We hope that this explicit
recipe can be a small first step towards a second quantised
description of particles with nonabelian statistics.

\subsection{Reproducing the results for the Pfaffian state}
\label{naywilrepro}

Now let us check that our results reproduce those of Nayak and Wilczek
\cite{naywil} for $k=2$, $n=2m$ even, $N$ even and $M=1$. In this
case, the relevant paths on the fusion diagram have to end at the
coordinates $(0,2m)$ in case $m$ is even and at the coordinates
$(2,2m)$ in case $m$ is odd. The fusion diagram for $k=2$ is given in
figure \ref{kistwofig}

\begin{figure}[h,t,b]

\begin{picture}(400,90)(-40,-20)
\put(-10,-10){\line(0,1){70}}
\multiput(-10,0)(0,20){3}{\line(1,0){5}}
\put(-20,0){\footnotesize 0}
\put(-20,20){\footnotesize 1}
\put(-20,40){\footnotesize 2}
\put(-43,50){\mbox{$\Lambda \uparrow$}}

\put(-10,-10){\line(1,0){260}}
\multiput(0,-10)(20,0){12}{\line(0,1){5}}
\put(0,-20){\footnotesize 0}
\put(20,-20){\footnotesize 1}
\put(40,-20){\footnotesize 2}
\put(60,-20){\footnotesize 3}
\put(80,-20){\footnotesize 4}
\put(235,-20){\mbox{$\lambda \rightarrow$}}

\multiput(0,0)(40,0){5}{\vector(1,1){20}}
\multiput(20,20)(40,0){5}{\vector(1,1){20}}
\multiput(20,20)(40,0){5}{\vector(1,-1){20}}
\multiput(40,40)(40,0){4}{\vector(1,-1){20}}
\end{picture}
\caption{\footnotesize fusion diagram for the quasiholes at $k=2$. The
diagram must be thought extended indefinitely in the
$\lambda$-direction}
\label{kistwofig}
\end{figure}
Each of these paths can be uniquely
characterised by stating whether or not it changes direction at each of
its odd numbered vertices. If $m$ is even, then the paths have to
change direction an even number of times in order to end up at the
point $(0,2m).$ If $m$ is odd then the number of changes of direction
also has to be odd in order for the path to end up at the point
$(2,2m).$ Thus, for $m$ even, we may represent any path of length $2m$
by a ket $\ket{s_1,s_2,\ldots s_m},$ where each of the $s_i$ is a
sign, a plus sign denoting a change of direction and a minus sign no
change. The physically relevant paths are then the paths for which the
product of all these signs is a plus sign. For $m$ odd, we may do the
same, but now with a minus sign denoting a change of direction and a
plus sign denoting no change. The relevant states are then once more
the ones whose overall sign is positive. Both for $m$ odd and for $m$
even, we thus describe a $2^{m-1}$ dimensional space whose basis
vectors are labelled in the same way as those of Nayak and
Wilczek. Just as Nayak and Wilczek have done, we will interpret this
space as a subspace of an $m$-fold tensor product of two dimensional
spaces, each of which has basis $\{\ket{+},\ket{-}\}.$ Now let us
check that the action of the braiding matrices on these states is also
the same as in \cite{naywil}.

For $k=2,$ the tensor product $V_{k,1}\otimes V_{k,2}$ contains four
states with the domino property: the states $v_{0,1}\otimes
v_{1,0},v_{0,1}\otimes v_{1,2},v_{2,1}\otimes v_{1,2}$ and
$v_{2,1}\otimes v_{1,0}.$ Using these as an ordered basis of relevant
states, the matrix $R_{k,1}$ can now be found by filling in
(\ref{brmats}). It is given by

\begin{equation}
R_{k,1}= \left( 
\begin{array}{cccc}
i&0&0&0\\
0&1&0&0\\
0&0&i&0\\
0&0&0&1
\end{array}\right)
\end{equation}
From this, we may read off that the action of the braid group
generator $\tau_{2l+1}$ on the sign states $\ket{s_1,\ldots,s_m}$ is
given by

\begin{equation}
\tau_{2l+1} \ket{s_1,\ldots,s_m} =
\left\{ 
\begin{array}{ll}
1 \ket{s_1,\ldots,s_m} & (s_{2l+1}=m+1 {\rm~mod~2})\\
i \ket{s_1,\ldots,s_m} & (s_{2l+1}=m  {\rm~mod~2})
\end{array}
\right.
\end{equation}
In this equation, we let the value $+$ of the symbol $s_{2l+1}$
correspond to $0 {\rm ~mod~} 2$ and we let the value $-$ correspond to
$1 {\rm ~mod~} 2.$ We see that $\tau_{2l+1}$ acts only on the
$(2l+1)^{\rm th}$ factor of the tensor product of sign spaces and on this
factor it is given by the following $2\times 2$ matrix:

\begin{equation}
\label{oddbr}
\tau_{2l+1}\equiv \left\{ 
\begin{array}{ll}
\left(
\begin{array}{cc}
1&0 \\
0&i \\
\end{array}
\right) & (m=1 {\rm~mod~2}) \\
\left(
\begin{array}{cc}
i&0 \\
0&1 \\
\end{array}
\right) & (m=0 {\rm~mod~2}) 
\end{array}
\right.
\end{equation}
In Nayak and Wilcek's work, the action of $\tau_{l+1}$ also
corresponds to the action of a diagonal matrix in the $(l+1)^{\rm th}$
tensor product factor. In this case, the matrix does not depend on $m$
and it is given by (cf. (\ref{nwmats}))

\begin{equation}
\tau^{NW}_{2l+1}\equiv e^{i\frac{\pi}{4}}e^{\frac{i\pi}{4}\sigma_3}=
\left(
\begin{array}{cc}
i&0 \\
0&1 \\
\end{array}
\right)
\end{equation}
Here, $\sigma_3$ denotes the third Pauli matrix. We see that the
Nayak-Wilczek matrix is the same as ours, up to a change in the order
of the basis when $m$ is odd.
 
The tensor product $V_{k,2}\otimes V_{k,3}$ contains only two states
with the domino property: the states $v_{1,0}\otimes v_{0,1}$ and
$v_{0,1}\otimes v_{1,0}.$ Using this as an ordered basis for the
relevant states, the matrix $R_{k,2}$ can again be found from
(\ref{brmats}) and is given by

\begin{equation}
R_{k,2}= \left( 
\begin{array}{cc}
\frac{1+i}{2}&\frac{-1+i}{2}\\
\frac{-1+i}{2}&\frac{1+i}{2}
\end{array}\right)
\end{equation}
From this, we may read off the action of the braid group
generator $\tau_{2l}$ on the states $\ket{s_1,\ldots,s_m}.$ This
generator acts only on the $(2l)^{\rm th}$ and $(2l+1)^{\rm th}$ tensor
factors of the sign space and on those it is given by the matrix:

\begin{equation}
\label{evenbr}
\tau_{2l} \equiv
\left( 
\begin{array}{cccc}
\frac{1+i}{2}&0&0&\frac{-1+i}{2} \\
0&\frac{1+i}{2}&\frac{-1+i}{2}&0 \\
0&\frac{-1+i}{2}&\frac{1+i}{2}&0 \\
\frac{-1+i}{2}&0&0&\frac{1+i}{2}
\end{array}
\right)
\end{equation}
where the basis on which this matrix acts, is
$\{\ket{++},\ket{+-},\ket{-+},\ket{--}\}.$ The matrix above is
identical to Nayak and Wilczek's matrix for $\tau_{2l}$, which we gave
in (\ref{nwmats}). Hence, we have reproduced Nayak and Wilczek's
result.  Note that the change of order in the basis which was needed
for $\tau_{2l+1}$ when $m$ is odd has no effect on the matrix for
$\tau_{2l}$, which is why the result does not depend on the parity of
$m$ this time.

\section{Discussion and outlook}
\label{discosec}

In this paper, we have shown how quantum groups may be used to give an
algebraic description of the braiding and fusion properties of the
excitations of nonabelian quantum Hall systems. Due to the
well-established relationship between conformal field theory and
quantum groups, it is in principle be possible to find such a
description for any quantum Hall state that has a CFT-description. As
an application, we obtained the explicit braiding matrices for the
quasihole excitations over the Read-Rezayi series of states. In a
special case, these reduce to the matrices given by Nayak and Wilczek
in \cite{naywil}, as they should.

The obvious question to ask is now whether one can somehow make
predictions about physical quantities from the results we have
derived. The answer to this depends very much on what quantities one
considers as physical. For example, one may fairly easily calculate
amplitudes for Aharonov-Bohm scattering of quasiholes from the
braiding matrices we have given, but it seems unlikely that the
control over quasiholes that one needs to test these will soon be
reached in experiments. One would probably have better chances of
making contact with experiment if one could find effects of the
nonabelian braiding in some transport properties of the quantum Hall
state. To be able to make predictions about such quantities, one would
most probably need to have a better understanding of the relation
between the overcomplete set of states with localised quasiholes
which we deal with in this article and a basis of the Hilbert space of
the quantum Hall state. Suitable bases of the Hilbert space for the
Read-Rezayi-states are constructed in \cite{gurrez,kjs} and a logical
next step in the program of understanding the consequences of
nonabelian braiding in quantum Hall states would thus be to express
the states we deal with in this article in terms of these bases and
vice versa. Another important question is whether there is some
intuitive way of understanding which features of the underlying theory
cause the quantum symmetry exhibited by the effective theories at the
plateaus. If such an intuitive picture could be found it would
probably be very helpful in extracting physics from the effective
theories.

There are also some questions of a more mathematical nature which
arise naturally from our work. For example, one would like to
generalise the way we associated quantum groups to coset CFTs to more
general cosets than the ones we considered. A first generalisation
would be to look at the generalisations of the parafermions that were
defined by Gepner \cite{gepner87}. We expect that most arguments we
gave for the parafermions will go through unchanged for these
theories. In connection with this, there should be identities like
(\ref{x6jsyms}) for the ``$6j$-symbols'' of quantum universal
enveloping algebras more general than $U_{q}(sl(2))$; one identity for
each external automorphism of the corresponding Affine Lie
algebra. Another interesting mathematical issue is whether the groups
generated by the braiding matrices we have found are finite and/or can
be characterised in a nice way. For the case studied by Nayak and
Wilczek, this is certainly so: the group generated by the fundamental
exchange matrices is just a double cover of the permutation group in
this case. However, for the general case we have not yet obtained
interesting results. \vspace*{3mm}

\noindent{\large \bf Acknowledgements} 

\noindent The authors wish to thank Eddy Ardonne, Robbert Dijkgraaf,
J\"urgen Fuchs, Kareljan Schoutens and Christoph Schweigert for illuminating
discussions and comments. We thank Tom Koornwinder for useful remarks
on some of the quantum group theory we used.

\end{document}